\documentclass[journal,twocolumn,10pt]{IEEEtran}
\usepackage{amsmath,epsfig}
\usepackage{amssymb}
\usepackage{booktabs}
\usepackage{amsthm}
\usepackage{graphicx}
\usepackage{caption}
\usepackage{subcaption}
\usepackage{rotating}
\usepackage{floatflt} 
\usepackage{paralist} 
\usepackage{algorithm} 
\usepackage{xcolor}
\usepackage{color}
\usepackage{epstopdf}
\usepackage[normalem]{ulem}
\usepackage{float}
\usepackage{todonotes}
\usepackage{mathtools}
\usepackage{fancyhdr}
\usepackage{hyperref}

\theoremstyle{definition}
\newtheorem{defn}{Definition}
\newcommand{\mypar}[1]{{\bf #1.}}

\newtheorem{myThm}{Theorem}
\newtheorem{myCorollary}{Corollary}

\newcommand{\R}{\ensuremath{\mathbb{R}}}

\DeclareMathOperator{\Id}{I}

\def\a{\mathbf{a}}

\def\x{\mathbf{x}}

\def\g{\mathbf{g}}
\def\h{\mathbf{h}}

\def\uu{\mathbf{u}}
\def\vv{\mathbf{v}}
\def\w{\mathbf{w}}

\def\V{\mathcal{V}}
\def\E{\mathcal{E}}

\def\Z{\mathbb{Z} }

\DeclareMathOperator{\LL}{L}

\DeclareMathOperator{\PC}{PC}

\DeclareMathOperator{\PBL}{PBL}

\DeclareMathOperator{\Adj}{A}

\DeclareMathOperator{\D}{D}

\DeclareMathOperator{\Vm}{V}

\DeclareMathOperator{\W}{W}

\DeclareMathOperator{\one}{\bf{1}}

\begin{document}
\title{ Multiresolution Representations for Piecewise-Smooth Signals  on Graphs}
\author{Siheng~Chen, Aarti Singh,   Jelena~Kova\v{c}evi\'c
}
 \maketitle


\begin{abstract}
What is a mathematically rigorous way to describe the taxi-pickup distribution in Manhattan, or the profile information in online social networks?  A deep understanding of representing those data not only provides insights to the data properties, but also benefits to many subsequent processing procedures, such as denoising, sampling, recovery and localization. In this paper, we model those complex and irregular data as piecewise-smooth graph signals and propose a graph dictionary to effectively represent those graph signals. We first propose the graph multiresolution analysis, which provides a principle to design good representations. We then propose a coarse-to-fine approach, which iteratively partitions a graph into two subgraphs until we reach individual nodes. This approach efficiently implements the graph multiresolution analysis and the induced graph dictionary promotes sparse representations piecewise-smooth graph signals. Finally, we validate the proposed graph dictionary on two tasks: approximation and localization. The empirical results show that  the proposed graph dictionary outperforms eight other representation methods on six datasets, including traffic networks, social networks  and point cloud meshes.
\end{abstract}

\section{Introduction}
\label{sec:intro}
Today's data is being generated from a diversity of sources, all
residing on complex and irregular structures; examples include profile
information in social networks, stimuli in brain connectivity networks
and traffic flow in city street networks~\cite{Jackson:08,
  Newman:10}. The need for understanding and analyzing such complex
data has led to the birth of signal processing on
graphs~\cite{ShumanNFOV:13,SandryhailaM:13}, which generalizes
classical signal processing tools to data supported on graphs; the
data is the graph signal indexed by the nodes of the underlying graph.

\begin{figure}[htb]
  \begin{center}
    \begin{tabular}{cc}
 \includegraphics[width=0.35\columnwidth]{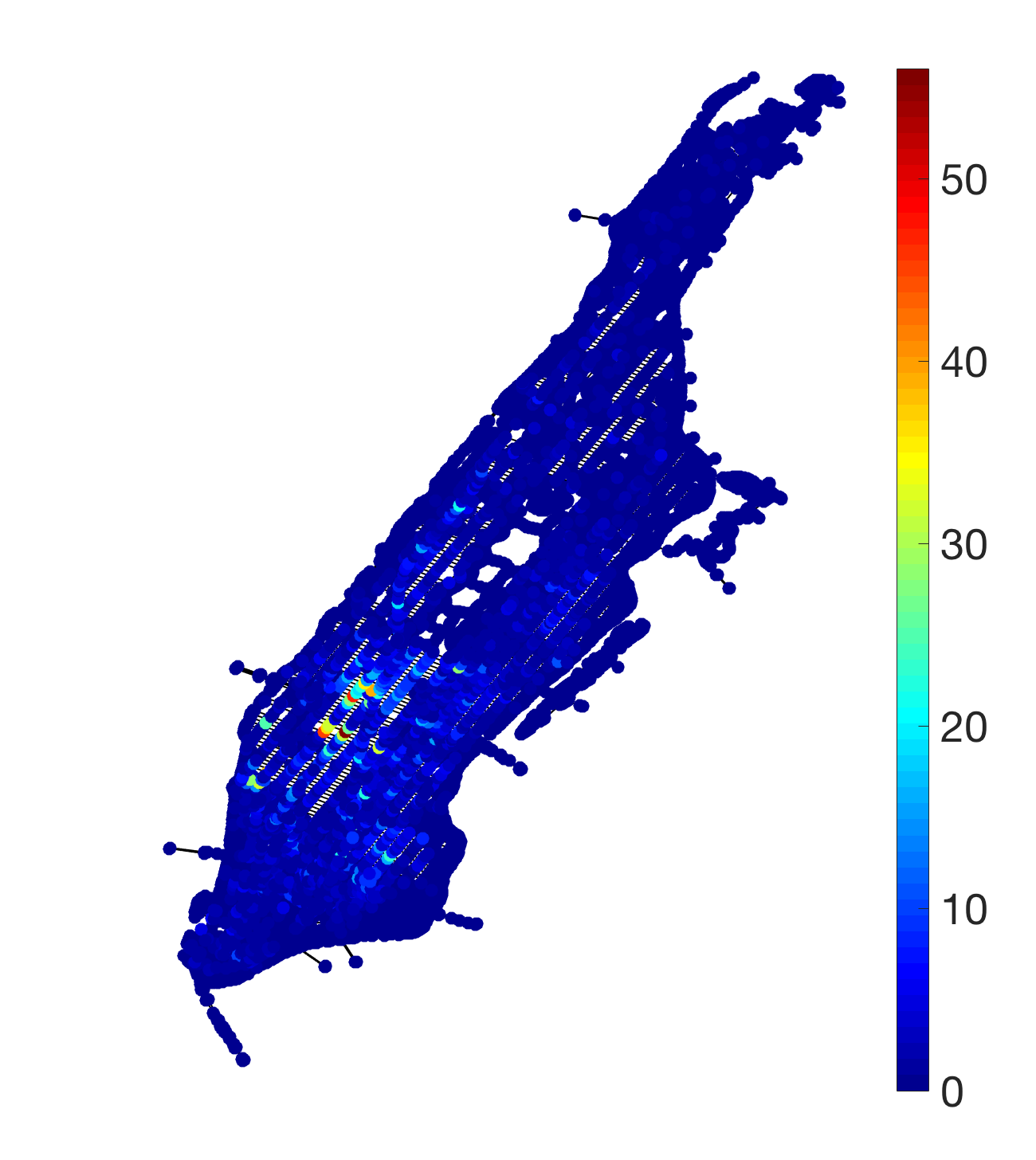} &
 \includegraphics[width=0.35\columnwidth]{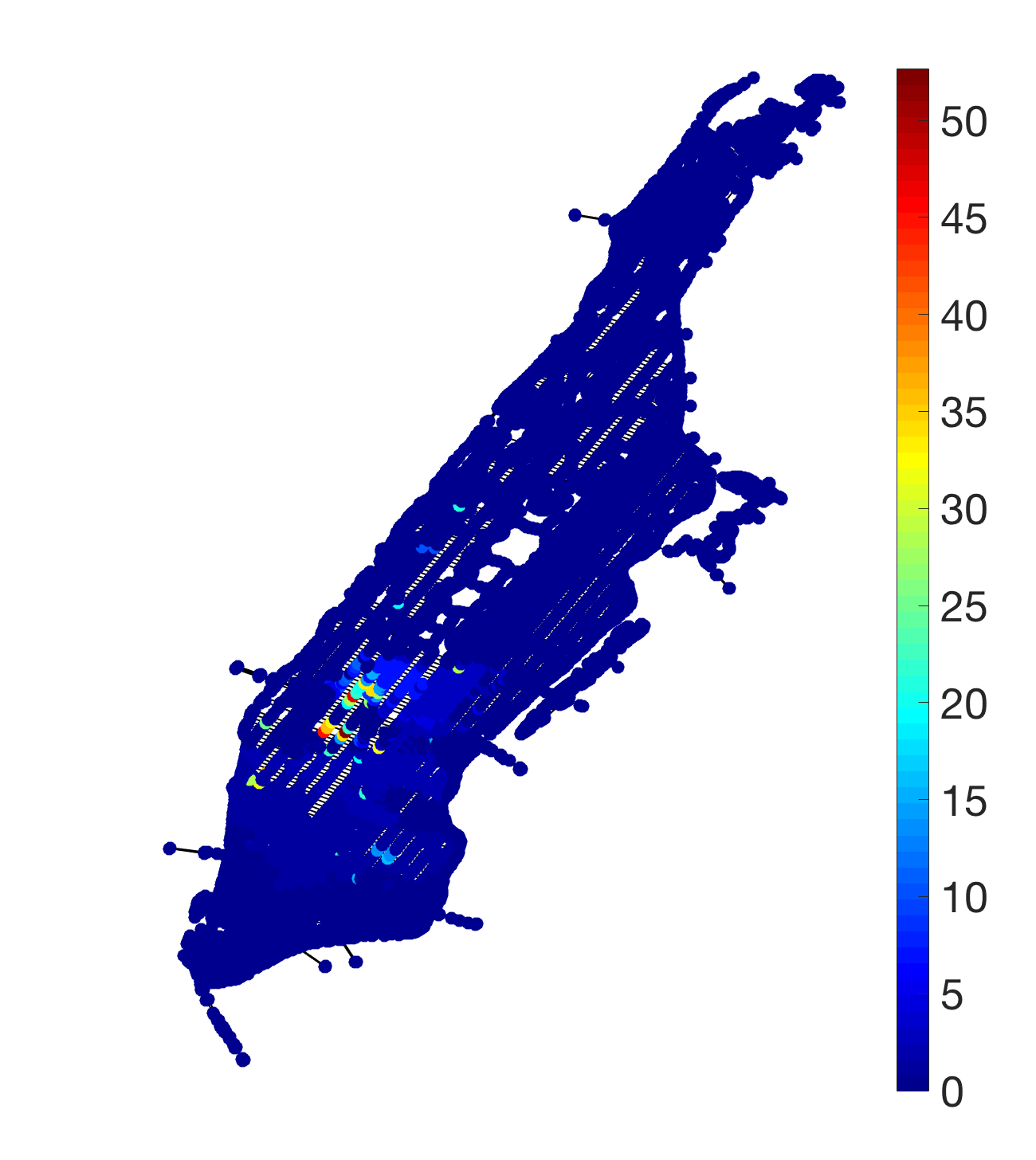} 
      \\
      {\small (a) Taxi-pickup distribution} &  {\small (b) Piecewise-smooth}
            \\
   {\small  in Manhattan (13,679 intersections).}    &  {\small  approximation (50 coefficients).} 
      \\
  \includegraphics[width=0.35\columnwidth]{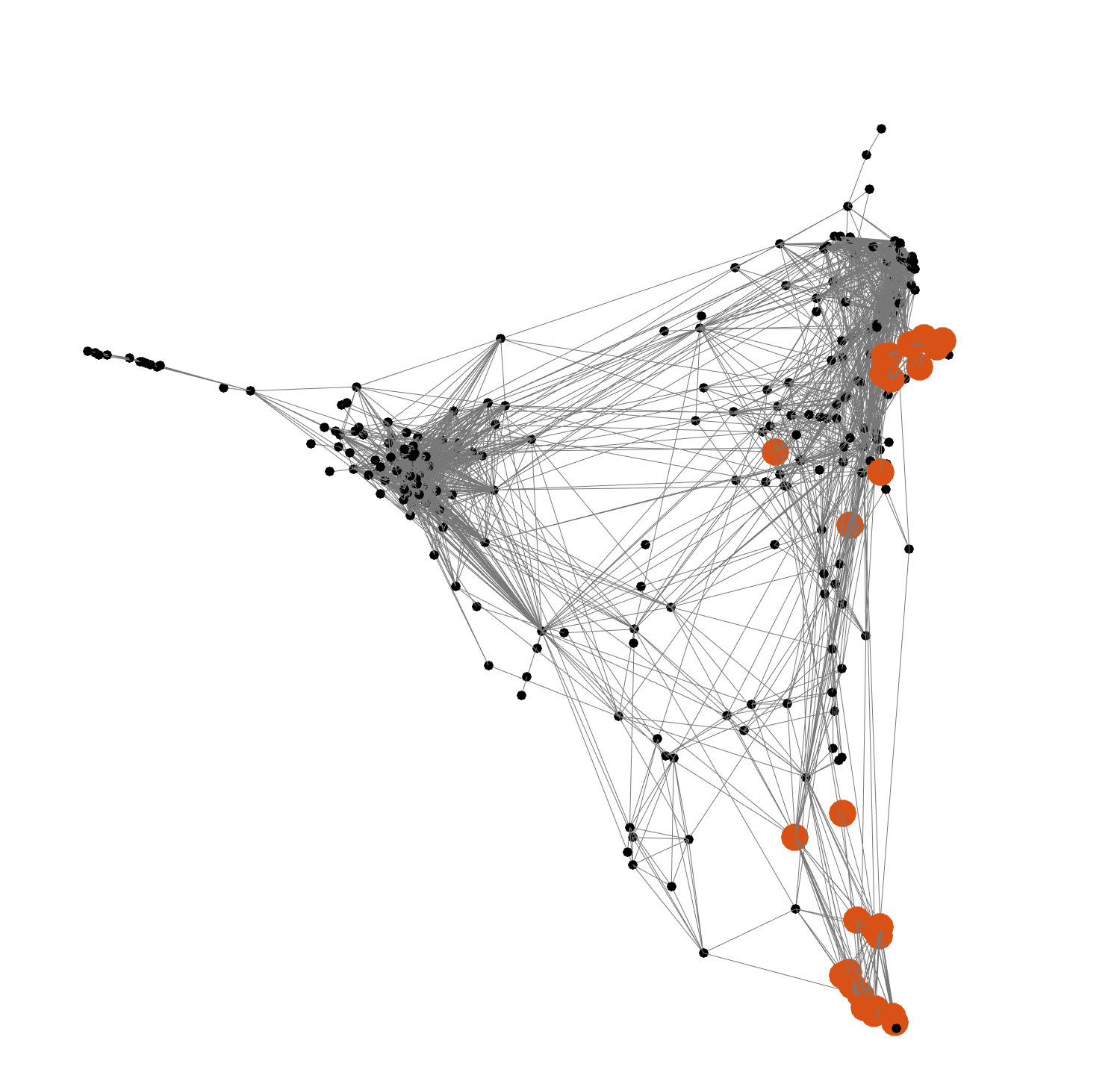} &
  \includegraphics[width=0.35\columnwidth]{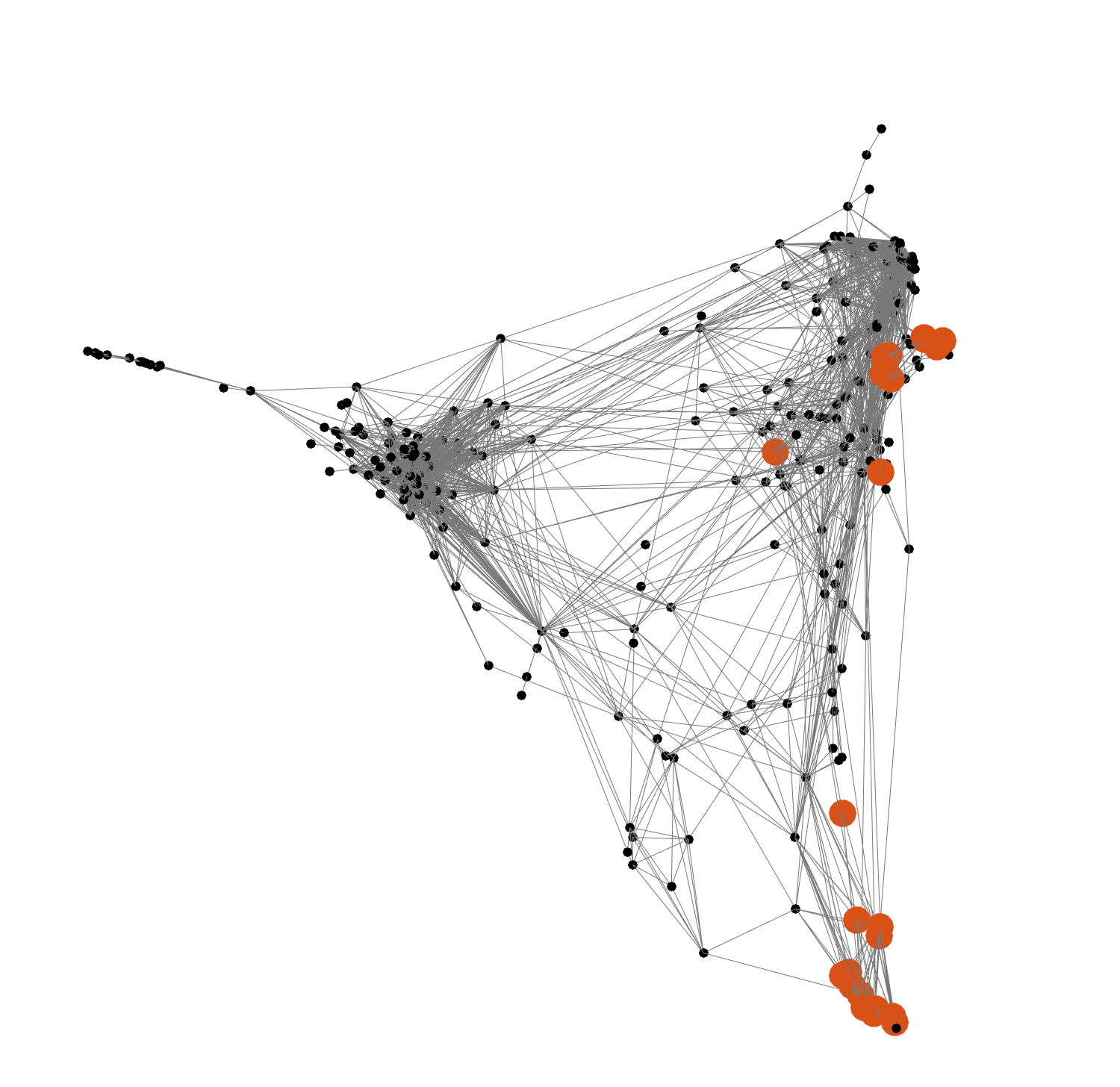} 
      \\
      {\small (c) Profile information}  & 
      {\small (d) Piecewise-smooth} 
      \\
     {\small  on Facebook (277 users).}  &  {\small  approximation (5 coefficients).} 
    \end{tabular}
  \end{center}
  \caption{\label{fig:toy_pc} Piecewise-smooth graph signals approximate irregular, nonsmooth graph signals by capturing both large variations at boundaries as well as small variations within pieces.}
\end{figure}

\mypar{Modeling real-world data using piecewise-smooth graph signals}
In urban settings, the intersections around shopping areas will
exhibit homogeneous mobility patterns and life-style behaviors, while
the intersections around residential areas will exhibit different, yet
still homogeneous mobility patterns and life-style
behaviors. Similarly, in social networks, within a given social circle
users' profiles tend to be homogeneous, while within a different
social circle they will be different, yet still homogeneous. We can
model data generated from both cases as piecewise-smooth graph
signals, as they capture large variations between pieces and small
variations within pieces.  Figure~\ref{fig:toy_pc} illustrates how a
piecewise-smooth signal model can be used to approximate the
taxi-pickup distribution in Manhattan and users' profile information
on Facebook (hard thresholding is applied for better visualization).

The piecewise-smooth signal model has been intensely studied and
widely used in classical signal processing, image processing and
computer graphics~\cite{PrandoniV:99,
  WakinRCB:06,ChandrasekaranWBB:09}. Multiresolution analysis and
splines are standard representation tools to analyze piecewise-smooth
signals~\cite{Unser:99}.  The idea of using piecewise-smooth graph
signals is not novel either; in~\cite{ShumanNFOV:13}, the authors show
that graph wavelets can capture discontinuities in a piecewise-smooth
graph signal and in~\cite{WangSST:15}, the authors proposed denoising
for piecewise-polynomial graph signals through minimizing a
generalized total-variation term. There are two gaps in the previous
literature we address here: (1) define piecewise-smooth graph signals
precisely and find appropriate representations and (2) provide
theoretical results on sparse representations for piecewise-smooth
graph signals.





\mypar{Representations for piecewise-smooth graph signals} Signal
representations are at the heart of most signal processing
techniques~\cite{VetterliKG:12}, allowing for targeted signal models
for tasks such as denoising, compression, sampling, recovery and
detection~\cite{Mallat:09}.  Our aim in this paper is to find an
appropriate and efficient approach to represent piecewise-smooth graph
signals.


We define a mathematical model for piecewise-smooth graph signals and propose a
graph dictionary to sparsely represent piecewise-smooth graph signals.
 Inspired by
classical signal processing, we generalize the idea of multiresolution
analysis to graphs as a representation tool for piecewise-smooth
signals~\cite{VetterliK:95}.  We implement the graph multiresolution
analysis  by using a
coarse-to-fine decomposition approach; that is, we iteratively partition a graph
into two connected subgraphs until we reach individual nodes. We show that the process leads
to an efficient construction of a graph wavelet basis that satisfies the graph multiresolution analysis, and the induced graph dictionary
promotes sparse representations for piecewise-smooth graph signals.
We validate the proposed graph dictionaries on two tasks:
approximation and localization. We show that the proposed graph dictionaries outperform eight
other representation methods on six graphs, including traffic
networks, citation network, social networks and point cloud meshes.

\mypar{Contributions}  The main contributions of this paper are:
\begin{itemize}
\item A novel and explicit definition for piecewise-smooth graph
  signals; see Section~\ref{sec:PS}.
\item A novel graph multiresolution analysis that is implemented by a
  coarse-to-fine decomposition approach; see Section~\ref{sec:GMRA}.
\item A novel graph dictionary that promotes the sparsity for piecewise-smooth graph signals and is well-structured and
  storage-friendly; see Section~\ref{sec:R_PS}; and
\item An extensive empirical study on a number of real-world graphs,
  including traffic networks, citation networks, social networks and
  point cloud meshes; see Section~\ref{sec:experiment}.
\end{itemize}  

\mypar{Outline of the paper} Section~\ref{sec:background} reviews the
background materials; Sections~\ref{sec:PS} defines piecewise-smooth
graph signals; Section~\ref{sec:GMRA} proposes the graph
multiresolution analysis, which provides a principled way to represent
graph signals.
Sections~\ref{sec:R_PS} show that the proposed graph dictionary
promotes sparse representations for piecewise-smooth graph signals. We
validate the proposed methods in Section~\ref{sec:experiment} and
conclude in Section~\ref{sec:conclusions}.

\section{Background}
\label{sec:background}

We briefly introduce the framework of graph signal processing. We then overview related works on graph
signal representation.

\subsection{Graph Signal Processing}
\label{sec:GSP}
\mypar{Graphs} Let $\mathcal{G} = (\V,\E, \Adj)$ be an undirected,
irregular and non-negative weighted graph, where $\V = \{v_i \}_{i=1}^N$ is the set
of nodes (vertices), $\E = \{e_i \}_{i=1}^E$ is the set of weighted
edges and $\Adj \in \R^{N \times N}$ is the weighted adjacency matrix
whose element $\Adj_{i,j}$ is the edge weight between the $i$th and
the $j$th nodes.  Let $\D \in \R^{N \times N}$ be a diagonal degree
matrix with $\D_{i,i} = \sum_j \Adj_{i,j}$. The graph Laplacian matrix
is $\LL = \D - \Adj \in \R^{N \times N}$, which is a second-order
difference operator on graphs~\cite{BelkinN:03}. Let $\Delta \in \R^{E
  \times N}$ be the graph incidence matrix; its rows correspond to
edges. If $e_i$ is an edge that connects the $j$th node to the $k$th
node ($j < k$), the elements of the $i$th row of $\Delta$ are
\begin{equation*}
  \label{eq:Delta}
  \Delta_{i, \ell} = 
  \left\{ 
    \begin{array}{rl}
      -\sqrt{ \Adj_{j,k} }, & \ell = j;\\
      \sqrt{ \Adj_{j,k} }, & \ell = k;\\
      0, & \mbox{otherwise}.
  \end{array} \right.
\end{equation*}
The graph incidence matrix measures the first-order difference and
$\Delta^T \Delta = \LL$.

\mypar{Graph signals} Given a fixed ordering of nodes, we assign a
signal coefficient to each node; a~\emph{graph signal} is then defined
as a vector,
\begin{displaymath}
  \x = [x_1,x_2,\cdots,x_N]^T \in \R^N,
\end{displaymath}
with $x_n$ the signal coefficient corresponding to the node $v_n$.  

We say that the graph is \emph{smooth} when adjacent nodes have
similar signal coefficients~\cite{ShumanNFOV:13, SandryhailaM:14}.  
\begin{itemize}
\item \emph{Smoothness in the vertex domain.} Consider $\Delta \x \in
  \R^E$ as an~\emph{edge signal} representing the first-order difference
  of $\x$. The $i$th element of $\Delta \x$, $\left( \Delta \x
  \right)_i = \sqrt{ \Adj_{j,k} } \left( \x_k - \x_j \right)$, assigns
  the difference between two adjacent signal coefficients to the $i$th
  edge, which connects the $j$th node to the $k$th node ($j <
  k$). When the variation $\left\| \Delta \x \right\|_2^2 = \x^T \LL
  \x$ is small, the differences are small and $\x$ is smooth.

\item \emph{Smoothness in the spectral domain.} We typically call this
  type of smoothness bandlimitedness~\cite{AnisGO:14,ChenVSK:15}. Let
  the graph Fourier basis $\Vm \in \R^{N \times N}$ be the eigenvector
  matrix of $\LL$\footnote{The graph Fourier basis can also be
    defined based on adjacency matrix~\cite{SandryhailaM:13}.}, with
  $\LL = \Vm \Lambda \Vm^T$ and  the diagonal elements of
  $\Lambda$ are arranged in ascending order.  The graph spectrum is
  then $\widehat{\x} = \Vm^T \x \in \R^N$.  Let $\Vm_{(K)} \in \R^{N
    \times K}$ be the first $K$ columns of $\Vm$, which span the
  lowpass bandlimited subspace.  For $K \ll N$, when ${ \left\|
      \Vm_{(K)}^T \x \right\|_2^2}/{ \left\| \x \right\|_2^2} = 1$,
  the energy concentrates in the lowpass band and $\x$ is smooth.
\end{itemize}

In practice, graph signals may not satisfy the smoothness constraint
as above (as shown in Figure~\ref{fig:toy_pc}), which serves as
motivation to further develop graph signal models and tools to
represent them, the topic of this paper.

\subsection{Related Works}
Ideally, a good representation should be efficient, have some
structure such as orthogonality and promote sparse representations for
graph signals (at least in some subspaces). To deal with large-scale
graphs, we may also need the representation itself to be sparse and
storage-friendly.  We categorize the previous work on graph signal
representations as follows:\footnote{The categorization follows
  from~\url{https://www.macalester.edu/~dshuman1/Talks/Shuman_GSP_2016.pdf}.}.

\subsubsection{Graph Filter Banks} Here, representations are
constructed by generalizing classical filter banks to
graphs. \cite{NarangO:12,NarangO:13} designs critically-sampled filter
banks via bipartite subgraph decomposition;~\cite{EkambaramFAR:13,
  KotzagiannidisD:16} design critically-sampled filter banks for
circulant graphs;~\cite{TanakaS:14} designs oversampled filter
banks;~\cite{ShumanFV:16} designs iterative filter
banks;~\cite{TremblayB:16} designs critically-sampled filter banks via
community detection; and ~\cite{JinS:17} designs each channel via
sampling and recovery.

\subsubsection{Graph Vertex-Domain Designs} Here, representations are
constructed by designing each basis vector (atom) in the graph vertex
domain. \cite{CrovellaK:03} designs spatial wavelets via
neighborhoods; and~\cite{SzlamMCB:05, GavishNC:10,Rustamov:11,
  IrionS:14} considers coarse-to-fine approaches.
 
\subsubsection{Graph Spectral-Domain Designs} Here, representations
are constructed by designing graph filters in the graph spectral
domain.~\cite{HammondVG:11} designs graph
wavelets;~\cite{LeonardiV:13,ShumanWHV:15} design tight frames;
and~\cite{ShumanRV:16} designs the windowed graph Fourier transforms
by generalizing translation.

\subsubsection{Graph Diffusion-Based Designs} Here, representations
are constructed based on the polynomial of the transition matrix.
\cite{CoifmanM:06} designs diffusion wavelets; and
~\cite{BremerCMS:06} designs diffusion wavelet packets;

\subsubsection{Graph Dictionary Learning} Here, representations are
constructed by learning from the given graph signals. The
representations in the other branches depend on the graph structure
only; \cite{ZhangDF:12, ThanouSF:14} learn graph dictionaries that
provide smoothness for given graph signals, which is adaptive and
biased to the observed graph signals.

In this paper, we consider connecting graph filter banks and
graph vertex domain designs. Similarly to~\cite{SzlamMCB:05,
  GavishNC:10,Rustamov:11, IrionS:14,TremblayB:16}, the proposed
representation considers the coarse-to-fine decomposition in the
graph vertex domain.  Our goal is to implement the graph
multiresolution analysis, where the coarse-to-fine approach is more
efficient and straightforward than the local-to-global approach (graph
filter banks). We further show that the proposed representation is
efficient, orthogonal and storage-friendly; it also satisfies the
graph multiresolution analysis and promotes the sparsity for
piecewise-constant and piecewise-smooth graph signals.

The representations of smooth graph signals have been thoroughly studied in the graph spectral domain~\cite{ZhuR:12, RicaudSV:13}. In this paper, we emphasize the representations of piecewise-smooth graph signals in the graph vertex domain. As a continuous counterpart of graph signal representations, some other works study on manifold data representations~\cite{AllardCM:17, AllardCM:12, LiaoM:16}.

\section{Graph Signal Models}
\label{sec:PS}
Piecewise-smooth graph signals can model a number of real-world cases
as they capture large variations between pieces and small variations
within pieces.  In this section, we mathematically define
piecewise-smooth graph signals. We start with piecewise-constant graph
signals, an important subclass, and then extend it to piecewise-smooth
graph signals.

\subsection{Piecewise-Constant Model}
In classical signal processing, a piecewise-constant signal is a
signal that is locally constant over connected regions separated by
lower-dimensional boundaries. Such a signal is often related to step
functions, square waves and Haar wavelets and is widely used in image
processing~\cite{VetterliKG:12}. Piecewise-constant graph signals have
been used in many applications without having been explicitly defined;
for example, in community detection, community labels form a
piecewise-constant graph signal for a given social network and in
semi-supervised learning, classification labels form a
piecewise-constant graph signal for a graph constructed from the
dataset. While smooth graph signals emphasize slow transitions,
piecewise-constant graph signals emphasize fast transitions
(corresponding to boundaries) and localization in the vertex domain
(corresponding to signals being nonzero in a local neighborhood).

We define a piecewise-constant graph signal through the concept of a piece that has been implicitly used before~\cite{Luxburg:07, WangLG:14}.  The definition is
intuitive; a piecewise-constant graph signal partitions a graph into
several pieces; within each piece, signal coefficients are constant.
\begin{defn}
  \label{df:localset}
  Let $S$ be a subset of the node set $\V$. We call $S$ a
  \emph{piece} when its corresponding subgraph $G_{S}$ is connected.
\end{defn}
We can represent a piece $S$ by using a one-piece graph signal, ${\bf 1}_{S} \in \R^N$. A piecewise-constant graph signal is a linear combination of several one-piece graph signals.
\begin{defn}
  \label{df:pc_gen}
  Let $\{S_c\}_{c=1}^C$ be a partition of the node set $\V$, where each $S_c$ is a piece. A graph
  signal $\x \in \R^N$ is piecewise-constant with $C$ pieces when
  \begin{equation*}
    \x \ = \  \sum_{c=1}^C a_c {\bf 1}_{S_c},
  \end{equation*}
  where and $a_c \in \R$ is
  the piece coefficient for the piece $S_c$. Denote this class by $\PC(C)$.
\end{defn}

For most graph signals, two adjacent signal coefficients are typically the same; that is, $\left\| \Delta \x  \right\|_0$ may be close to the number of edges $E$.  For a piecewise-constant graph signal $\x$ with a few number of pieces, however, $
\left\| \Delta \x \right\|_0$ is usually small. Within a piece, it is
0 while across pieces,  $\left\| \Delta \x
\right\|_0$ is the cut cost to separate the pieces. For example,
in an unweighted graph, 
\begin{displaymath}
  \left\| \Delta \x  \right\|_0 \leq \#~\text{edges~across~pieces~} \{S_c\}_{c=1}^C,
\end{displaymath}
for all $\x \in \PC(C)$,
where the equality is achieved when all $a_c$ are different. Thus, $\left\| \Delta \x  \right\|_0 \ll E$ when $C \ll N$; see a quick summary in Table~\ref{tab:property}.

\begin{table}[htbp]
  \footnotesize
  \begin{center}
    \begin{tabular}{@{}lll@{}}
      \toprule
 Graph signal models  & \multicolumn{2}{c}{Properties} \\
   & $\left\| \Delta \x \right\|_0$ &  $\left\| \Delta \x \right\|_2$  \\
      \midrule \addlinespace[1mm]
 Arbitrary graph signal &    $O(E)$  &  large \\
 Smooth graph signal &   $O(E)$ &   small  \\
 Piecewise-constant graph signal &   $O(1)$ & large \\
\bottomrule
\end{tabular} 
\caption{\label{tab:property} Property summary of some typical graph signals. }
\vspace{-5mm}
\end{center}
\end{table}

\subsection{Piecewise-Smooth Model}
\label{sec:R_PS}
Piecewise-smooth signals are widely used to represent images, where
edges are captured by the piece boundaries and smooth content is
captured by the pieces themselves. Piecewise-smooth graph signals
arise naturally from piecewise-constant graph signals with more
flexibility to model real-world data, such as taxi-pickup distribution supported
on the city-street networks and 3D point cloud information supported
on the meshes.

We define a piecewise-smooth graph signal as a generalization of a
piecewise-constant graph signal. For a piecewise-constant signal,
signal coefficients within a piece are constant; for a
piecewise-smooth signal, signal coefficients within a piece form a
smooth graph signal over that piece.

Let $S$ be a piece, $G_S$ be the corresponding subgraph, $\Vm_{S} \in \R^{|S| \times |S|}$ be the corresponding graph Fourier basis. Given a graph signal $\x \in \R^N$,  $\x_S \in \R^{|S|}$ denotes  the signal coefficients supported on $G_S$ and $\x_{\bar{S}} \in \R^{N-|S|}$ denotes the rest signal coefficients.

\begin{defn}
  \label{def:loc_and_band}
  A graph signal $\x \in \R^N$ is localized and bandlimited over the piece $S$ with
  bandwidth $K (K \leq |S|)$ when $\x_{\bar{S}} \ = \ 0$ and
  \begin{equation*}
    {\Vm_{S}}_{(K)} {\Vm_{S}}_{(K)}^T  \x_{S} \ = \ \x_{S}  \in \R^{|S|}, 
  \end{equation*}
  where ${\Vm_{S}}_{(K)} \in \R^{|S| \times K} $ contains the first $K$
  columns of the graph Fourier basis $\Vm_{S}$.
\end{defn}
Definition~\ref{def:loc_and_band} shows a class of graph signals that is localized in both the vertex and the graph spectral domains. Since these signals are bandlimited over a piece, we consider them lowpass and smooth with in this piece.  A similar definition has also been proposed in~\cite{TsitsveroBL:15}; the difference is that the bandlimitedness in~\cite{TsitsveroBL:15} is defined for the entire graph, while the bandlimitedness in Definition~\ref{def:loc_and_band} is defined for a subgraph only. We then consider piecewise-bandlimited graph signals as a linear combination of  localized and bandlimited graph signals.
\begin{defn}
  A graph signal $\x \in \R^N$ is piecewise-bandlimited with $C$
  pieces and bandwidth $K$ when $\x \ = \ \sum_{c=1}^C \x^{(c,K)}$, where each $\{S_c\}$, $c=1, \ldots, C$ is a valid piece and
  $\x^{(c,K)} \in \R^N$ is bandlimited over the piece
  $S_c$ with bandwidth $K$. Denote this class by $\PBL(C, K)$.
\end{defn}

\section{Graph Multiresolution}
\label{sec:GMRA}
Having defined a piecewise-smooth model for the data we are interested
in, we now embark upon looking for the appropriate
representations. Inspired by classical signal processing, we
generalize the multiresolution analysis to graph signals and propose a
coarse-to-fine approach to implement it.




\subsection{Graph Multiresolution Analysis}
\begin{defn}
  \label{df:MRA}
  \emph{A multiresolution analysis on graphs} consists of a sequence
  of embedded closed subspaces $ V^{(L)} \subset \cdots \subset
  V^{(1)} \subset V^{(0)}, $ such that
  \begin{itemize}
  \item it satisfies upward completeness, $ V^{(0)} = \R^N$;
  \item it satisfies downward completeness, $ V^{(L)} = \{ c {\bf 1}_{\V}, c \in \R
    \}$;

  \item there exists an orthonormal basis $ \{ \vv^{(\ell)}_k
    \}_{k=0}^{K^{(\ell)}-1}$ for $V^{(\ell)}$;

  \item it satisfies generalized shift invariance, that is,
    for any $\x \in V^{(\ell)}$, there exists an nontrivial permutation operator
    $\Phi \in \{0, 1\}^{N \times N}$ ($\Phi \one = \Phi^T \one= \one$)
    such that $\Phi \x \in V^{(\ell)}$. The permutation operator
    $\Phi$ only allows for swapping signal coefficients in two nonoverlapping pieces.
    
  \item it satisfies generalized scale invariance; that is,
    for any $\x \in V^{(\ell)}$, there exists an nontrivial permutation operator, $\Phi \in \{0, 1\}^{N \times N}$ such that $\Phi \x \in V^{(\ell)}$.  When the permutation operator $\Phi$ swaps signal coefficients in two nonoverlapping pieces,  each piece has at most $2^{\ell}$ nodes.
  \end{itemize}
\end{defn}

\begin{figure}[t]
  \begin{center}
    \includegraphics[width= 0.9\columnwidth]{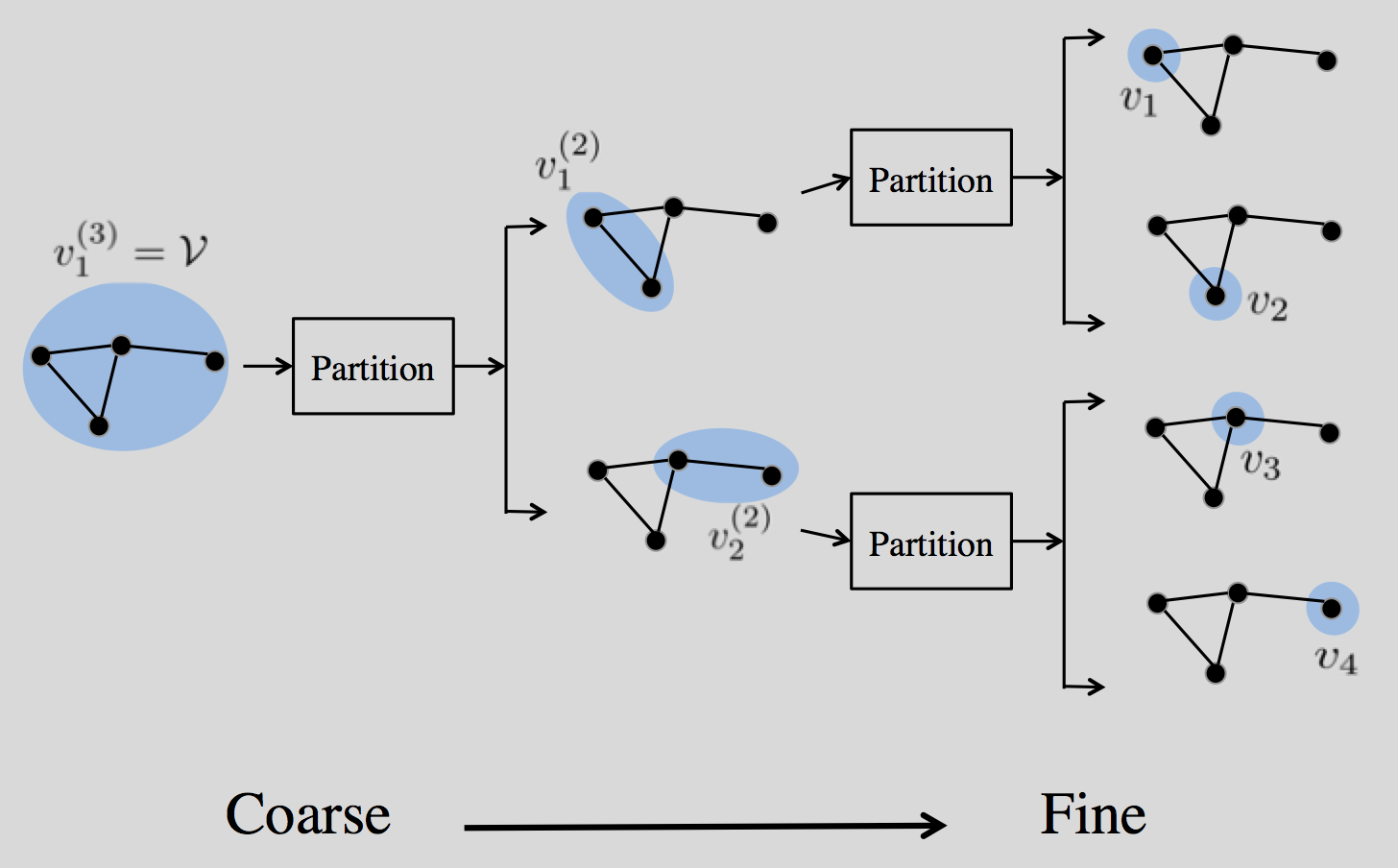}
  \end{center}
  \caption{\label{fig:decomposition} Coarse-to-fine decomposition approach. At each
    step, we partition a larger piece into two smaller disjoint
    pieces and generate a pair of lowpass/highpass basis
    vectors. Piece $v_{1}^{(3)} = \V$ is at Level 3; Pieces
    $v_{1}^{(2)}, v_{2}^{(2)}$ are at Level 2; and Pieces $v_{1}, v_{2},
    v_{3}, v_{4}$ are at Level~1.}
\end{figure}

\begin{figure}[t]
  \begin{center}
    \includegraphics[width= 0.9\columnwidth]{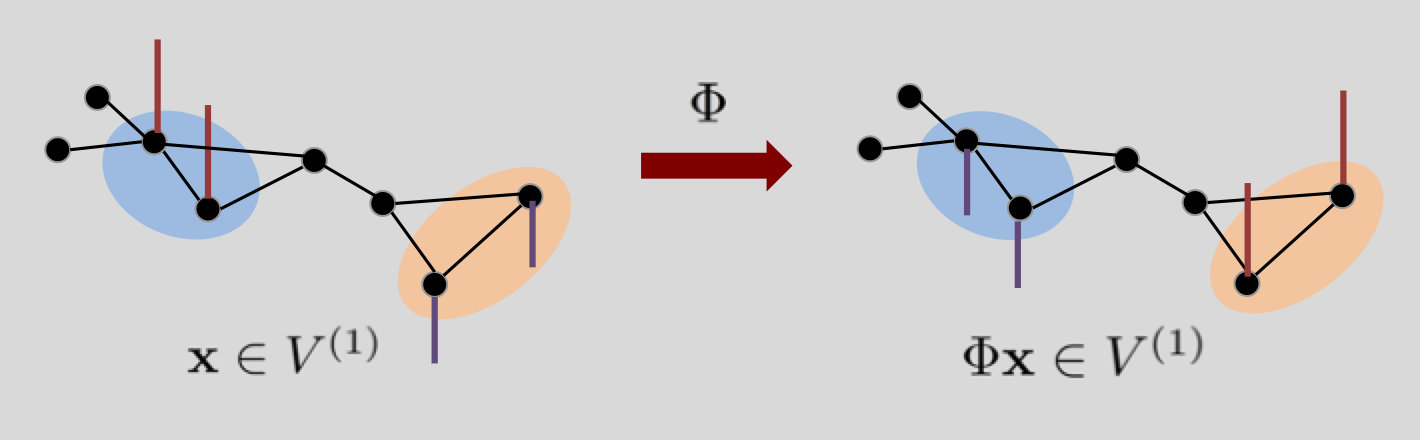}
    \\
    {\small (a) Permutation in $V^{(1)}$. We swap the signal coefficients between the blue piece and the yellow piece. The total number of swaps is 2.}
    \\
	\vspace{1mm}
        \includegraphics[width= 0.9\columnwidth]{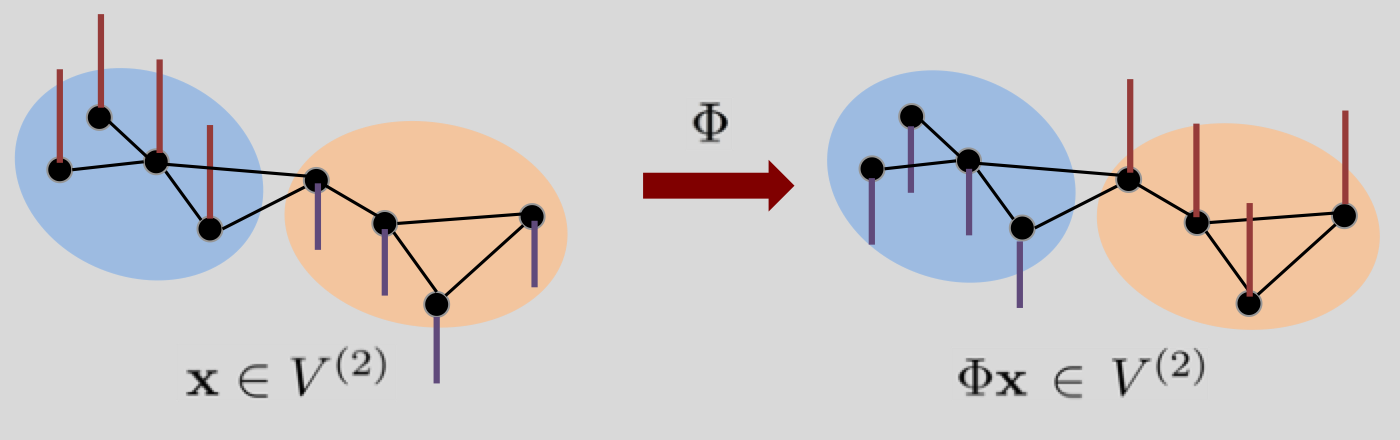}
        \\
    {\small (b) Permutation in $V^{(2)}$. We swap the signal coefficients between the blue piece and the yellow piece. The total number of swaps is 4.}
  \end{center}
  \caption{\label{fig:permutation} Permutation leads to the generalized shift and scale invariances. The permutation operator $\Phi$ shifts a graph signal  $\x \in V^{(\ell)}$ to  another graph signal $\Phi \x \in V^{(\ell)}$ by swapping signal coefficients supported on two difference pieces, which leads to the generalized shift invariance; the permutation operator needs twice as many swaps to permute a graph signal in a coarser space, which leads to the generalized scale invariance. }
\end{figure}

While similar in spirit, the proposed graph multiresolution analysis
is different from the original one~\cite{VetterliK:95}. For example,
the complete space here is $\R^N$ instead of $\mathcal{L}_2(\R)$
because of the discrete nature of the graph. We unify the shift and
scale invariance axioms via a permutation operator, which reshapes a
graph signal by swapping signal coefficients. The standard shift
invariance axiom ensures that the input signal shape is preserved
during shifting; here, this is accomplished by requiring that the
permutation operator swap the signal coefficients supported on two nonoverlapping pieces only.  The standard scale invariance axiom ensures that the input signal shape is preserved
during scaling; here, this is accomplished by requiring that the number of swaps scale exponentially as the multiresolution level $\ell$ grows; see Figure~\ref{fig:permutation} for illustration.


\vspace{-2mm}
\subsection{Coarse-to-Fine Construction}
\label{sec:global}

Our goal now is to implement the graph multiresolution analysis. In classical signal processing, this is  typically accomplished by using filter banks, which involves a series of downsampling and shifting. Filter banks start with building filters in a fine space, which captures local information, and gradually building them in coarser spaces, which captures global information.  For discrete-time signals, filter banks happen to be an efficient way to implement the multiresolution analysis because the downsampling and shifting operators follow naturally.
For graph signals, however, there is no recipe to permute the nodes; thus, it is hard to obtain efficient downsampling and shifting operators; see details in Appendix~\ref{sec:one_layer_haar}.

Instead,  we consider implementing graph multiresolution analysis
using a~\emph{coarse-to-fine approach}.  The main idea is to
recursively partition each piece into
two smaller disjoint child pieces as follows: Given a connected
graph $G_0( \V_0, \E_0, \Adj_0)$ with $|\V_0| >1$, 
partition $G_0$ into two smaller graphs $G_1( \V_1, \E_1, \Adj_1)$ and
$G_2( \V_2, \E_2, \Adj_2)$ by solving
\begin{subequations}
 \label{eq:partition}
\begin{eqnarray}
 \label{eq:objective}
  \min_{\V_1, \V_2} &&  | |\V_1| - |\V_2|  |
  \\   
  \label{eq:constraint}
  {\rm subject~to} &&  \V_1 \cap \V_2 = \emptyset,~\V_1 \cup \V_2 = \V_0,
  \\ \nonumber
  &&  G_1, G_2~{\rm are~connected}.
\end{eqnarray}
\end{subequations}
In other words, we want (1) each of the two child pieces to be
connected; and (2) the partition to be close to a bisection; that is,
the difference between cardinalities of two child pieces is as
small as possible. These properties ensure that the coarse-to-fine
approach implements the graph multiresolution analysis. We
solve~\eqref{eq:partition} in Section~\ref{sec:partition}.

We start with the coarsest lowpass subspace $V^{(0)} = \{ c {\bf 1}_{\V}, c \in \R \}$ and  partition the largest piece $\V$ into two disjoint and connected child pieces $v^{(1)}_1, v^{(1)}_2 \subseteq \V$; that is, $v^{(1)}_1 \cup v^{(1)}_2 = \V, v^{(1)}_1 \cap v^{(1)}_2 = \emptyset$, where  the  subscript denotes the index at each level. The lowpass/highpass basis vectors are, respectively,
\begin{eqnarray*}
  \vv^{(1)}_1  & = & g( v^{(1)}_{1}, v^{(1)}_{2} ), 
  \\
  \uu^{(1)}_1  & = & h( v^{(1)}_{1}, v^{(1)}_{2} ), 
\end{eqnarray*}
where 
\begin{eqnarray*}
  g(S_1, S_2) & = &  \sqrt{ \frac{S_1| |S_2|}{|S_1|+|S_2|}} \left(  \frac{ {\bf 1}_{S_1} }{|S_1|}  + \frac{ {\bf 1}_{S_2 }}{|S_2|}  \right)  \in \R^N,
  \\
  h(S_1, S_2) & = &  \sqrt{ \frac{S_1| |S_2|}{|S_1|+|S_2|}} \left(  \frac{ {\bf 1}_{S_1} }{|S_1|}  - \frac{ {\bf 1}_{S_2 }}{|S_2|}  \right)  \in \R^N,
\end{eqnarray*}
with $S_1, S_2 \subset \V$ two nonoverlapping pieces.  The normalization ensures that each basis vector is of unit norm and $\one^T \uu^{(1)}_1 = 1$. The highpass subspace is $U^{(1)} = \{ c \uu^{(1)}_1 , c \in \R \}.$

We now partition pieces $v^{(1)}_1$ and $v^{(1)}_2$ to obtain $v^{(1)}_1, v^{(2)}_2$ and $v^{(2)}_3, v^{(2)}_4$, respectively. The lowpass/highpass basis vectors are
\begin{eqnarray*}
  \vv^{(2)}_k  & = &  g( v^{(2)}_{2k-1}, v^{(2)}_{2k} ), 
  \\
  \uu^{(2)}_k  & = & h( v^{(2)}_{2k-1}, v^{(2)}_{2k} ), 
\end{eqnarray*}
for $k = 1, 2$.  The lowpass subspace is $V^{(2)} = {\rm span} \left(
  \{ \vv^{(2)}_k \}_{k=1}^{K^{(2)}} \right)$ and the highpass subspace
is $U^{(2)} = {\rm span} \left( \{ \uu^{(2)}_k \}_{k=1}^{K^{(2)}}
\right)$, where $K^{(2)} = 2$. We keep on partitioning, building the
lowpass and highpass subspaces and their corresponding bases in the
process.

At the $\ell$th level, we partition $v^{(\ell+1)}_k$ to obtain $v^{(\ell)}_{2k-1}, v^{(\ell)}_{2k}$. When both $v^{(\ell)}_{2k-1}, v^{(\ell)}_{2k}$ are nonempty,  
\begin{displaymath}
\vv^{(\ell)}_k = g( v^{(\ell)}_{2k-1}, v^{(\ell)}_{2k})
\end{displaymath}
is a lowpass basis vector and
\begin{displaymath}
  \uu^{(\ell)}_k  = h( v^{(\ell)}_{2k-1}, v^{(\ell)}_{2k})
\end{displaymath} 
is a highpass basis vector; when one of them is empty, the cardinality
of $v^{(\ell+1)}_k$ is $1$ and we cease partitioning in this
branch. At the finest resolution, each piece corresponds to an
individual node. Since we promote bisection, the total decomposition
depth $L$ is around $1+\log_2 N$.

At the end of the process, we collect all highpass basis vectors into a Haar-like graph wavelet basis (see Algorithm~\ref{alg:wavelet}). A toy example is shown in Figure~\ref{fig:decomposition}.

\begin{algorithm}[h]
  \footnotesize
  \caption{\label{alg:wavelet} Haar-like Graph Wavelet Basis
    Construction }
  \begin{tabular}{@{}lll@{}}
    \addlinespace[1mm]
   {\bf Input} 
      & $G(\V, \E, \Adj )$~~graph \\
     {\bf Output}  
      & $\W$~~~~~~~~~~~~wavelet basis \\
    \addlinespace[2mm]
    {\bf Function} & &\\
    & 1) initialize a stack of pieces sets $\mathbb{S}$ and a set of vectors $\W$ \\
    & 2) push $S = \V$ into $\mathbb{S}$ \\ 
    & 3) add $\w = \frac{1}{ \sqrt{|S|}} {\bf 1}_S$ to $\W$ \\
    & 4) while the cardinality of the largest element in $\mathbb{S}$ is larger than $1$  \\ 
    &~~4.1) pop up one element from $\mathbb{S}$ as $S$ \\
    &~~4.2) evenly partition $S$ into two disjoint pieces $S_1, S_2$ \\
    &~~4.3) push $S_1, S_2$ into $\mathbb{S}$ \\
    &~~4.4) add $\w = \sqrt{ \frac{|S_1| |S_2|}{|S_1|+|S_2|}} \left(  \frac{1}{|S_1|} {\bf 1}_{S_1} - \frac{1}{|S_2|} {\bf 1}_{S_2}\right)$ to $\W$\\
    & {\bf return} $\W$ \\  
     \addlinespace[1mm]
  \end{tabular}
\end{algorithm}

\subsection{Graph Wavelet Basis Properties}

\subsubsection{Efficiency}
This coarse-to-fine approach involves $(N-1)$ partitions.  The
overall computational complexity is approximately
$\sum_{\ell=1}^{1+\log_2 N} 2^{\ell} f({N}/{2^{\ell}})$, where $f(N)$
is the computational complexity of partitioning an $N$-node graph. For
a sparse graph ($E = O(N)$), when we use a standard graph partitioning algorithm, METIS~\cite{KarypisK:98}, to partition the graph,
$f(N) = O(N)$ and the overall computational complexity is $O(N \log_2
N)$.

\subsubsection{Graph multiresolution} 
When the number of nodes $N = 2^L$ for some $L \in \Z_{+}$, the
proposed graph wavelet basis in Algorithm~\ref{alg:wavelet} satisfies
the axioms of the graph multiresolution analysis. When the number of
nodes cannot be partitioned equally, the proposed graph wavelet basis
may not exactly satisfy the generalized shift and scale invariance axioms due
to the residual condition, but still comes close to the spirit of
multiresolution.

\subsubsection{Orthogonality}
Orthogonality also implies efficient perfect reconstruction.
\begin{myThm}
  \label{thm:basis}
  The proposed graph wavelet basis $\W \in \R^{N \times N}$ in
  Algorithm~\ref{alg:wavelet} is orthonormal; that is, for any graph
  signal $\x \in \R$, we have $ \x = \W \W^T \x.  $
\end{myThm}
The proof is given in Appendix~\ref{app:basis}.

\subsection{Graph Partition Algorithm}
\label{sec:partition}
An ideal graph partitioning results in two connected subgraphs with
the same number of nodes; however, connectivity and bisection may
conflict in practice. Many existing graph partition algorithms can be
used in graph partition.  For example, METIS provides an efficient
bisection, but does not ensure that two resulting subgraphs are
connected. In~\eqref{eq:partition}, we consider the
connectivity-first approach, as the constraints~\eqref{eq:constraint} requires that the resulting subgraphs be connected. The objective function~\eqref{eq:objective} promotes a
bisection; that is, two subgraphs have similar number of nodes. 
The optimization problem~\eqref{eq:partition} is combinatorial and we aim to obtain a suboptimal solution with certain theoretical guarantee.

To solve~\eqref{eq:partition}, we consider finding two nodes with the longest geodesic distance as two hubs and then compute the geodesic
distances from each nodes to two hubs. We rank all the nodes based on
the difference between the geodesic distances to two hubs and record
the median value. We partition the nodes according to this median
value. All the nodes falling into the median value forms the boundary
set. We further partition the boundary set to ensure connectivity and
promote bisection. The details are summarized in
Algorithm~\ref{alg:bisection}.

\begin{algorithm}[h]
  \footnotesize
  \caption{\label{alg:bisection} Graph Partition with Connectivity Guarantee }
  \begin{tabular}{@{}lll@{}}
    \addlinespace[1mm]
   {\bf Input} 
      & $G_0$~~~~~~~~original graph \\
     {\bf Output}  
      & $\V_1, \V_2$~~~~two node sets \\
    \addlinespace[2mm]
    {\bf Function} & &\\
& 1) compute the geodesic distance matrix $\D \in \R^{|\V_0| \times |\V_0|}$; \\
& 2) select $v_i, v_j \in \V_0$, such that $\D_{v_i, v_j}$ is maximized; \\
&  3) let $p$ be median value of $\D_{v_i,:} - \D_{v_j,:} $; \\
&  4) let $S_1 = \{ v | \D_{v_i, v} - \D_{v_j, v} > p \}$ and \\
&~~~boundary set $S_2 = \{ v | \D_{v_i, v} - \D_{v_j, v} = p \}$; \\
&  5) partition $S_2$ into connected components $C_1, C_2, \hdots C_M$ \\
& ~~~~~~~~~~~~~~~~~~~~~~~~~~~~~~~with $|C_1| <  |C_2| < \hdots < |C_M|$.  \\
&  6) set $q_m = |S_1 \cup C_1 \cup \hdots \cup C_m |$ for $m = 1, 2, \cdots, M$; \\
&  7) set $m^* = \arg \min_m |q_m - |\V_0|/2 |$; \\
&  8) $\V_1 = S_1 \cup C_1 \cup \hdots \cup C_{m^*}$ and $\V_2 = \V_0 \backslash  \V_1$ \\
    & end \\
    & {\bf return} $\V_1, \V_2$ \\  
     \addlinespace[1mm]
  \end{tabular}
\end{algorithm}

We can show that Algorithm~\ref{alg:bisection} provides a near-optimal solution; see Appendix~\ref{app:partition} for the proof.
 \begin{myThm}
   \label{thm:partition}
   Let $\widehat{\V}_1, \widehat{\V}_2$ be the solution given by
   Algorithm~\ref{alg:bisection}.  Then, $\widehat{\V}_1,
   \widehat{\V}_2$ is a feasible solution of the optimization
   problem~\eqref{eq:partition} and
   \begin{equation*}
     | |\widehat{\V}_1| - |\widehat{\V}_2|  | \leq 2 |C_{m^*}|,
   \end{equation*}
   where $C_{m^*}$ is the $m^*$th smallest connected component in the
   boundary set, following from the Steps $5$-$7$ in
   Algorithm~\ref{alg:bisection}.
 \end{myThm}

\section{Graph Dictionaries}
\label{sec:R_PC}
We now use the graph dictionary induced by the graph multiresolution analysis
from the previous section to represent piecewise-smooth graph
signals. As before, we start with piecewise-constant graph signals and
then generalize to the piecewise-smooth ones.

\subsection{Piecewise-Constant Graph Dictionary}
Representing piecewise-constant graph signals is difficult because the
geometry of the pieces is arbitrary. We now show that the graph
wavelet basis in Algorithm~\ref{alg:wavelet} can effectively parse the
pieces and promote the sparse representations for piecewise-constant
graph signals.

\begin{myThm} 
  \label{thm:sparse}
  Let $\W \in \R^{N \times N}$ be the graph wavelet basis in
  Algorithm~\ref{alg:wavelet}. For a piecewise-constant graph signal
  $\x \in \R^N$, 
  \begin{eqnarray*}
    \left\| \W^T \x \right\|_0 \ \leq  \ 1+   \left\|  \Delta \x \right\|_0  L.
  \end{eqnarray*}
  where $L$ is the decomposition depth.
\end{myThm}
The proof is given in Appendix~\ref{app:sparse}. Since we promote the
bisection scheme, $L$ is roughly $1+\log_2
N$. Theorem~\ref{thm:sparse} shows an upper bound on the sparsity of
graph wavelet coefficients, which depends on the cut cost $\left\|
  \Delta \x \right\|_0$ and the size of the graph.  As shown in Table~\ref{tab:property}, $\left\|
  \Delta \x \right\|_0$ is usually small when $\x$ is a piecewise-constant signal.  In~\cite{ChenYZSK:17}, we also show that this graph wavelet basis can be used to detect localized graph
signals.

We can expand the graph wavelet basis from Algorithm~\ref{alg:wavelet}
to a redundant graph dictionary, allowing for more flexibility.  Each piece $v_{k}^{(\ell)}$ obtained from the graph partition  is a column vector  (called an atom) $\one_{v_{k}^{(\ell)}}$ in the graph dictionary; we collect all the pieces at all levels to obtain a dictionary. In other words, the~\emph{piecewise-constant
  graph dictionary} is
\begin{equation}
  \label{eq:pc_dict}
  \D_{\rm PC} = \{ {\bf 1}_{v^{(\ell)}_k} \}_{\ell=1, k = 1}^{\ell= L, k = 2^\ell}.
\end{equation}
There are $2N-1$ pieces in total; thus, $\D_{\rm PC} \in \R^{N \times (2N-1)}$ and the proposed graph dictionary $\D_{\rm PC}$ contains a series of atoms with different sizes
activating different positions. Each graph wavelet basis vector in
Algorithm~\ref{alg:wavelet} can be represented as a linear combination
of two atoms in the piecewise-constant graph dictionary.

 Since most atoms are sparse, the number of nonzero
elements in the piecewise-constant dictionary is small, allowing for efficient storage. For example, when $N = 2^L$ for some $L = \Z_+$, the
number of nonzero elements is exactly $NL$.
\begin{myCorollary}
  \label{cor:sparse}
  Let $\D_{\rm PC} \in \R^{N \times (2N-1)}$ be the piecewise-constant
  graph dictionary. Let the sparse coefficients of a piecewise-constant
  graph signal $\x \in \PC(C)$ be
  \begin{eqnarray}
    \label{eq:sp_rep}
    \a^* &  = &   \arg \min_{\a \in \R^{2N-1}}   \left\| \a \right\|_0,
    \\ 
    \nonumber
    && {\rm subject~to~} \x = \D_{\rm PC} \a.
  \end{eqnarray}
  Then, we have
  \begin{eqnarray*}
    \left\| \a^* \right\|_0 \leq  1+   \left\|  \Delta \x \right\|_0  L.
  \end{eqnarray*}
\end{myCorollary}

Corollary~\ref{cor:sparse} directly follows from Theorem~\ref{thm:sparse}, as the  graph wavelet basis can be linearly represented by the piecewise-constant graph dictionary. We expect the upper bound in Corollary~\ref{cor:sparse} is not tight. In practice, the corresponding sparsity is usually even smaller than the sparsity provided by the graph wavelet basis because of the redundancy and flexibility of the piecewise-constant graph dictionary.

\subsection{Piecewise-Smooth Graph Dictionary}
\label{ssec:Graph_D}
We now generalize the piecewise-constant graph dictionary to the piecewise-smooth graph dictionary. In the piecewise-constant graph dictionary, we use a single one-piece graph signal to activate a certain subgraph; in the piecewise-smooth graph dictionary, we can use multiple localized and bandlimited graph signals to activate the same subgraph. Since localized and bandlimited graph signals are smooth on the corresponding  subgraphs, the piecewise-smooth graph dictionary provides more redundancy and flexibility to capture the localized events within a graph signal. 


Let $v^{\ell}_k$ be the $k$th piece in the $\ell$th decomposition level, $G_{v^{\ell}_k}$  the corresponding subgraph and $\Vm_{v^{\ell}_k} \in \R^{|v^{\ell}_k| \times |v^{\ell}_k|}$  the corresponding graph Fourier basis. The subdictionary corresponding to the $k$th piece at the $\ell$th decomposition level is
\begin{equation*}
  \D_{v^{\ell}_k}   \ = \  { \Vm_{v^{\ell}_k} }_{(K)} \in \R^{ |v^{\ell}_k| \times \min( K,  |v^{\ell}_k|)},
\end{equation*}
which is the first $\min( K, |v^{\ell}_k|)$ columns of
$\Vm_{v^{\ell}_k}$.  We collect all  subdictionaries across all 
levels to obtain the~\emph{piecewise-smooth graph dictionary},
\begin{equation}
  \label{eq:ps_dict}
  \D_{\rm PS}~=~\{ \D_{v^{\ell}_k}  \}_{\ell=1, k = 1}^{\ell= L, k = 2^\ell}.
\end{equation}
The total number of atoms of $\D_{\rm PS}$ is upper bounded by $(2N-1)K$ with bandwidth $K$. The total number of nonzero elements of $\D_{\rm PS}$ is at most $O( NK \log_2 N )$, still storage friendly. 

We now show that $\D_{\rm PS}$ promotes sparsity for piecewise-bandlimited graph signals.
\begin{myThm}
  \label{thm:PPL}
  Let $\D_{\rm PS} $ be the piecewise-smooth graph dictionary. Let the
  sparse coefficient of a piecewise-bandlimited graph signal $x \in
  \PBL(C, K)$ be
  \begin{eqnarray*}
    \a^* &  = &   \arg \min_{\a}   \left\| \a \right\|_0,
    \\
    && {\rm subject~to~}  \left\| \x - \D_{\rm PS} \a \right\|_2^2 \leq \epsilon_{\rm par}  \left\| \x \right\|_2^2,
  \end{eqnarray*}
  where $\epsilon_{\rm par} $ is a constant determined by the graph
  partitioning algorithm.  Then, we have
  \begin{displaymath}
    \left\| \a^* \right\|_0 \ \leq \ 1 + 2 K \left\|  \Delta \x_{\PC} \right\|_0 L,
  \end{displaymath}
  where $L$ is the decomposition depth in the coarse-to-fine approach
  and $\x_{\PC}$ is a piecewise-constant signal that shares the same
  pieces with $\x$.
\end{myThm}
The proof is given in Appendix~\ref{app:PPL}.

\section{Experimental Results}
\label{sec:experiment}
A good representation can be  used in compression, approximation, inpainting, denoising and localization. Here we evaluate our proposed graph dictionaries on two tasks: approximation and localization. 

\begin{figure*}[htb]
  \begin{center}
    \begin{tabular}{ccccc}
      \includegraphics[width=0.28\columnwidth]{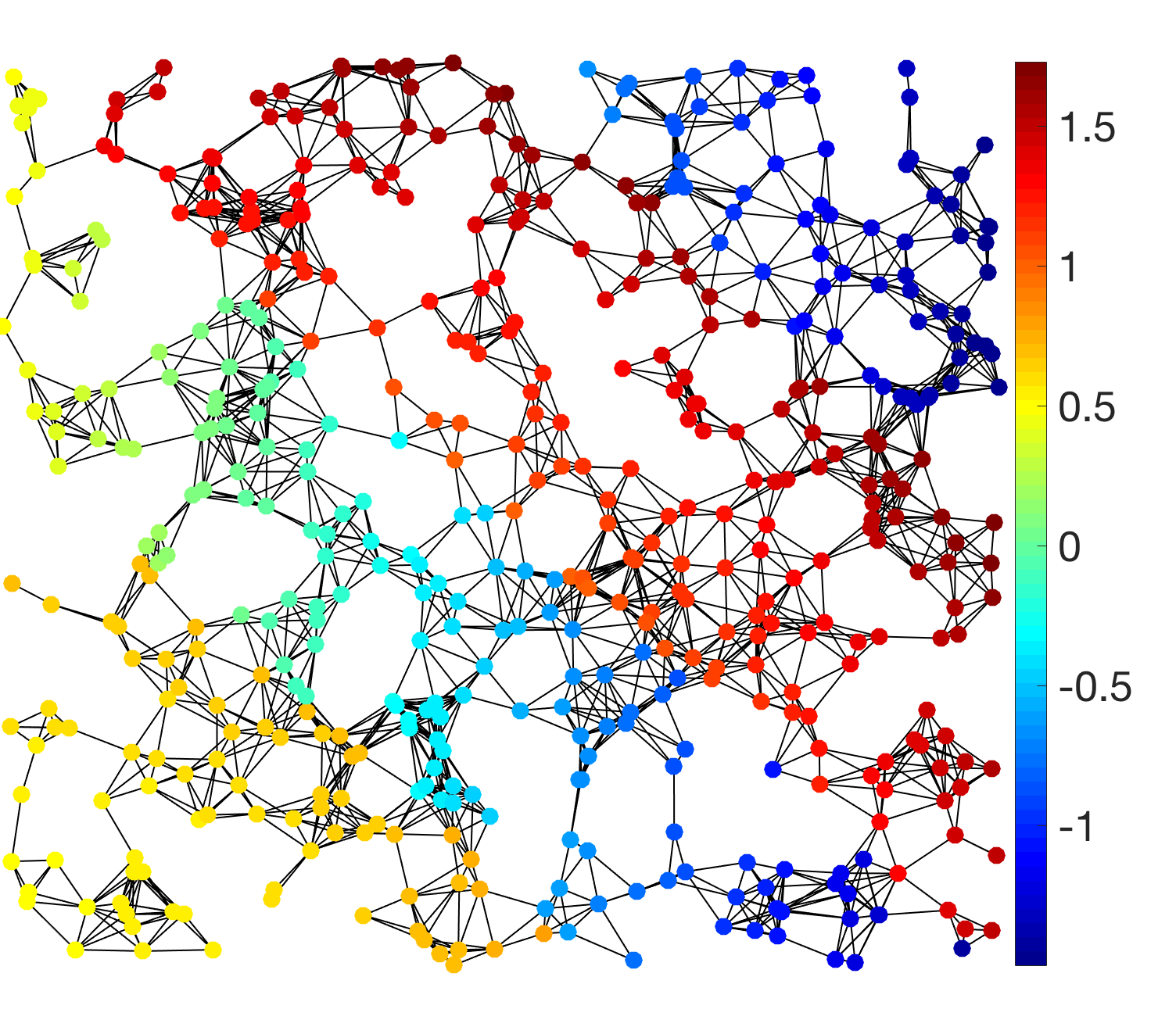} &
      \includegraphics[width=0.35\columnwidth]{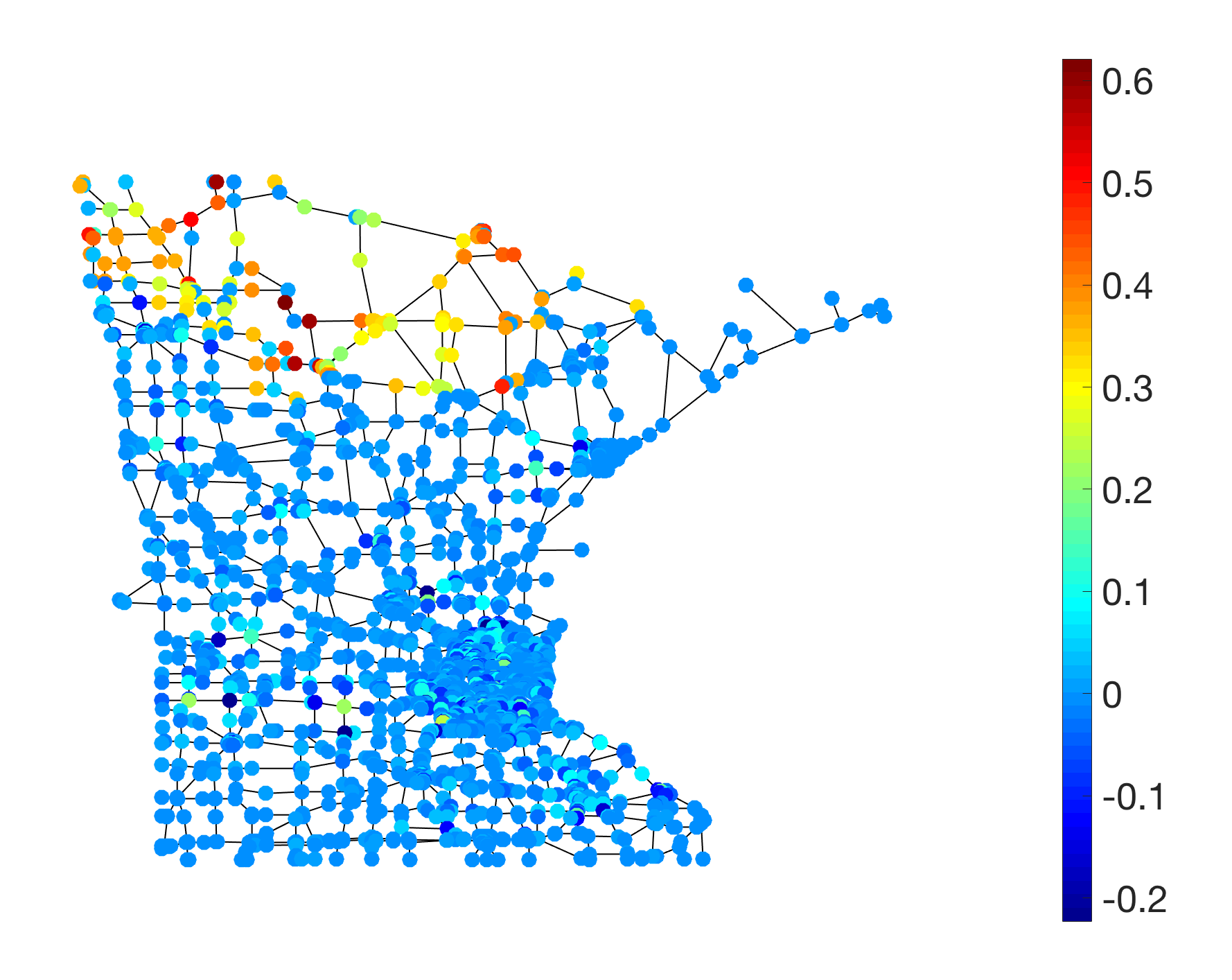} &
      \includegraphics[width=0.35\columnwidth]{figures/exp/1968/signal2.png} &
      \includegraphics[width=0.35\columnwidth]{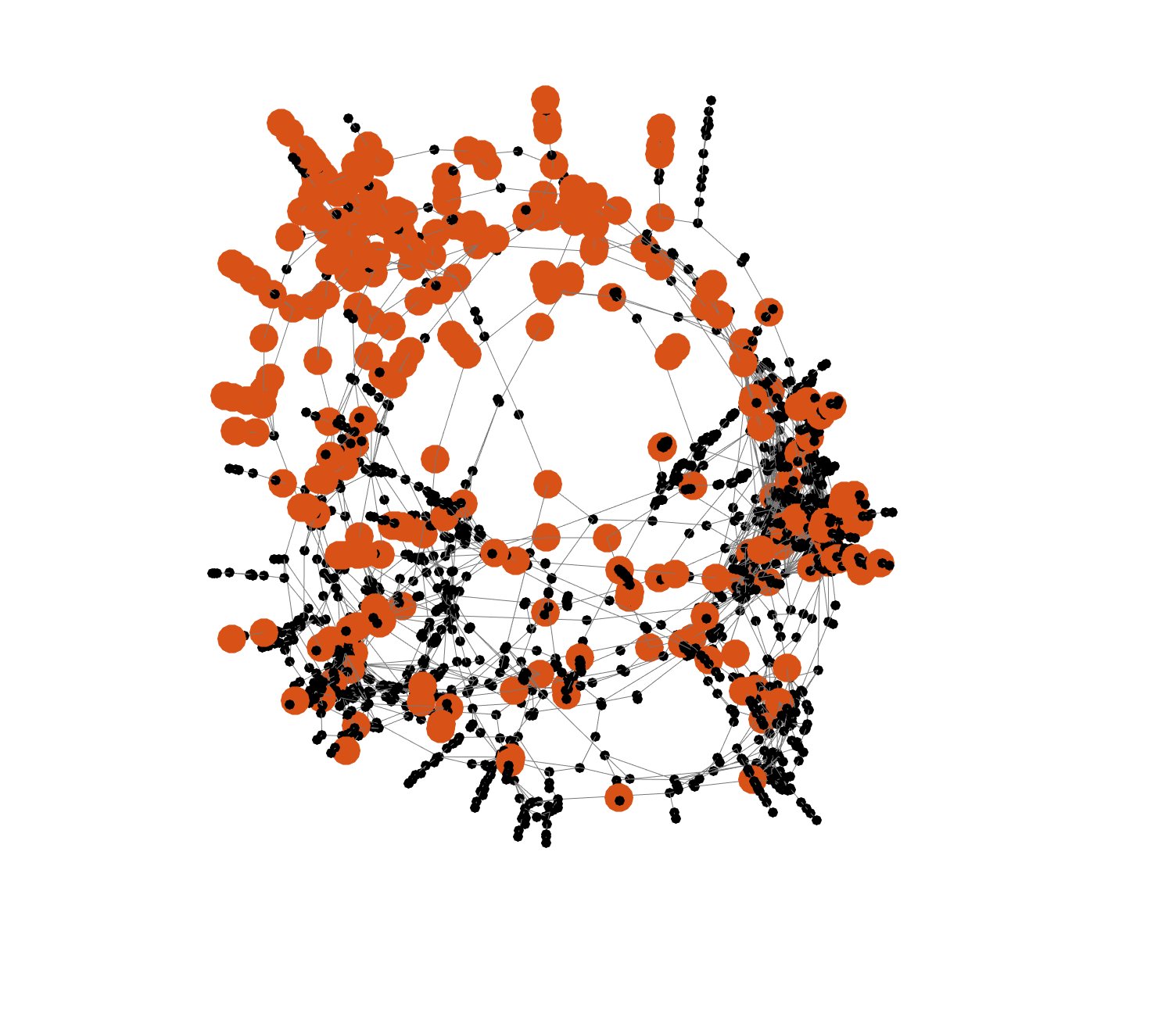} &
      \includegraphics[width=0.32\columnwidth]{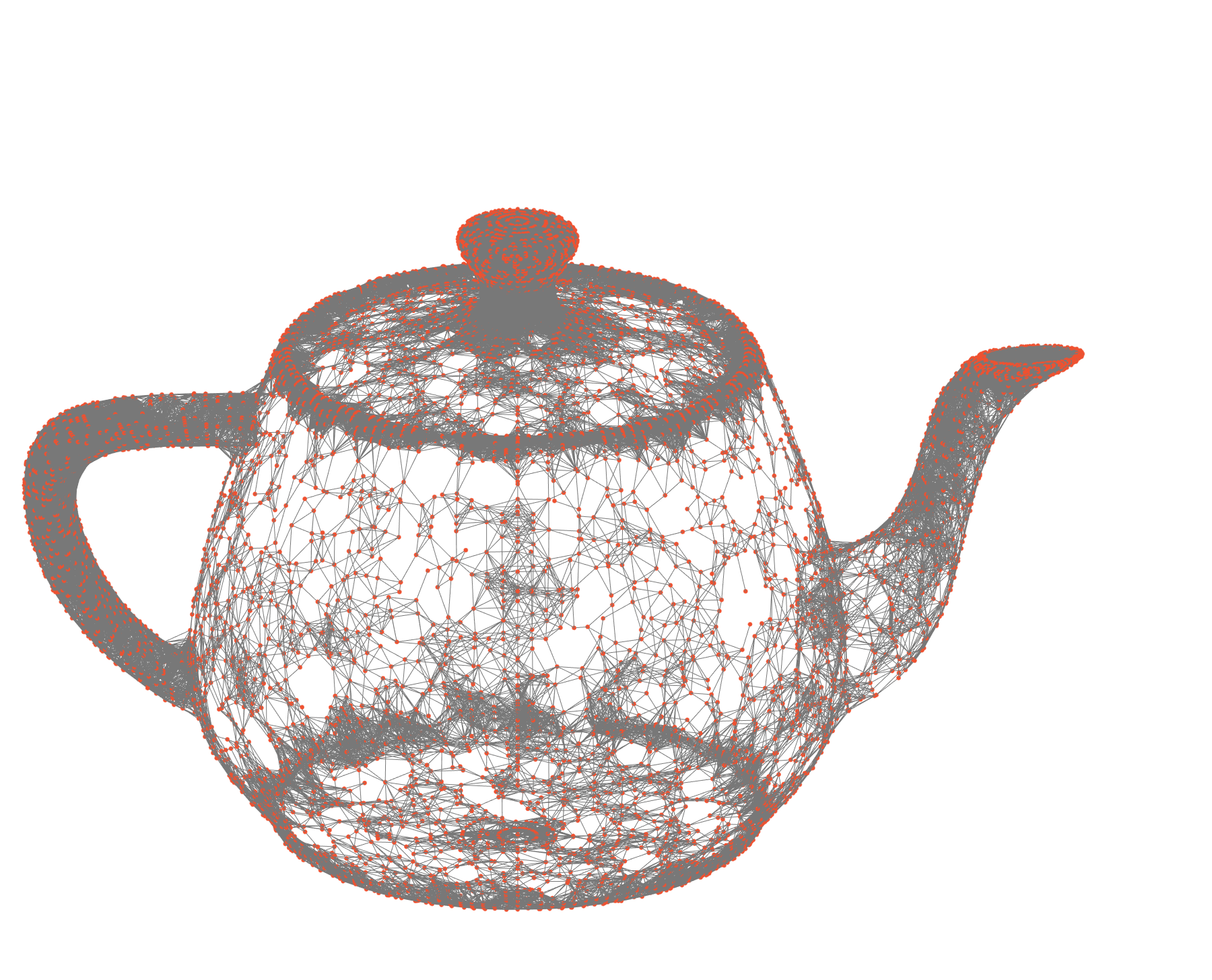}
      \\
      \includegraphics[width=0.35\columnwidth]{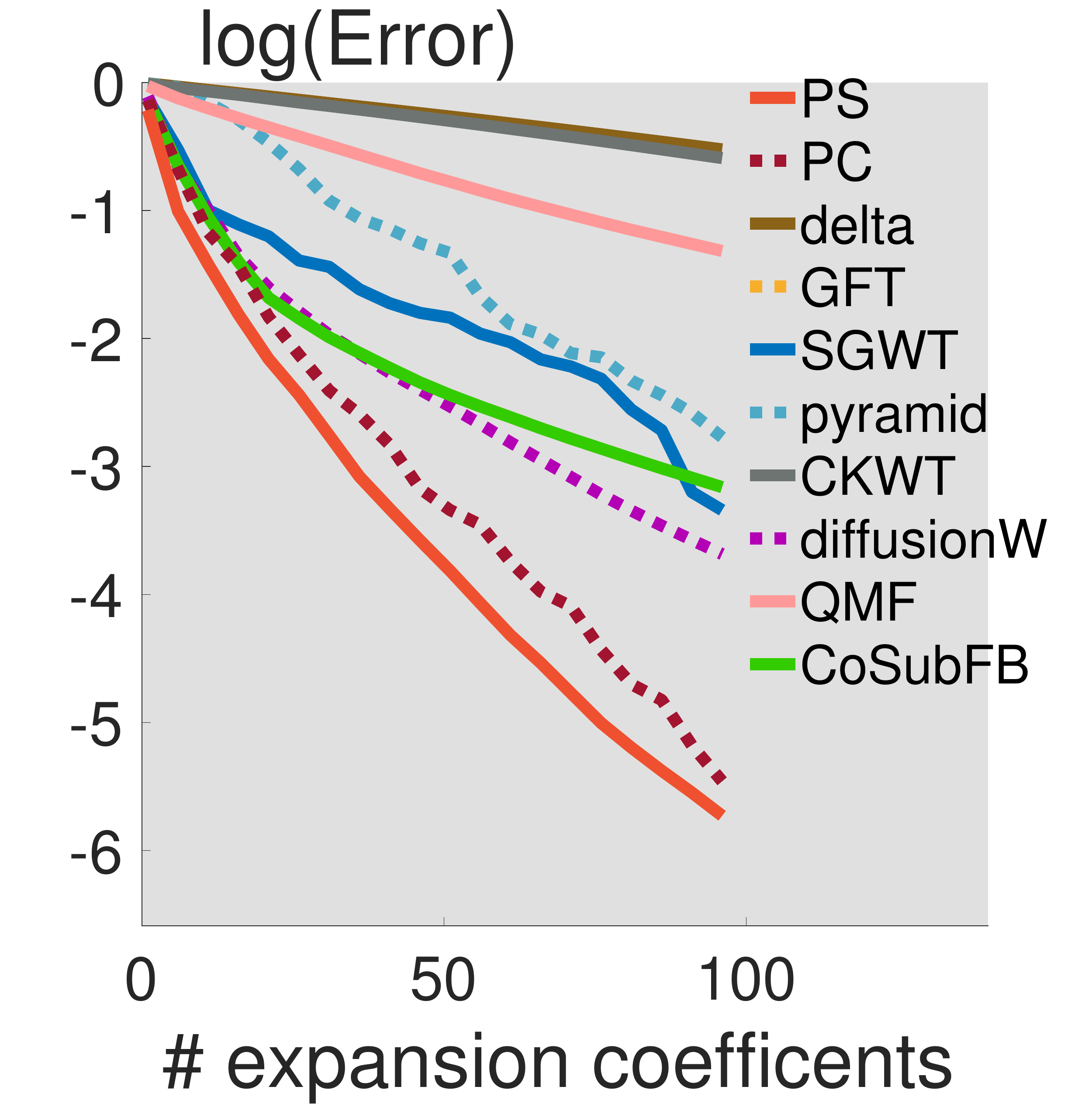} &
      \includegraphics[width=0.35\columnwidth]{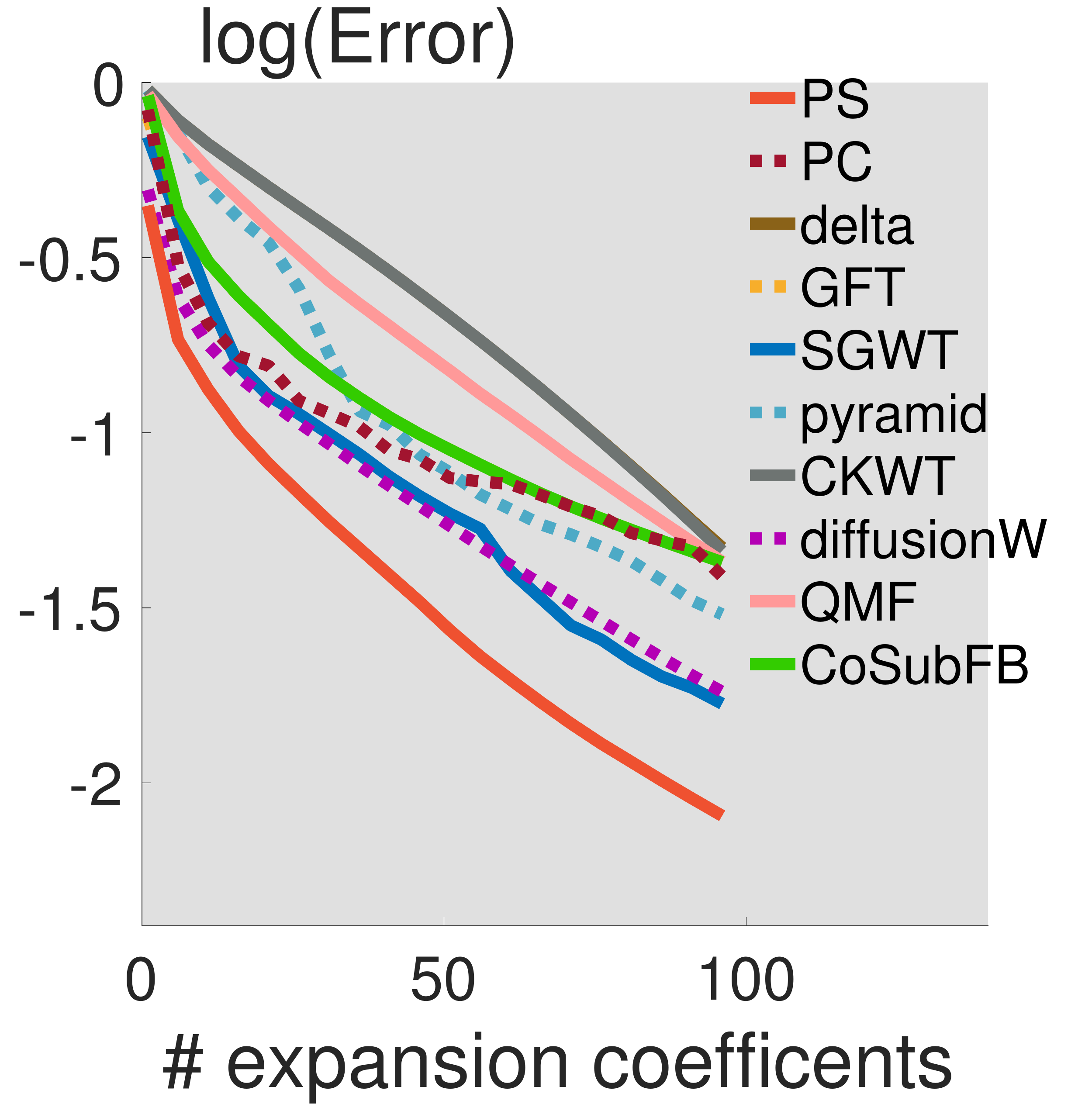} &
      \includegraphics[width=0.35\columnwidth]{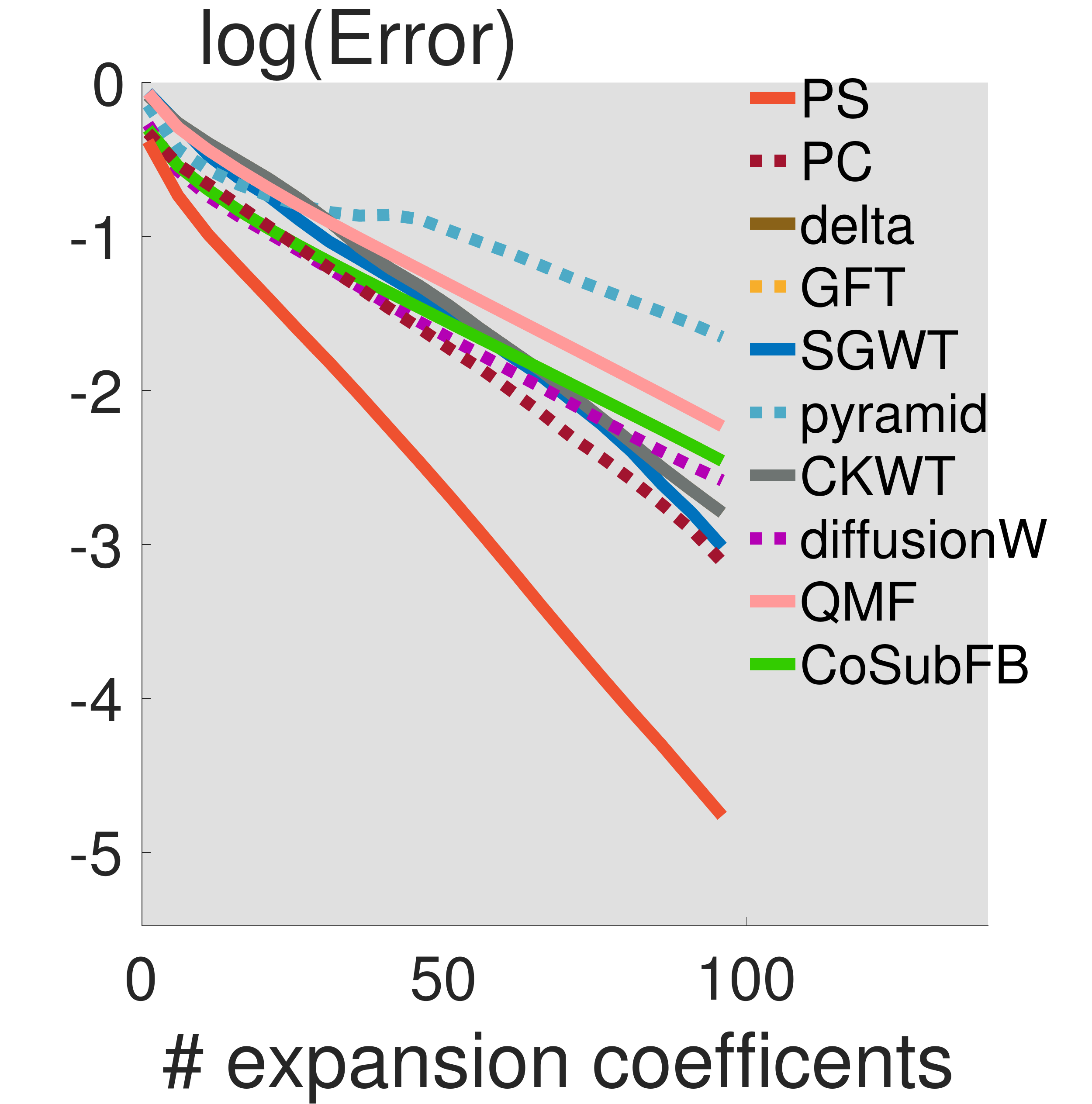} &
      \includegraphics[width=0.35\columnwidth]{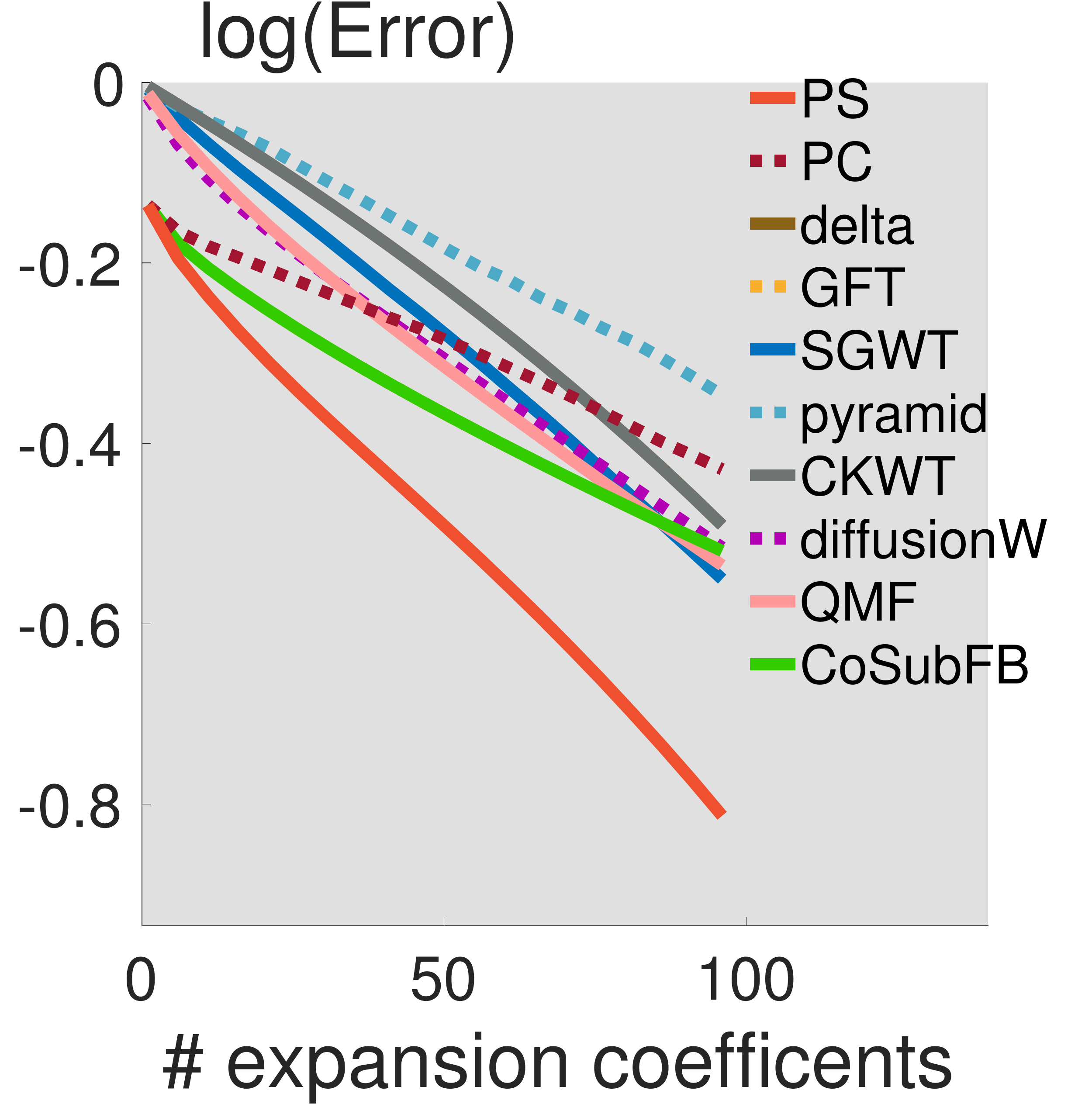} &
      \includegraphics[width=0.35\columnwidth]{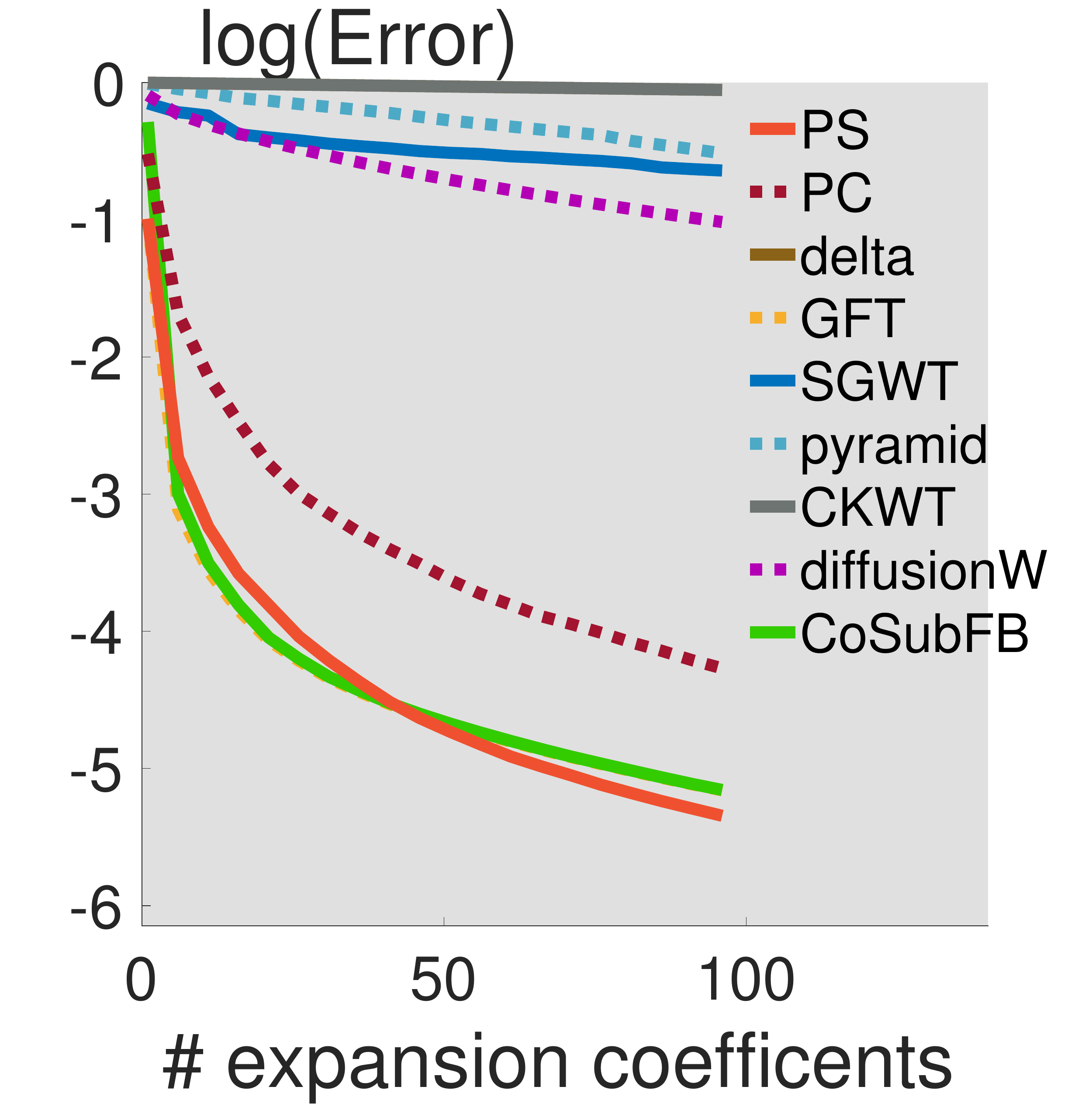} 
      \\
      {\small (a) Sensors.} &
      {\small (b)  Minnesota.} &
      {\small (c)  Kaggle 1968.} &
      {\small (d)  Citeseer.} &
      {\small (e)  Teapot.}
    \end{tabular}
  \end{center}
  \caption{\label{fig:ps_approximation} Piecewise-smooth graph
    dictionary (in red) outperforms the other competitive methods on five
    datasets.  The $x$-axis is the number of coefficients used in the approximation and the $y$-axis is the approximation error~\eqref{eq:error}, where lower means better. }
\end{figure*}

\subsection{Experimental Setup}
We consider six datasets summarized in Table~\ref{tab:dataset}.
\begin{itemize}
\item Sensors. This is a simulated geometric graph with 500 nodes and
  2,050 edges.  We simulate a piecewise-smooth graph signal
  following~\cite{ShumanFV:16}.

\item Minnesota.  This is the Minnesota road network with 2,642
  intersections and 3,304 road segments. We model each intersection as
  a node and each road segment as an edge. We simulate a localized
  smooth graph signal following~\cite{ShumanFV:16}.

\item Manhattan. This is the Manhattan street network with 13,679
  intersections and 34,326 road segments. We model each intersection
  as a node and each road segment as an edge. We model the restaurant
  distribution, and taxi-pickup positions as
  signals supported on the Manhattan street network.

\item Kaggle 1968.  This is a social network of Facebook users with
  277 nodes and 2,321 edges. It also contains 14 social circles, where
  each one can be modeled as a binary piecewise-constant signal
  supported on this social network.

\item Citeseer. This is a co-authorship network with 2,120 nodes and
  3,705 edges. It also contains 7 research groups, where each one can
  be modeled as a binary piecewise-constant signal supported on this
  co-authorship network.

\item Teapot.  This is a dataset with 7,999 3D points, representing
  the surface of a teapot. We construct a 10-nearest neighbor graph to
  capture the geometry. 3D coordinates can be modeled as three
  piecewise-smooth signals supported on this generalized mesh.
\end{itemize}

\begin{table}[htbp]
  \footnotesize
  \begin{center}
    \begin{tabular}{@{}l|llll@{}}
      \toprule
 Dataset  & Type & \# nodes  & \# edges   & Signals \\
      \midrule \addlinespace[1mm]
 Sensors &  Simulation  &   500 & 2,050  & Simulation  \\
 
  Minnesota &  Traffic net  &   2,642 & 3,304  & Simulation  \\

  Manhattan &  Traffic net  &  13,679 & 3,679 & Taxi  \\
  
   Kaggle 1968 &  Social net  &   277 & 2,321  & Circle \\

 Citeseer &  Citation net  &   2,120 & 3,705  & Attribute \\
 
  Teapot &  Mesh  &  7,999 & 198,035  & Coordinate  \\
\bottomrule
\end{tabular} 
\caption{\label{tab:dataset} Dataset description. }
\end{center}
\end{table}

We consider the following ten competitive representation methods:
\begin{itemize}
\item PC (dashed dark red line). This is our piecewise-constant graph
  dictionary~\eqref{eq:pc_dict}.

\item PS (solid red line).  This is our piecewise-smooth graph
  dictionary~\eqref{eq:ps_dict}. The bandwidth in each piece is $10$.

\item Delta (solid dark yellow line). This is the basis of Kronecker
  deltas.

\item GFT (dashed yellow line)~\cite{ShumanNFOV:13}. This is the graph
  Fourier basis.

\item SGWT (solid blue line)~\cite{HammondVG:11}.  This is the
  spectral graph wavelet transform with five wavelet scales plus the
  scaling functions for a total redundancy of 6.

\item Pyramid (dashed light blue line)~\cite{ShumanFV:16}. This is the
  multiscale pyramid transform.

\item CKWT (solid grey line)~\cite{CrovellaK:03}. These are spatial graph wavelets with wavelet functions based on the renormalized one-sided Mexican hat wavelet, also with five wavelet scales and  concatenated with the dictionary of Kronecker deltas.

\item DiffusionW (dashed purple line)~\cite{CoifmanM:06}. These are the
  the diffusion wavelets.

\item QMF (solid pink line)~\cite{NarangO:12}.  This is the graph-QMF
  filter bank transform.

\item CoSubFB (solid green line)~\cite{TremblayB:16}. This is the
  subgraph-based filter bank.

\end{itemize}

\subsection{Approximation}
Approximation is a standard task used to evaluate the quality of a representation. The goal here is to use a few expansion coefficients to approximate a graph signal.  We consider two approximation strategies: nonlinear approximation and orthogonal marching pursuit. Given the budget of $K$ expansion coefficients, nonlinear approximation chooses the $K$ largest-magnitude ones to minimize the approximation error while orthogonal marching pursuit greedily and sequentially selects $K$ expansion coefficients to minimize the residual error. For each representation method, we use both approximation strategies and report the results of the better one.  The evaluation metric is the normalized mean square error, defined as 
\begin{equation}
\label{eq:error}
  {\rm Error} \ = \  \frac{ \left\| \widehat{\x}  - \x \right\|_2^2 }{\left\| \x \right\|_2^2 }.
\end{equation}

Figure~\ref{fig:ps_approximation} compares the approximation
performances on five datasets. Five columns in
Figure~\ref{fig:ps_approximation} show the sensors, Minnesota, Kaggle
1968, Citeseer and Teapot, respectively. Each plot in the first row
shows the visualization of the graph signal; each plot in the second
row shows the approximation error on the logarithm scale, where the $x$-axis is the number of
expansion coefficients and the $y$-axis is the normalized mean square
error.
\begin{figure*}[htb]
  \begin{center}
    \begin{tabular}{cccccc}
      \includegraphics[width=0.3\columnwidth]{figures/exp/teapot/signal.png}  &
      \includegraphics[width=0.35\columnwidth]{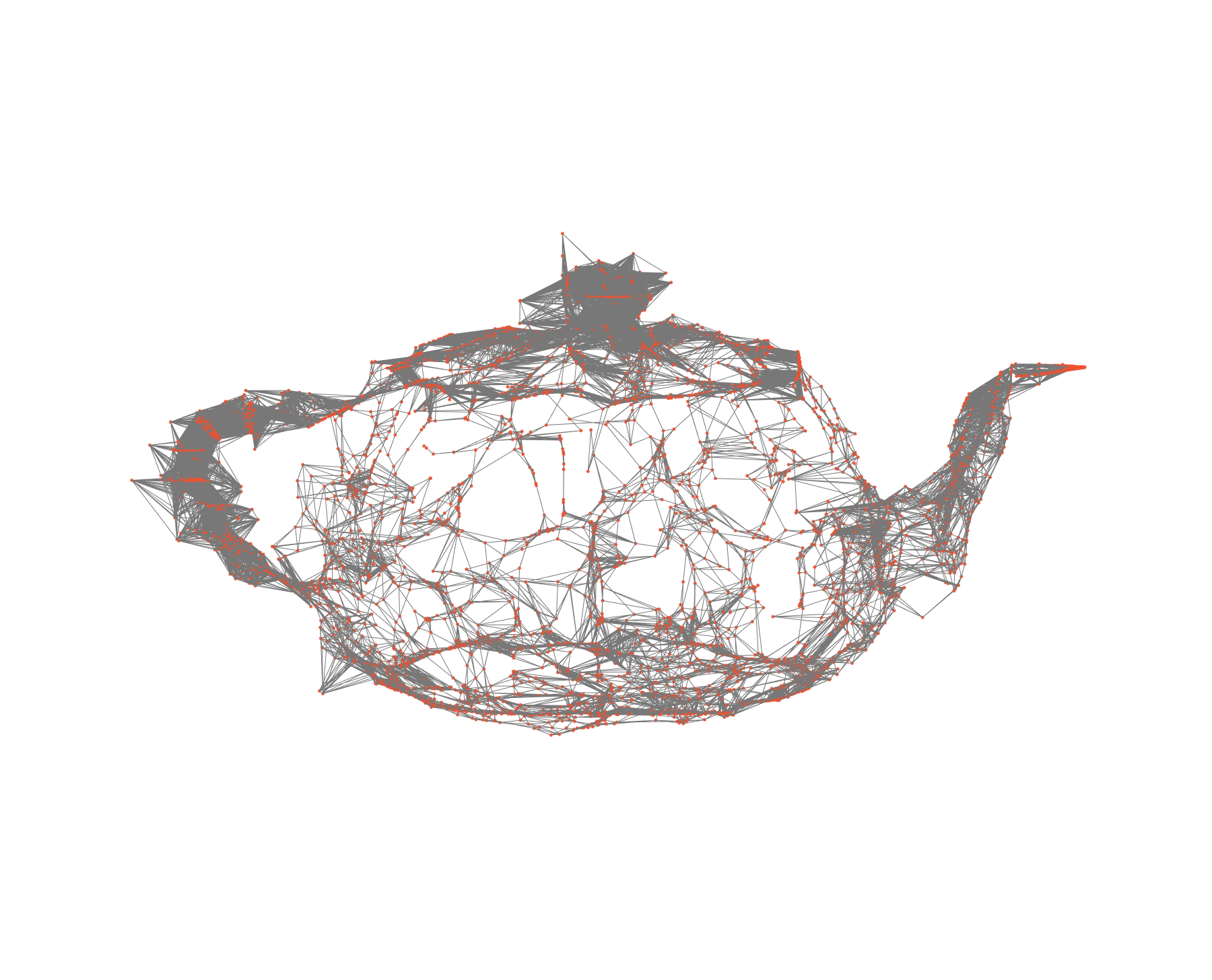} &
      \includegraphics[width=0.35\columnwidth]{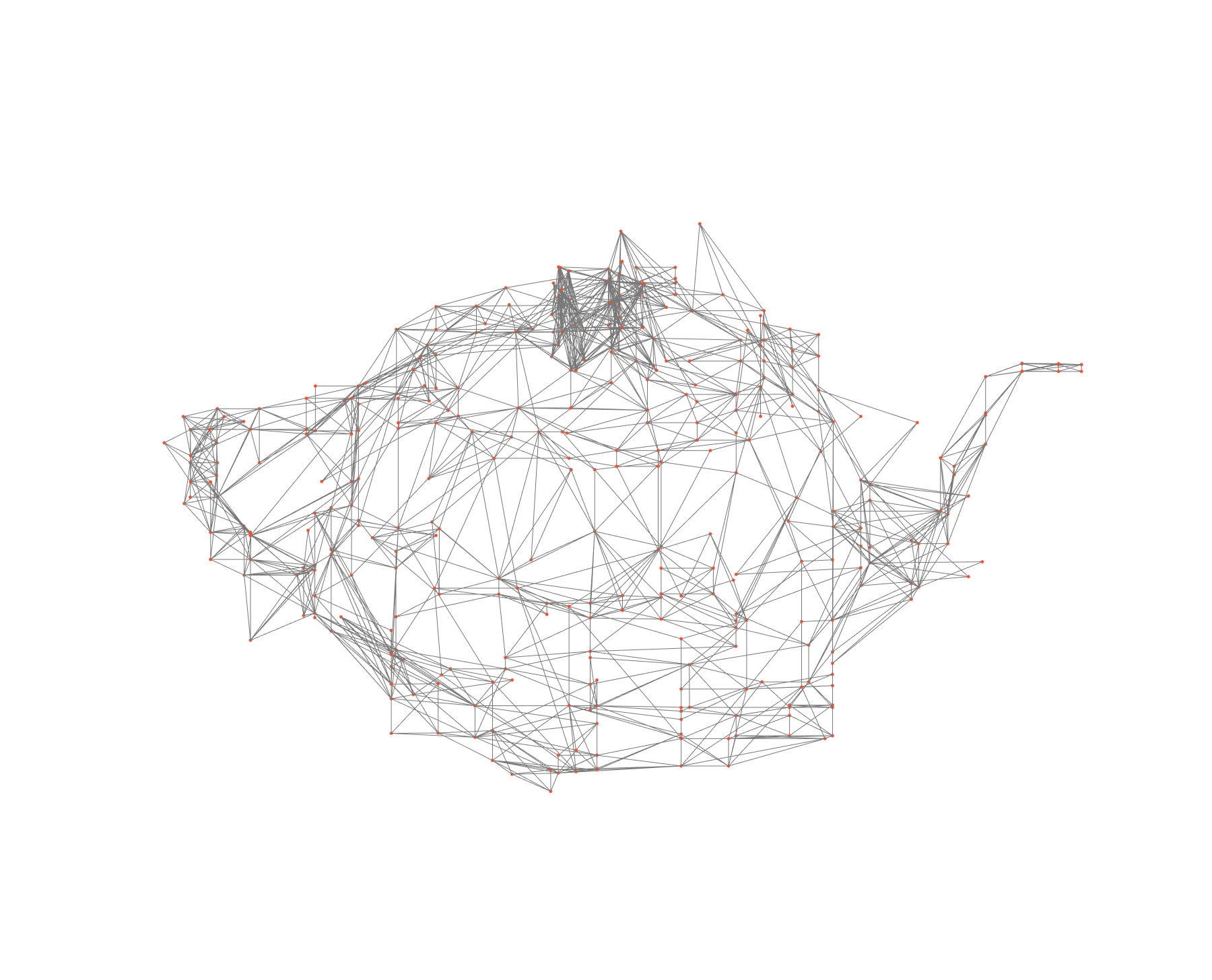} &
      \includegraphics[width=0.35\columnwidth]{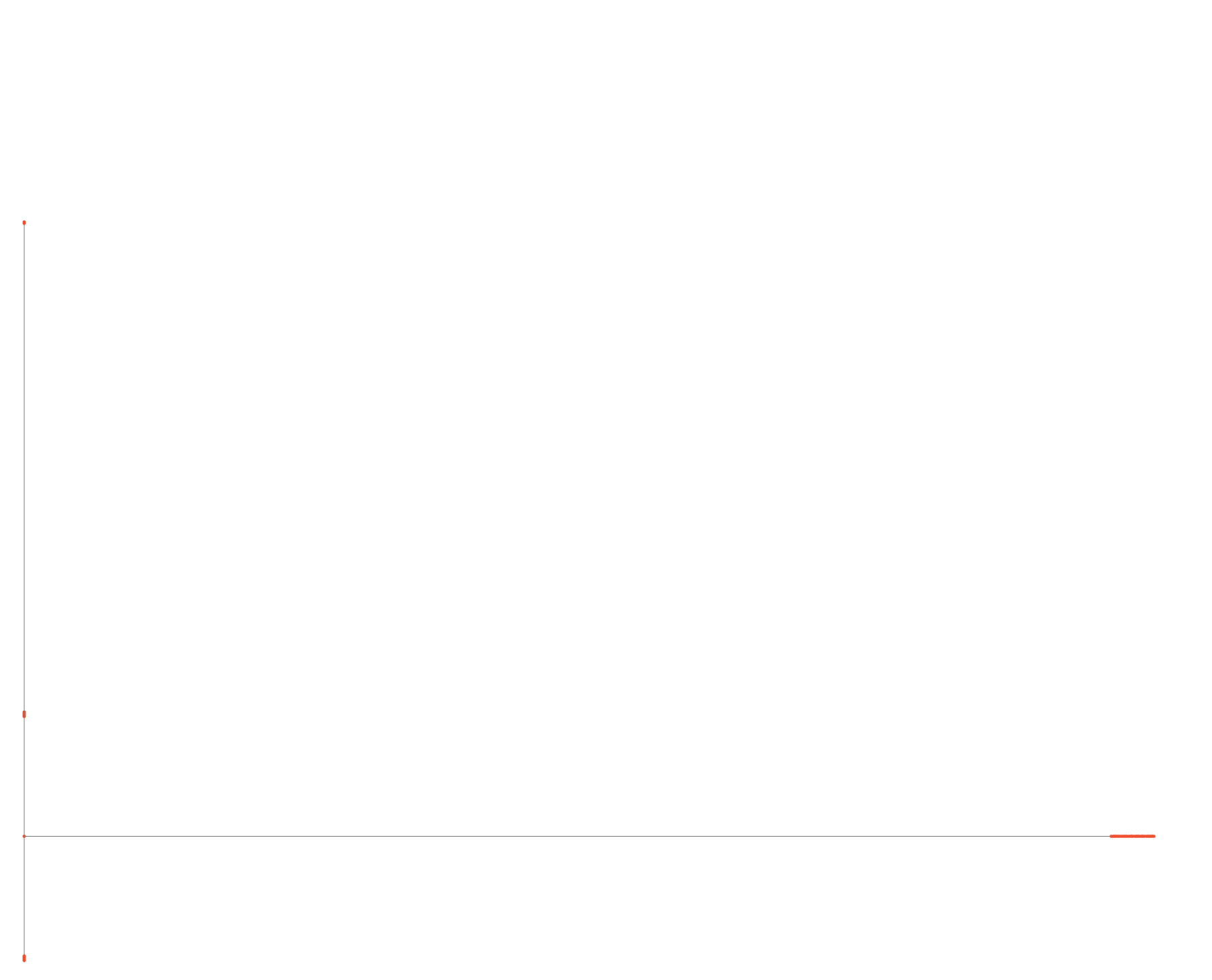} &
      \includegraphics[width=0.35\columnwidth]{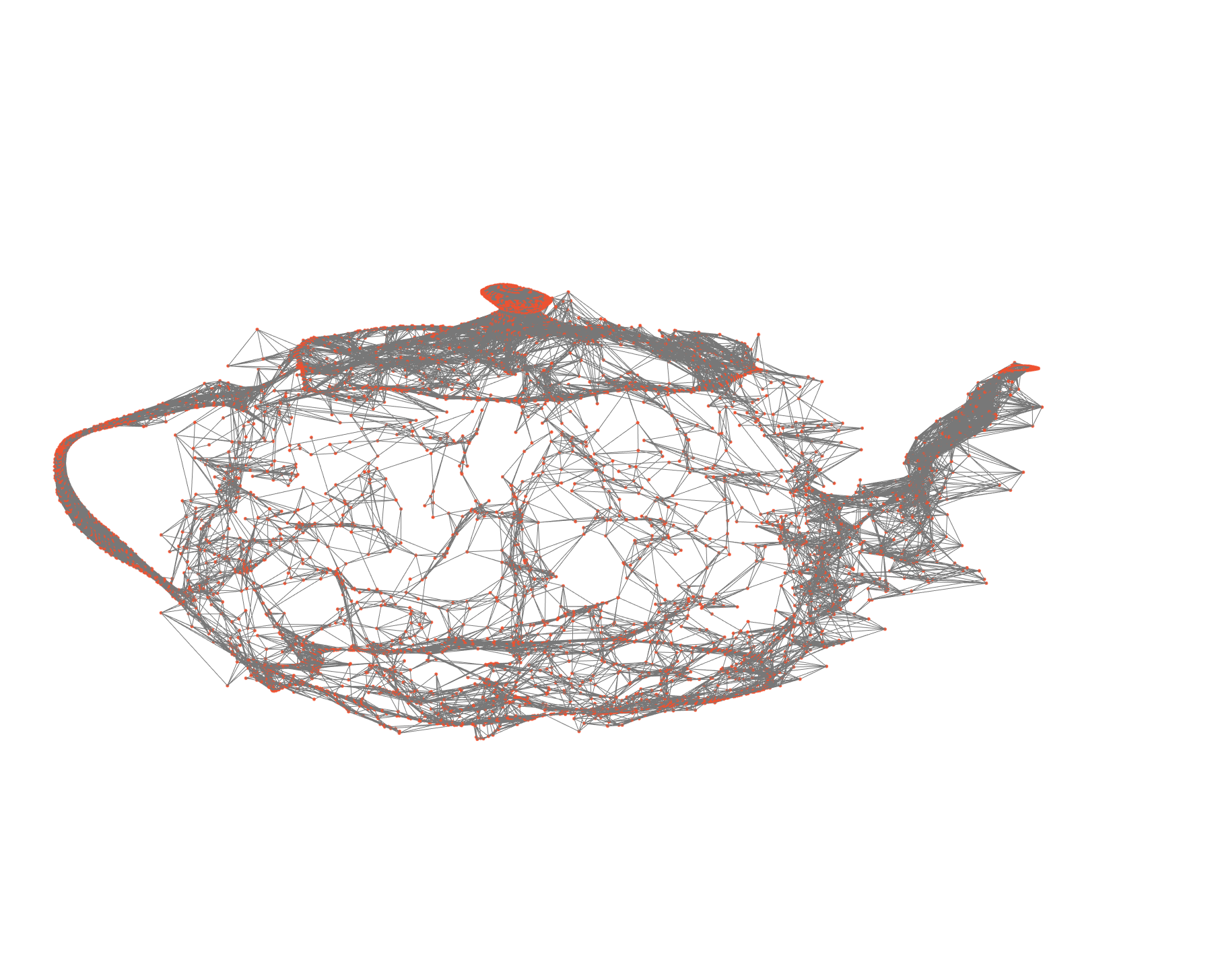} &
      \\
      {\small (a) Original.} & {\small (b) PS.} & {\small (c) PC.}  & {\small (d) Delta.} & {\small (e) GFT.} 
      \\   
      \includegraphics[width=0.35\columnwidth]{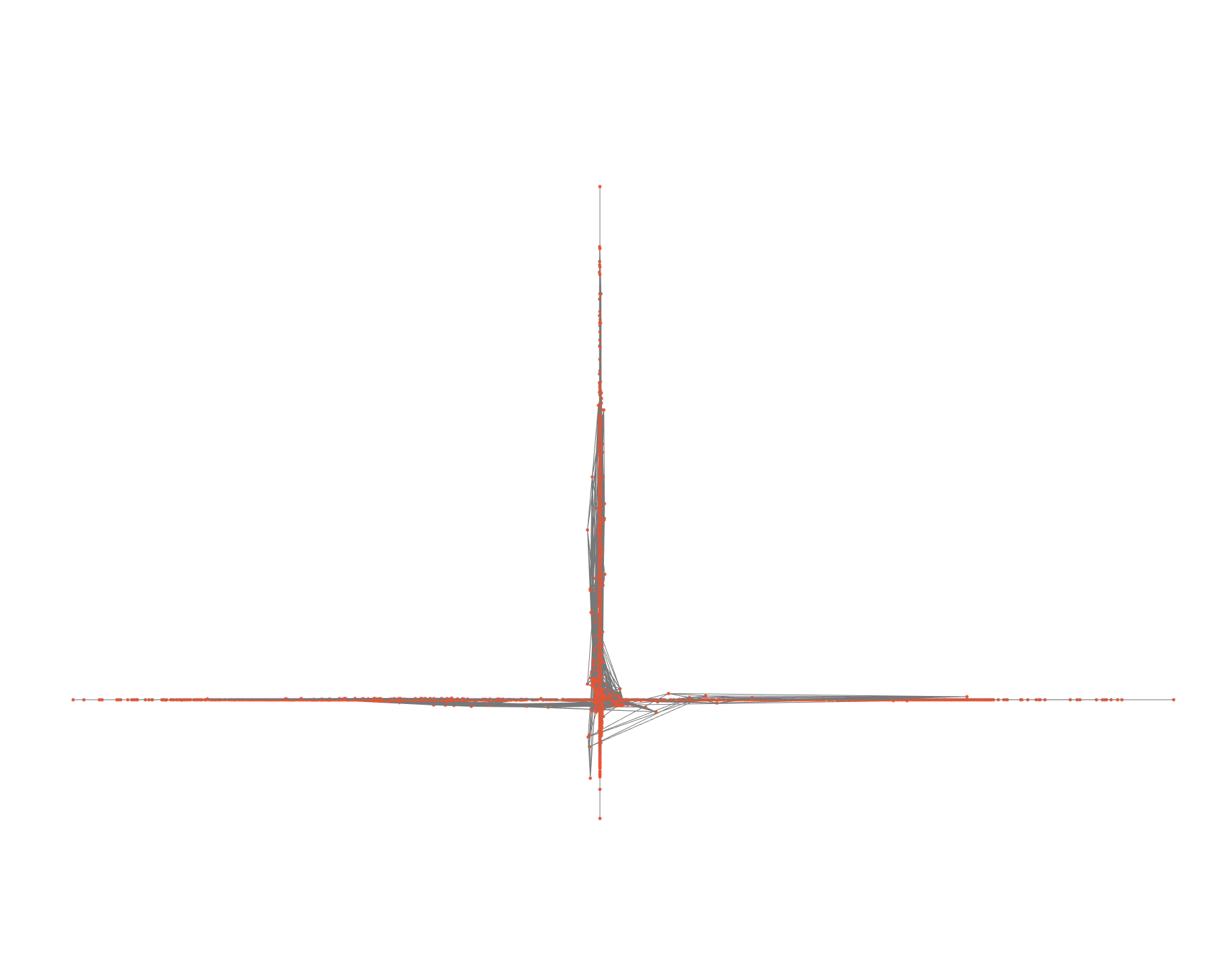} &
      \includegraphics[width=0.35\columnwidth]{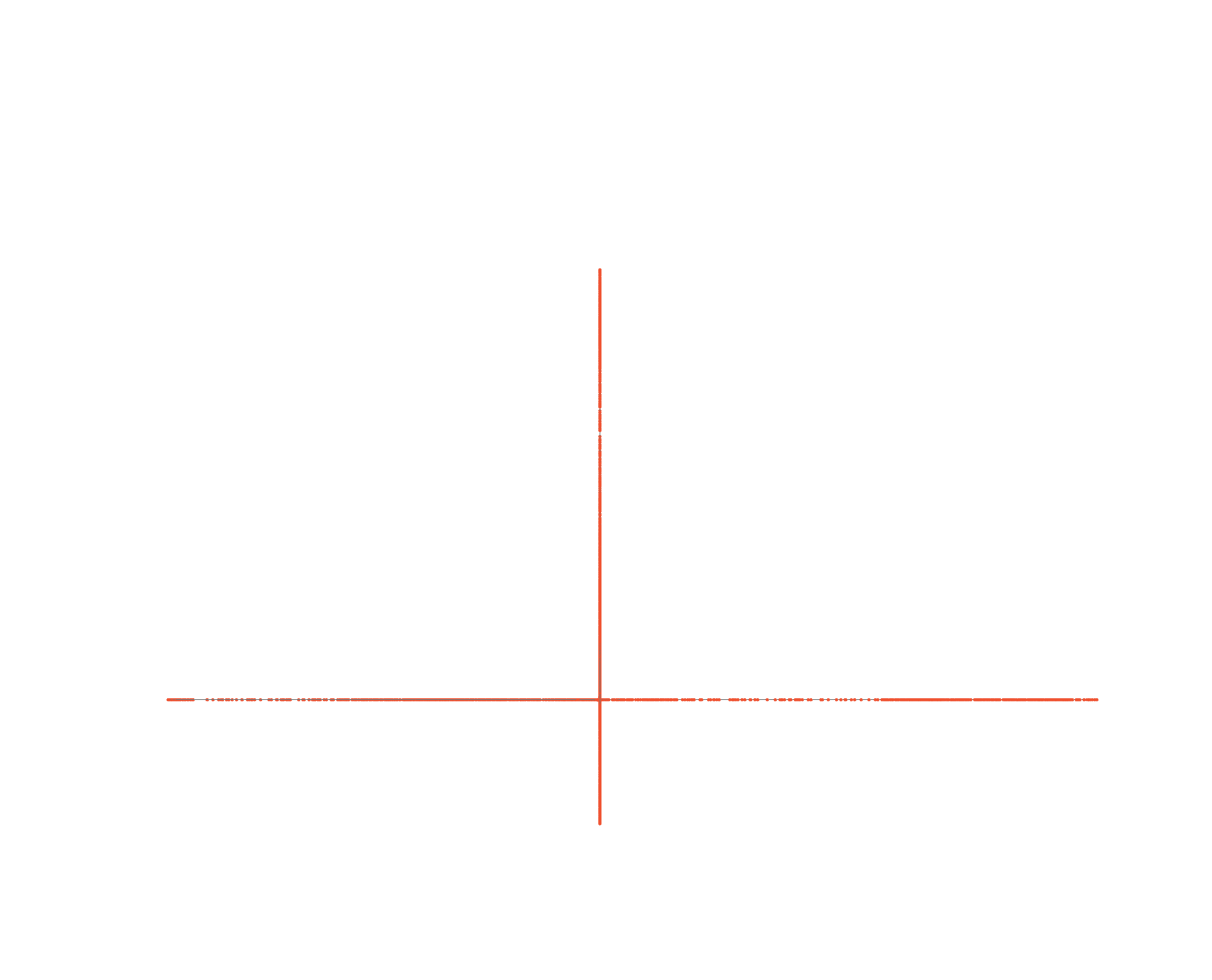} &
      \includegraphics[width=0.35\columnwidth]{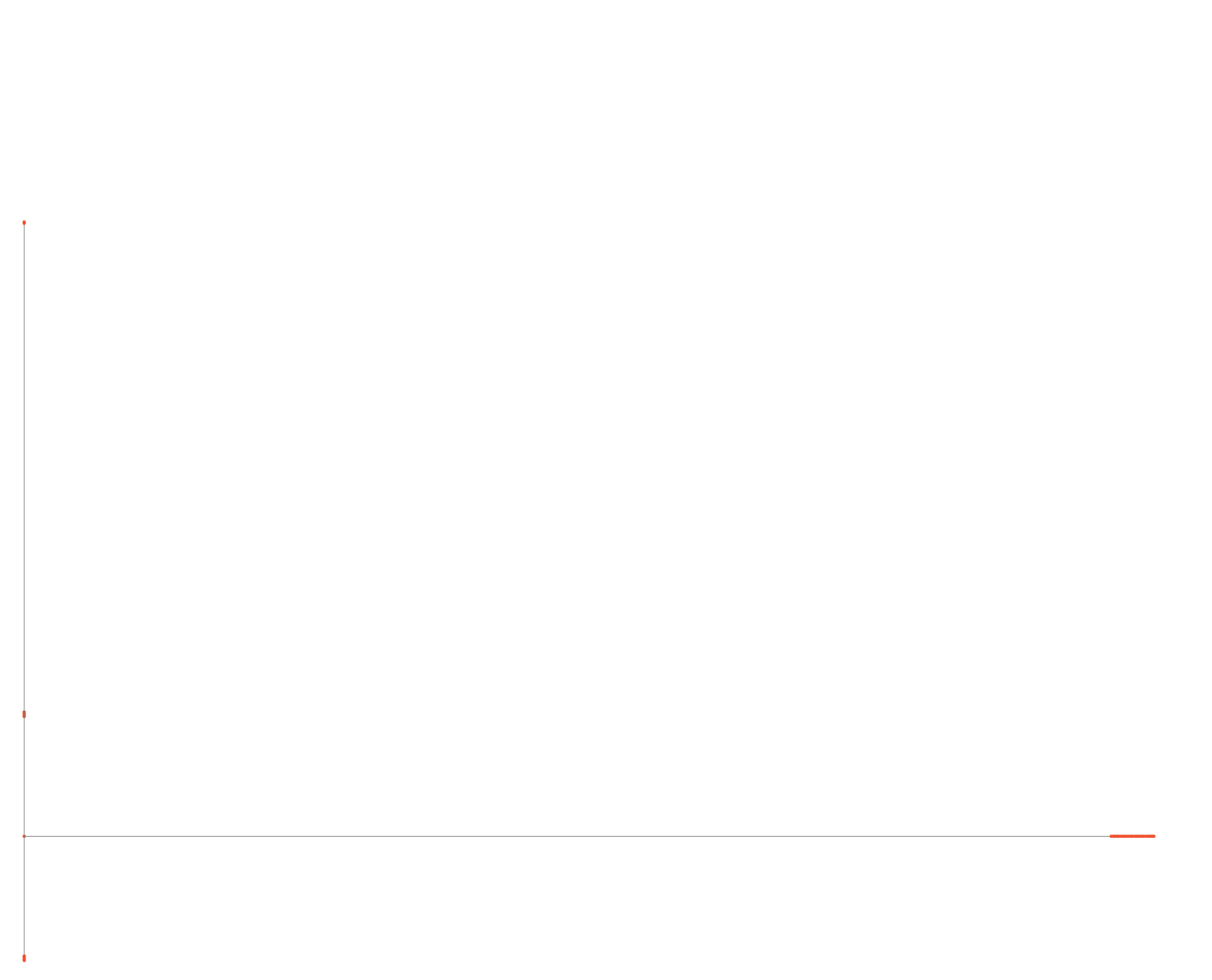} &
      \includegraphics[width=0.35\columnwidth]{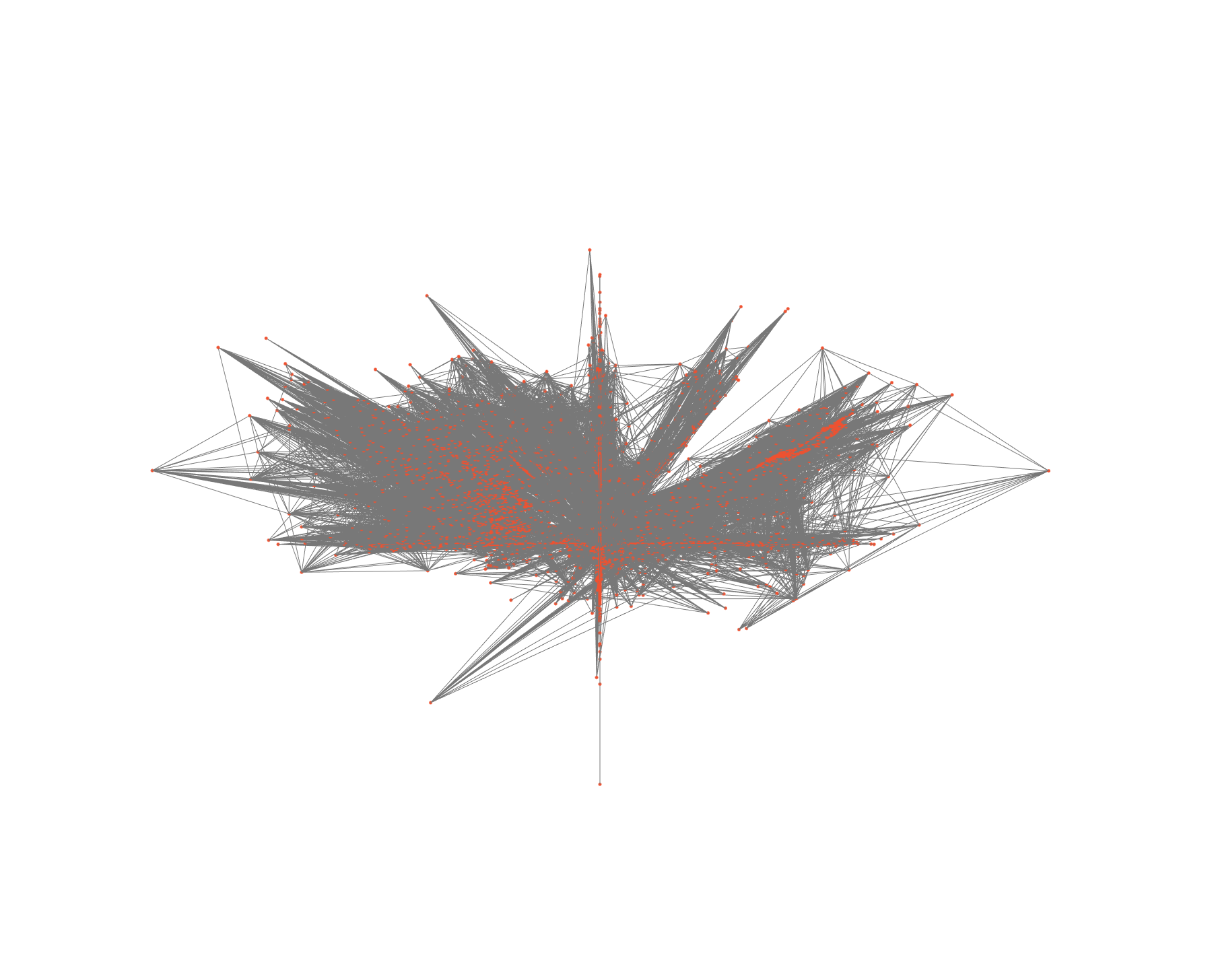} &
      \includegraphics[width=0.35\columnwidth]{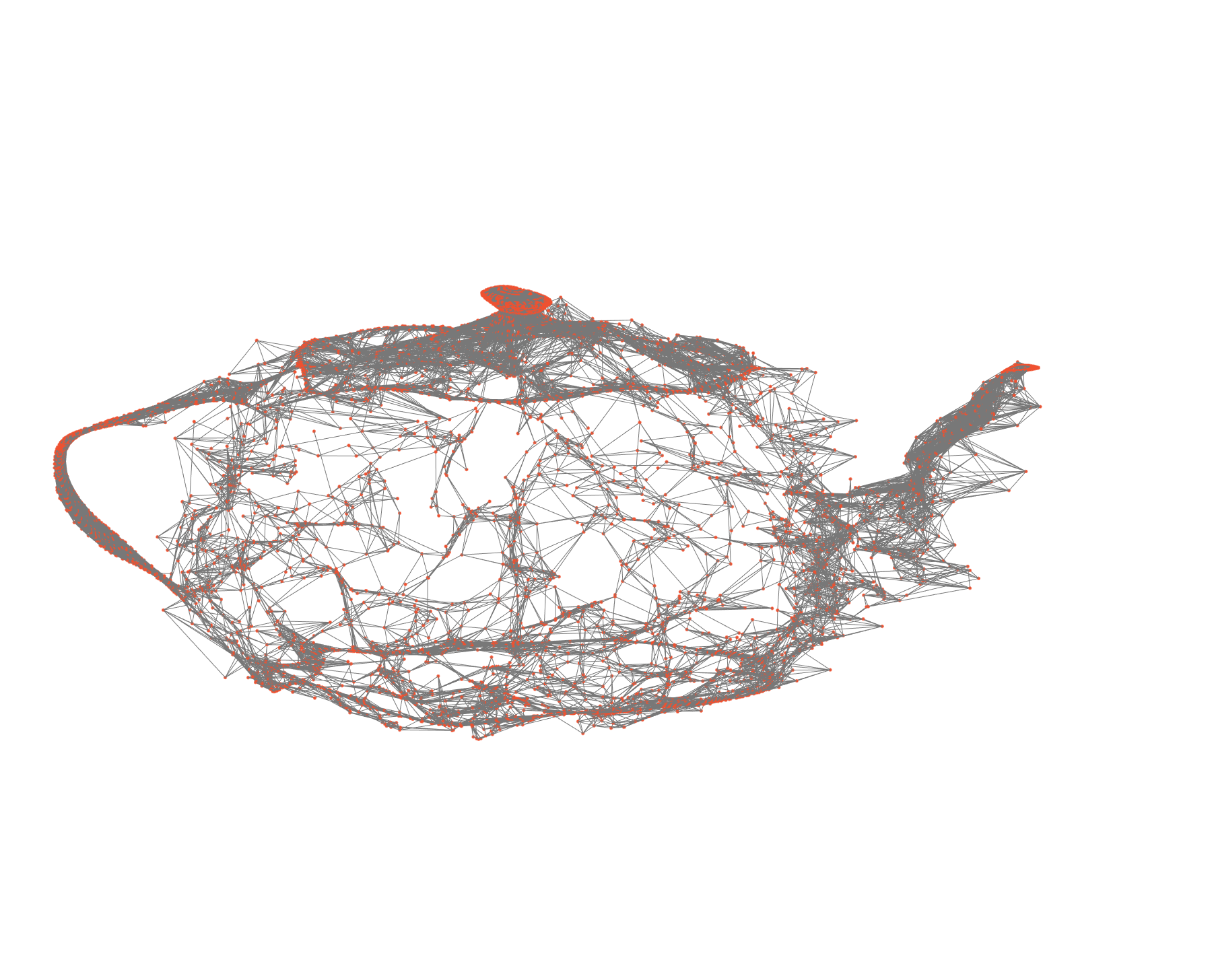} \\  
      {\small (f) SGWT.}  & {\small (g) Pyramid.} &  {\small (h) CKWT.}  & {\small (i) Diffusion wavelets.}  &  {\small (j) CSFB.} 
    \end{tabular}
  \end{center}
  \caption{\label{fig:teapot} Reconstruction visualization for Teapot. }
\end{figure*}

Overall, the proposed piecewise-smooth graph dictionary outperforms
its  competitors under various types of graphs and graph signals.

$\bullet$  Sensors. The graph signal is piecewise-smooth. The top three methods are the piecewise-smooth graph dictionary, the piecewise-constant graph dictionary and the diffusion wavelets; on the other end of the spectrum, the Kronecker deltas, which fit one signal coefficient at a time, fails.

$\bullet$   Minnesota. The graph signal is localized smooth. The top three methods are the piecewise-smooth graph dictionary, the diffusion wavelets and the spectral graph wavelet transform;  on the other end of the spectrum, the spatial graph wavelets fail.

$\bullet$  Kaggle 1968. The graph signal is binary and piecewise-constant with a few pieces. The top three methods are the piecewise-smooth graph dictionary, the piecewise-constant graph dictionary and the spectral graph wavelet transform;  on the other end of the spectrum, the multiscale pyramid transform fails.

$\bullet$  Citesser. The graph signal is binary, piecewise-constant  with a large number of pieces. None of the methods performs well due to the noisy input signal. The top three methods are the piecewise-smooth graph dictionary, the subgraph-based filter bank and the graph-QMF filter bank transform;  on the other end of the spectrum, the multiscale pyramid transform fails.

$\bullet$  Teapot. The graph signal is smooth. The top three methods are the piecewise-smooth graph dictionary, the subgraph-based filter bank and the graph Fourier basis;  on the other end of the spectrum, the Kronecker deltas and the spatial graph wavelets fail. To have an illustrative understanding,  we visualize the reconstructions in Figure~\ref{fig:teapot} where each plot shows the reconstruction by using 100 expansion coefficients.

\begin{figure*}[htb]
  \begin{center}
    \begin{tabular}{cccccc}
 \includegraphics[width=0.35\columnwidth]{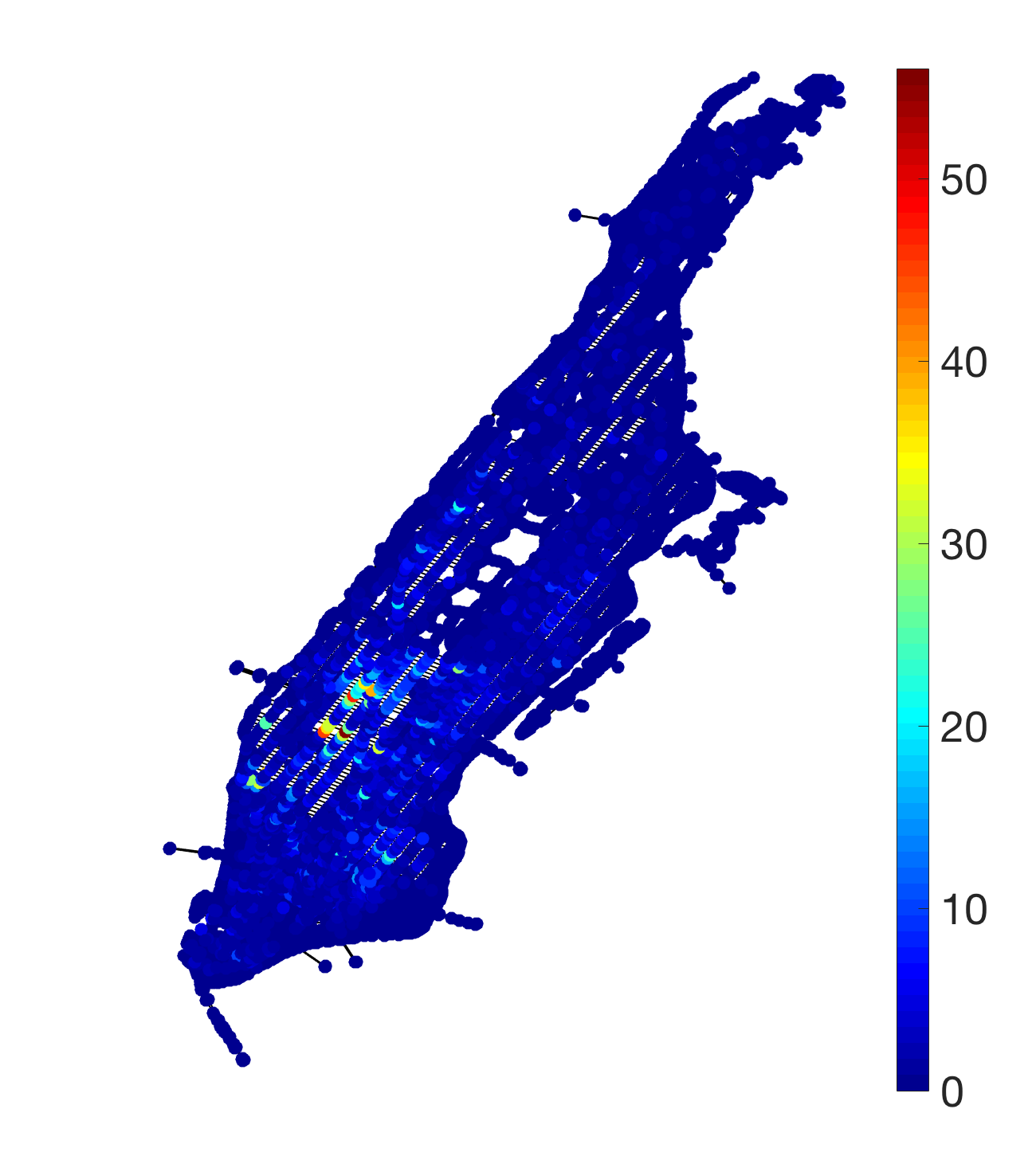} &
 \includegraphics[width=0.35\columnwidth]{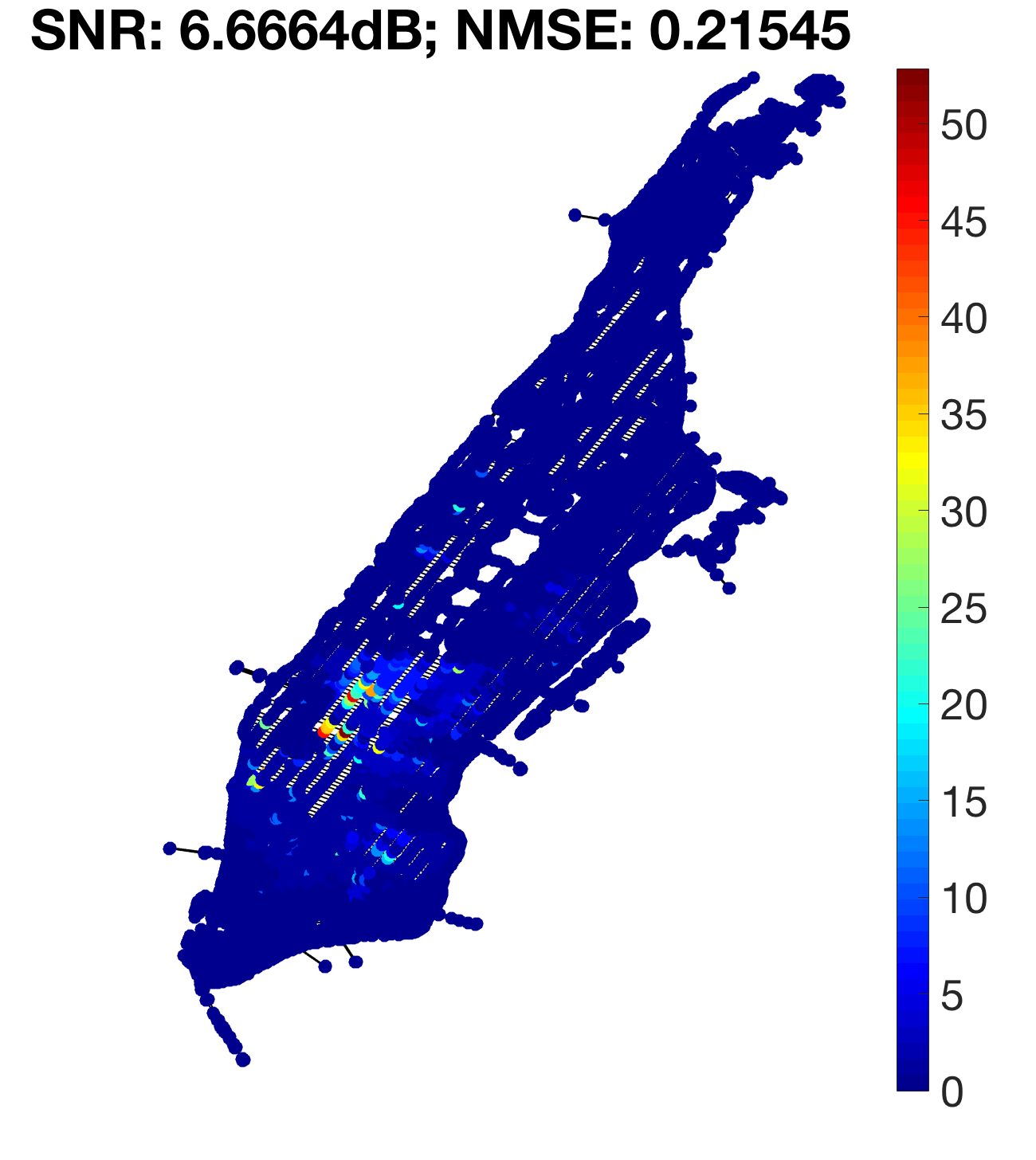}   &
 \includegraphics[width=0.35\columnwidth]{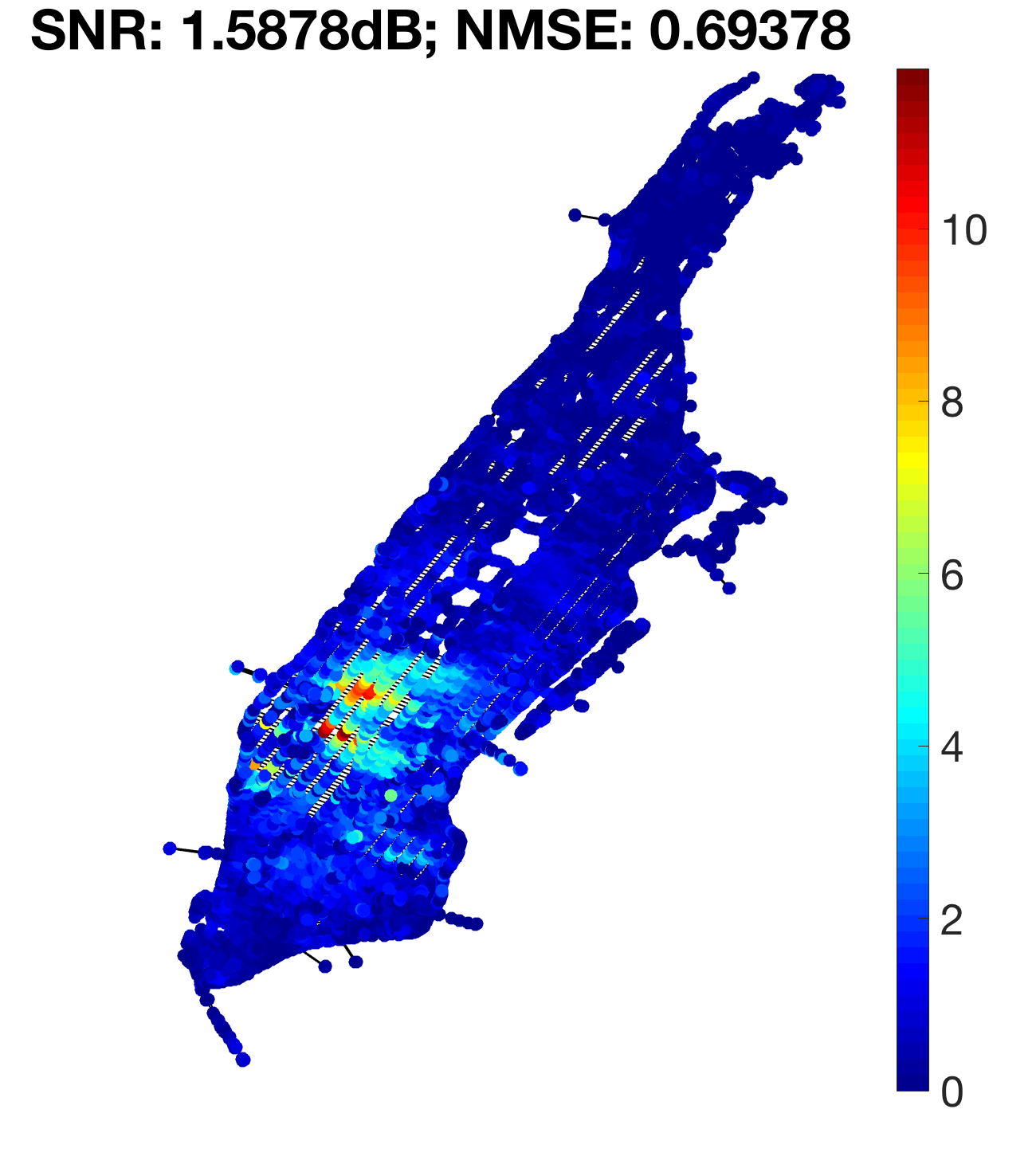}  &
 \includegraphics[width=0.35\columnwidth]{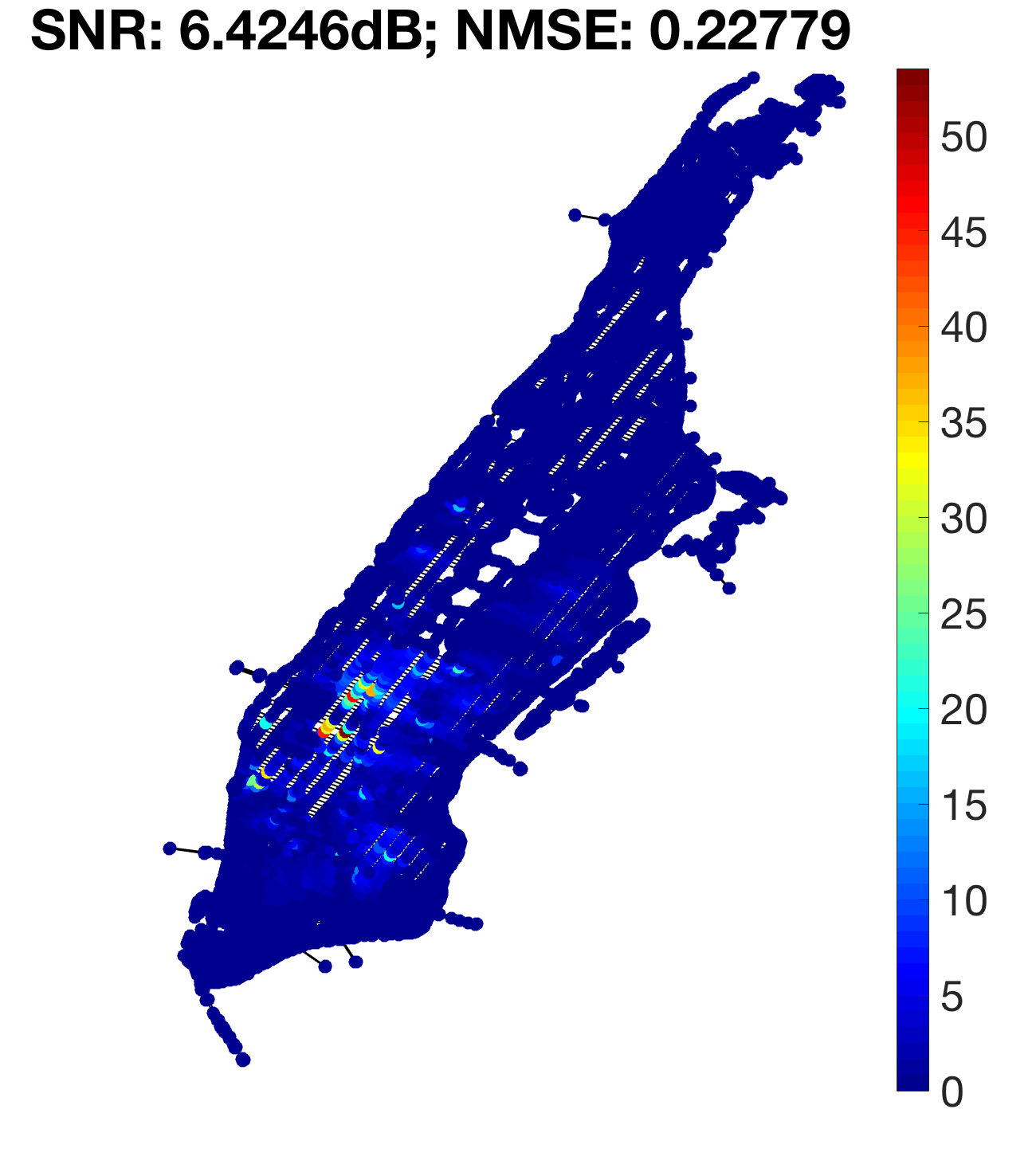} &
 \includegraphics[width=0.35\columnwidth]{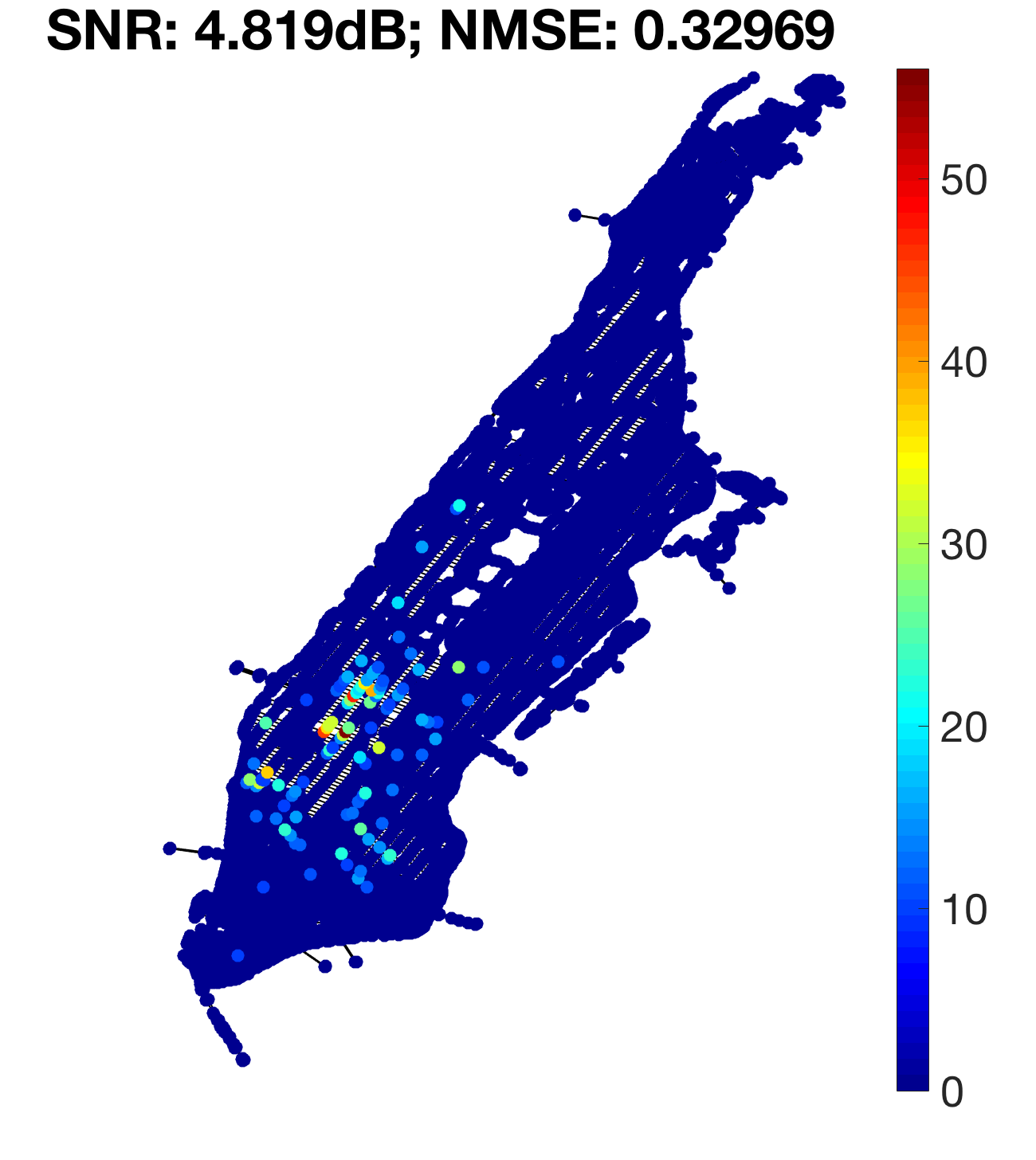}
\\
 {\small (a) Taxi-pickup distribution.} & {\small (b) PS.} & {\small (c) CSFB.}   & {\small (d) SGWT.} & {\small (e) CKWT.}
 \\
 \\
  \includegraphics[width=0.35\columnwidth]{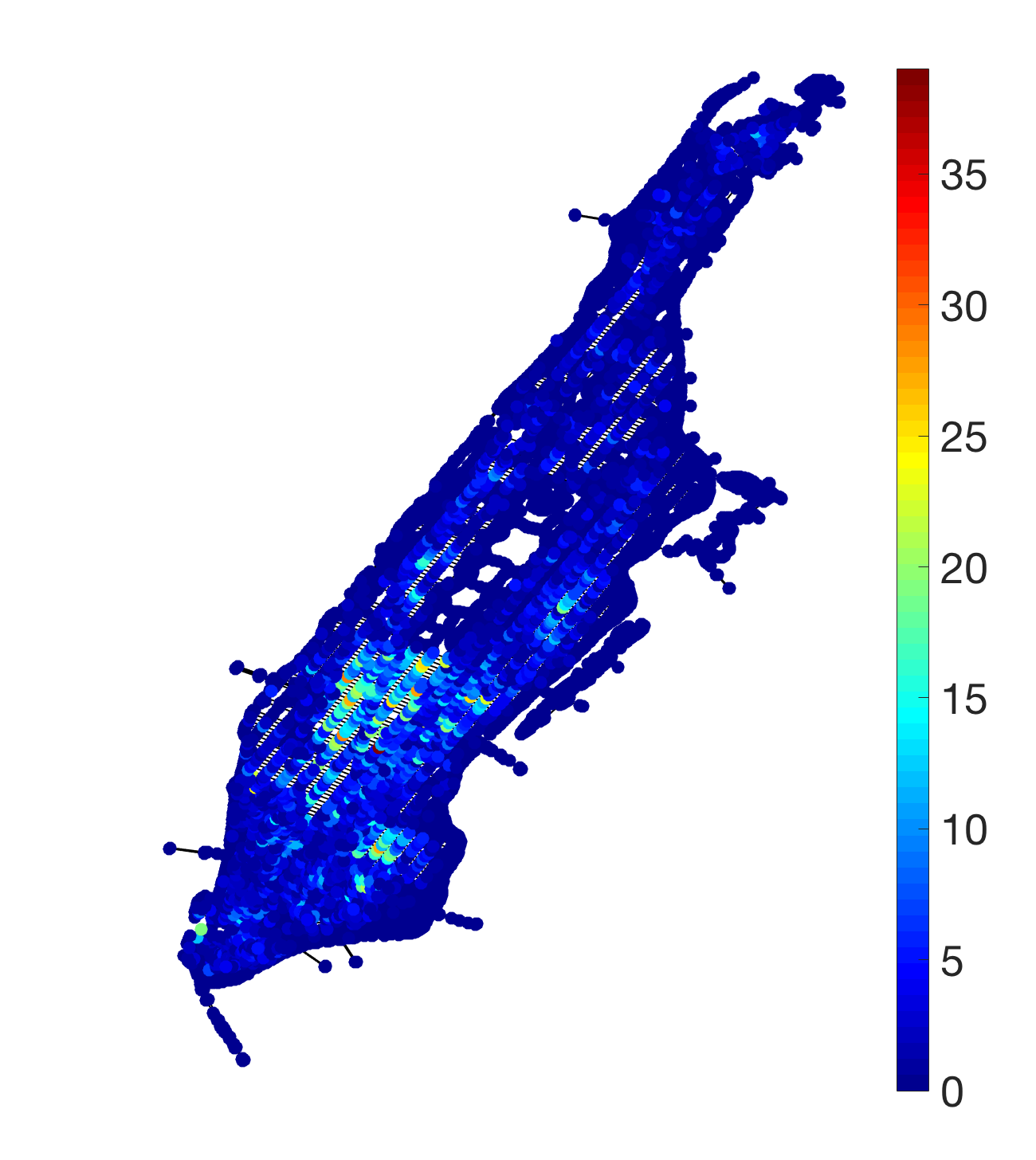} &
 \includegraphics[width=0.35\columnwidth]{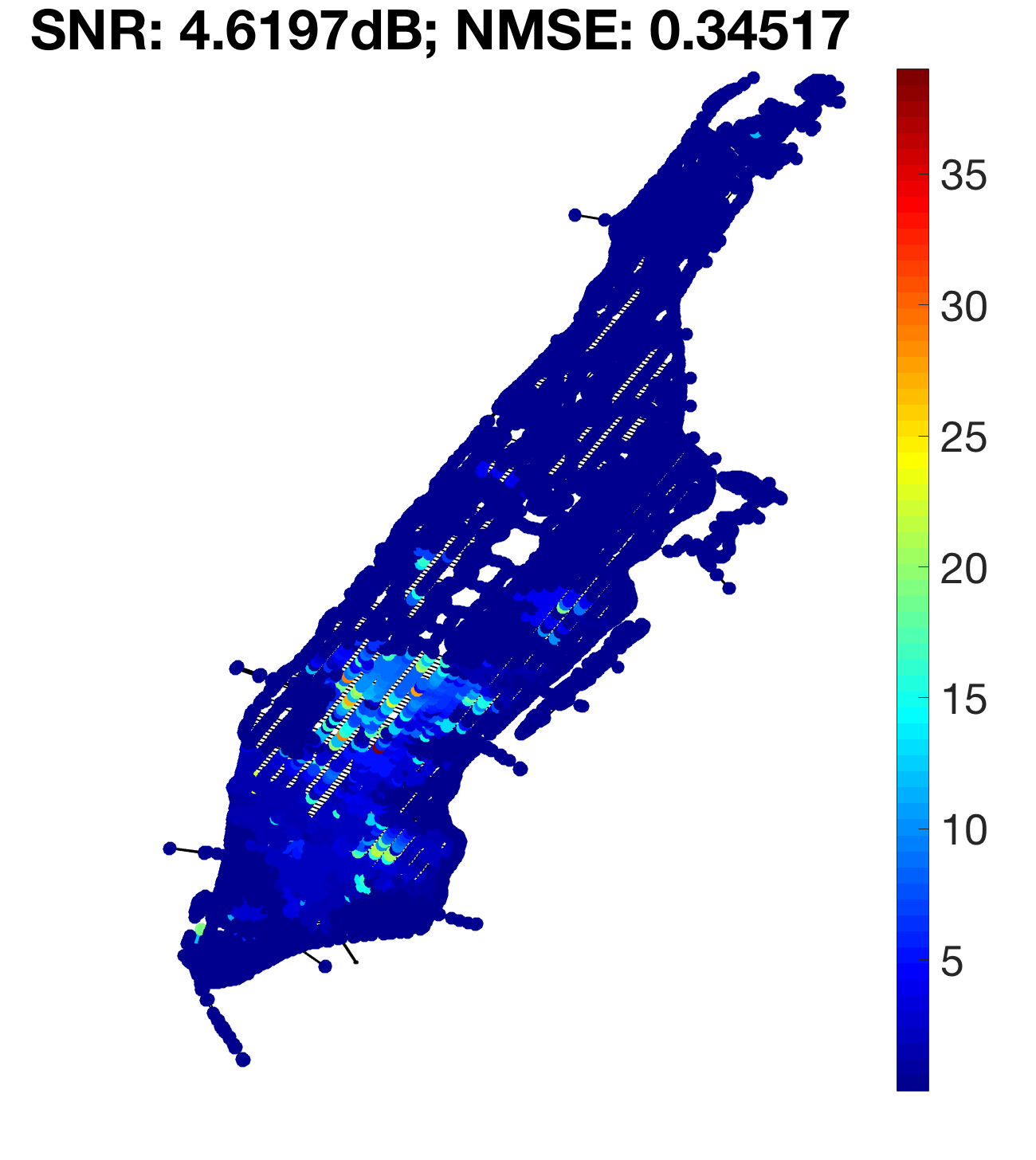}   &
 \includegraphics[width=0.35\columnwidth]{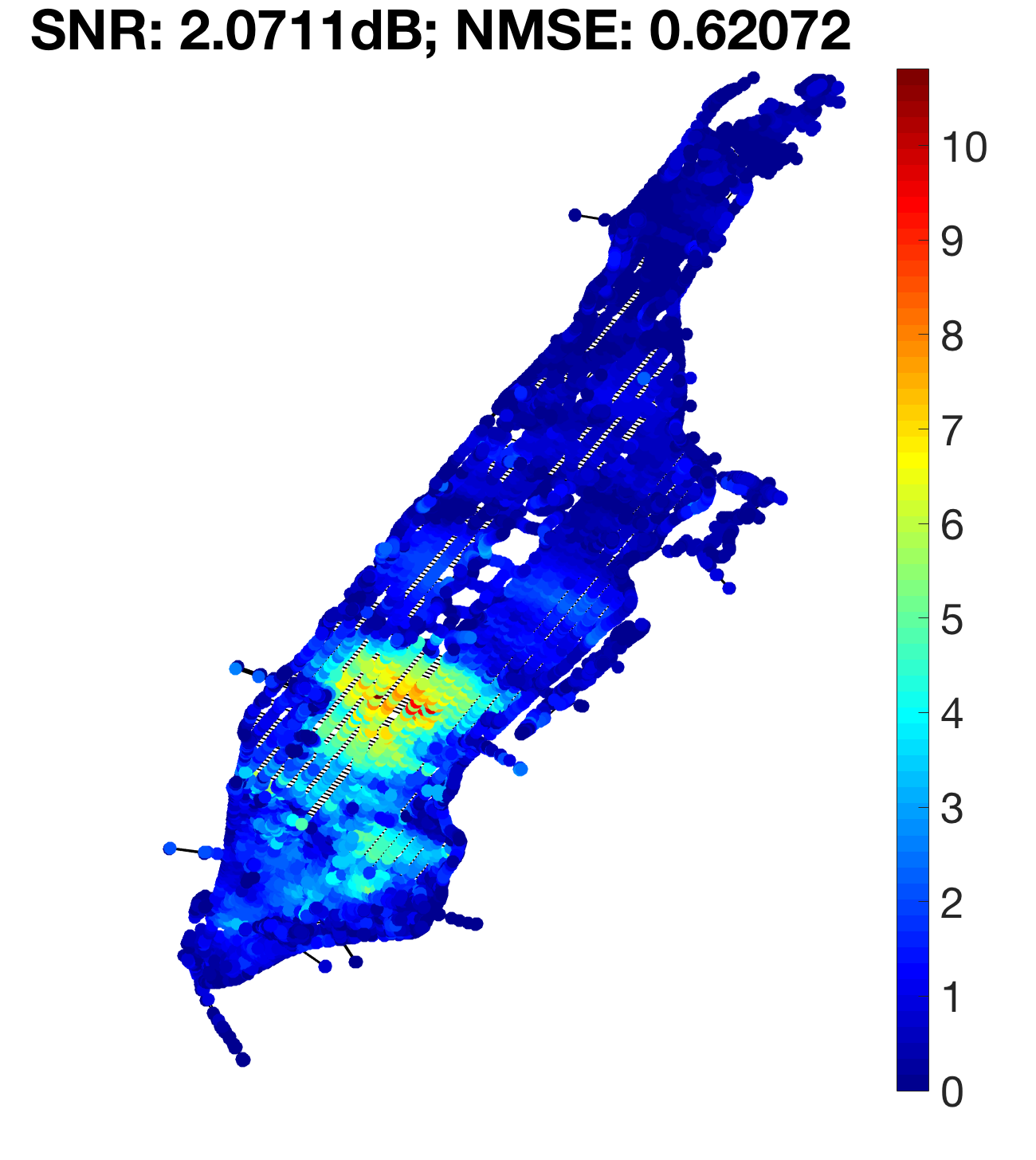}  &
 \includegraphics[width=0.35\columnwidth]{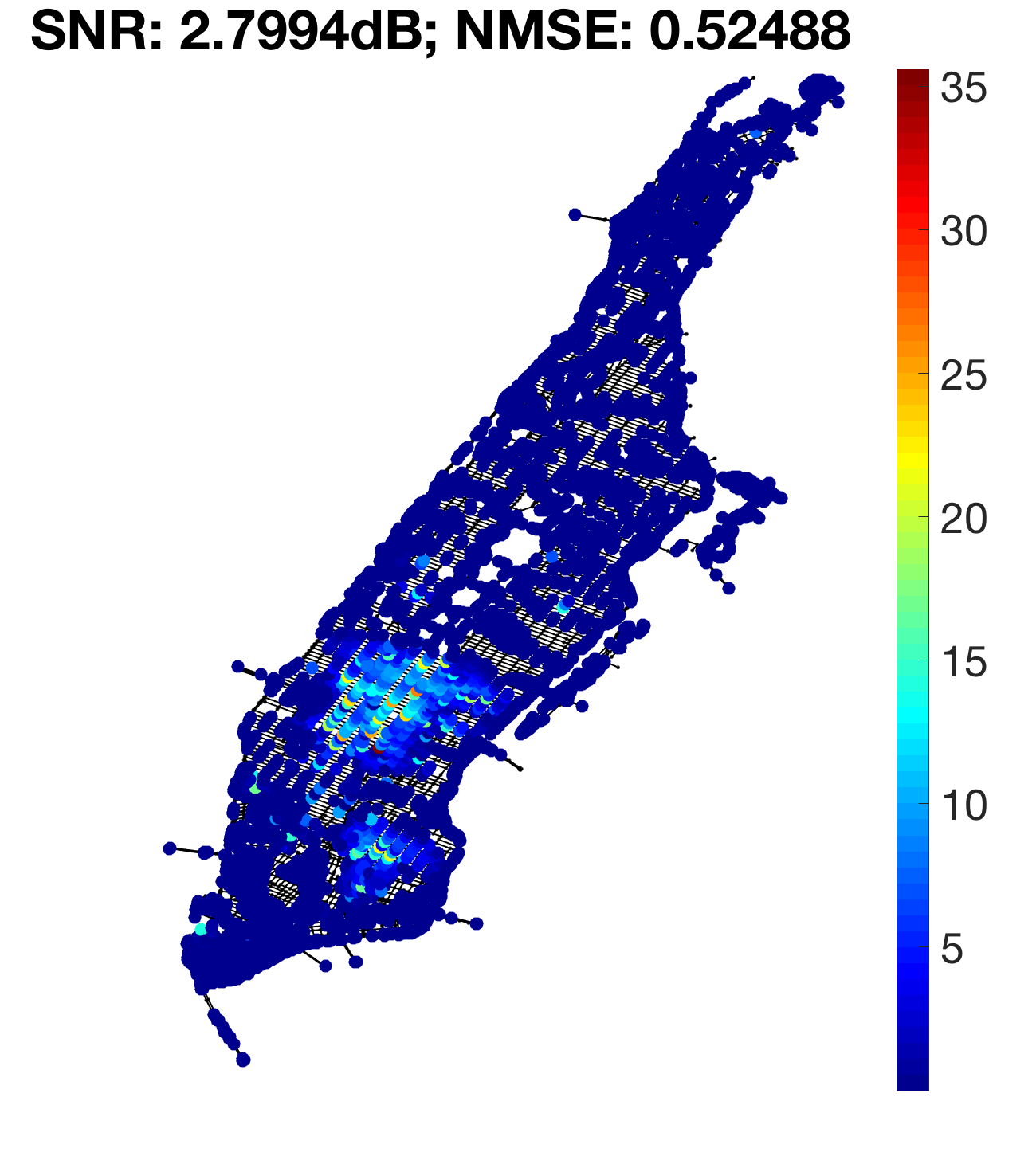} &
 \includegraphics[width=0.35\columnwidth]{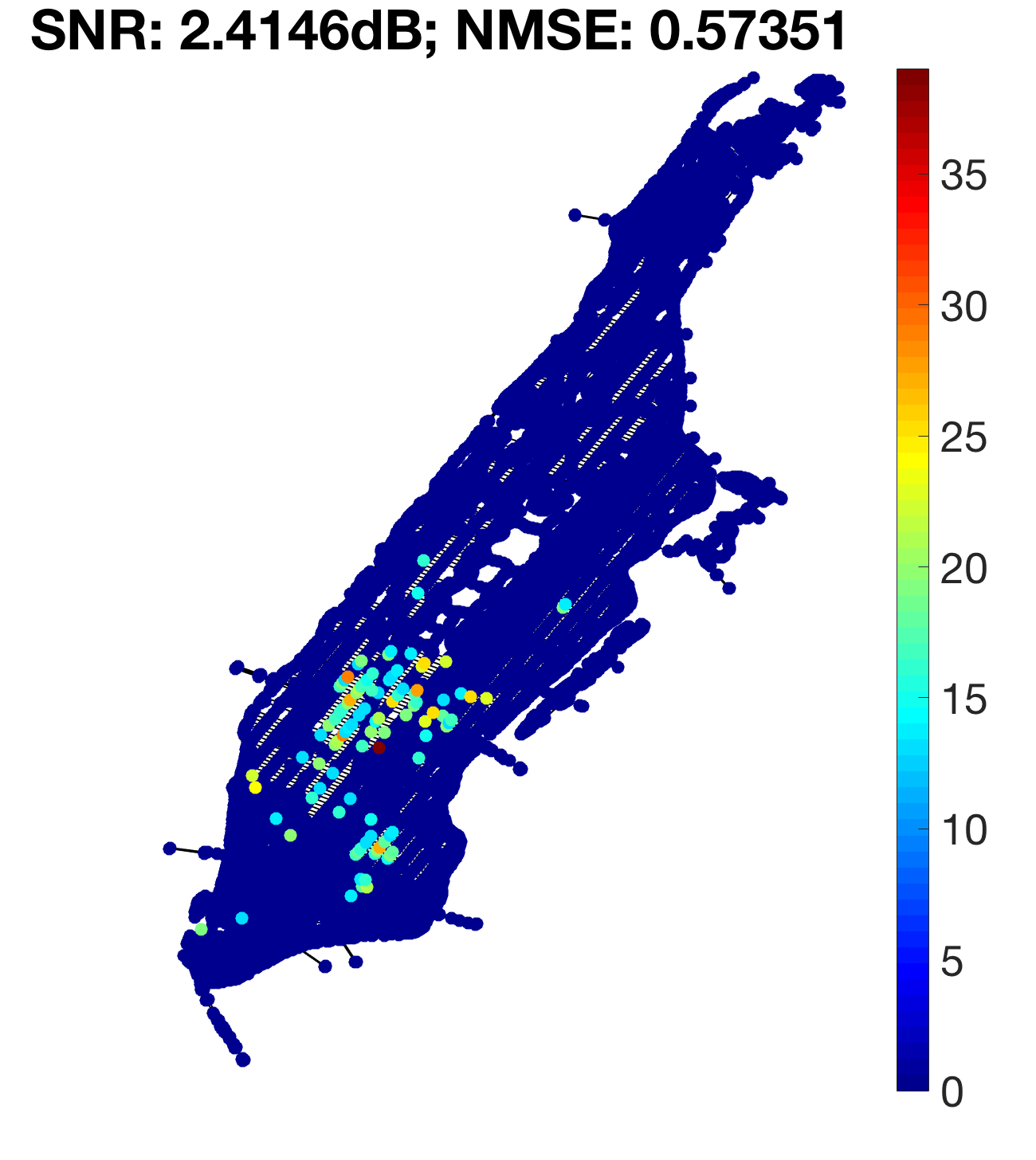}
\\
 {\small (f) Restaurant distribution.} & {\small (g) PS.} & {\small (h) CSFB.}   & {\small (i) SGWT.} & {\small (j) CKWT.}
  \\
\end{tabular}
  \end{center}
   \caption{\label{fig:Manhattan}  Reconstruction visualization for urban data. }
\end{figure*}

Additionally, Figure~\ref{fig:Manhattan} compares the approximations of urban data supported on the Manhattan street networks. The two rows show the reconstructions of the taxi-pickup distribution and restaurant distribution, respectively, by using 100 expansion coefficients.  We see that three graph signals are nonsmooth and inhomogeneous. For each of the three graph signals, the piecewise-smooth graph dictionary provides the largest signal-to-noise ratio (SNR) and smallest normalized mean square error.

The spectral graph wavelet transform is also competitive; the subgraph-based filter bank tends to be over smooth and the spatial graph wavelets tend to be less smooth.

\begin{figure}[htb]
  \begin{center}
    \begin{tabular}{cc}
 \includegraphics[width=0.45\columnwidth]{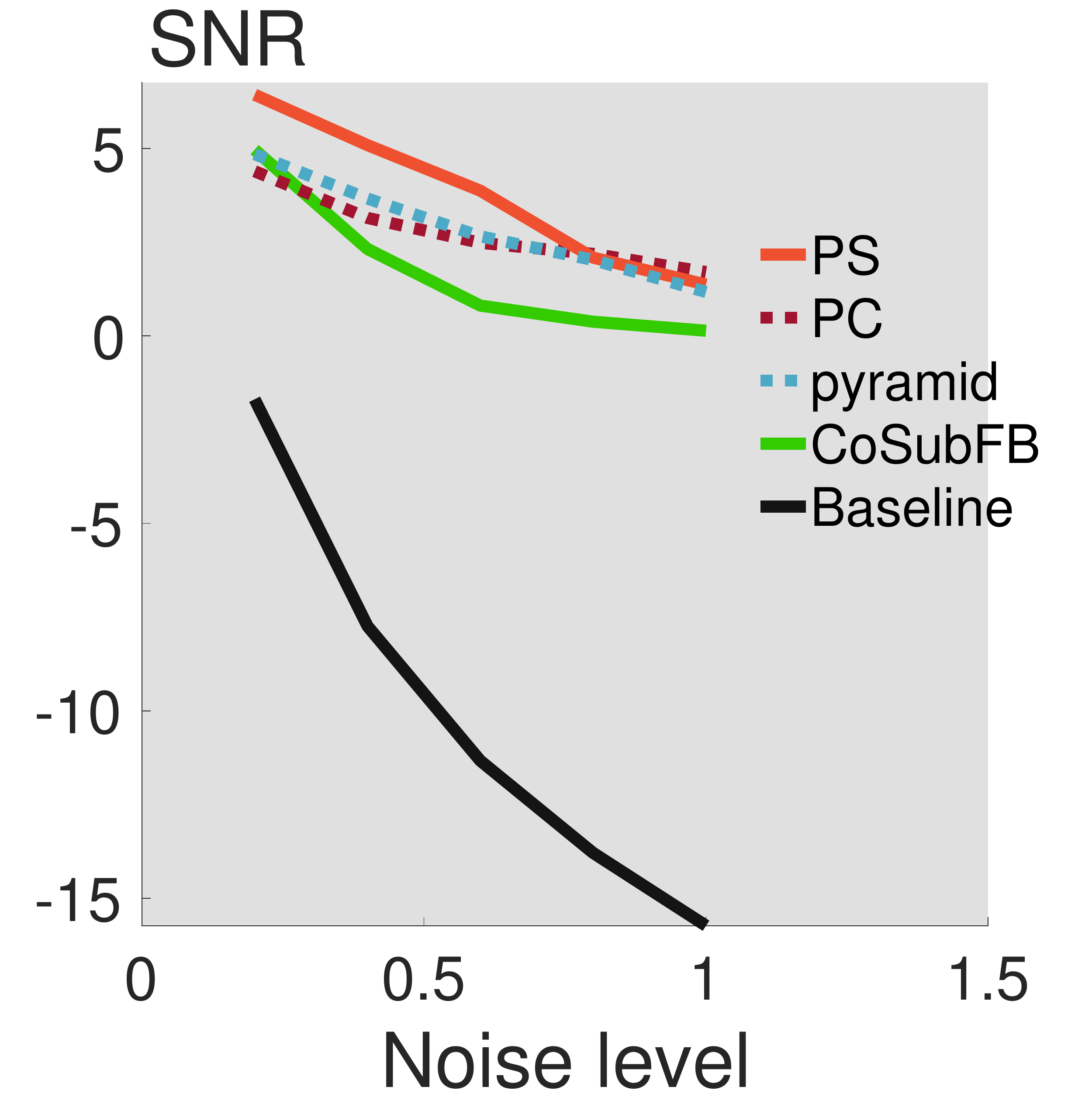} &
 \includegraphics[width=0.45\columnwidth]{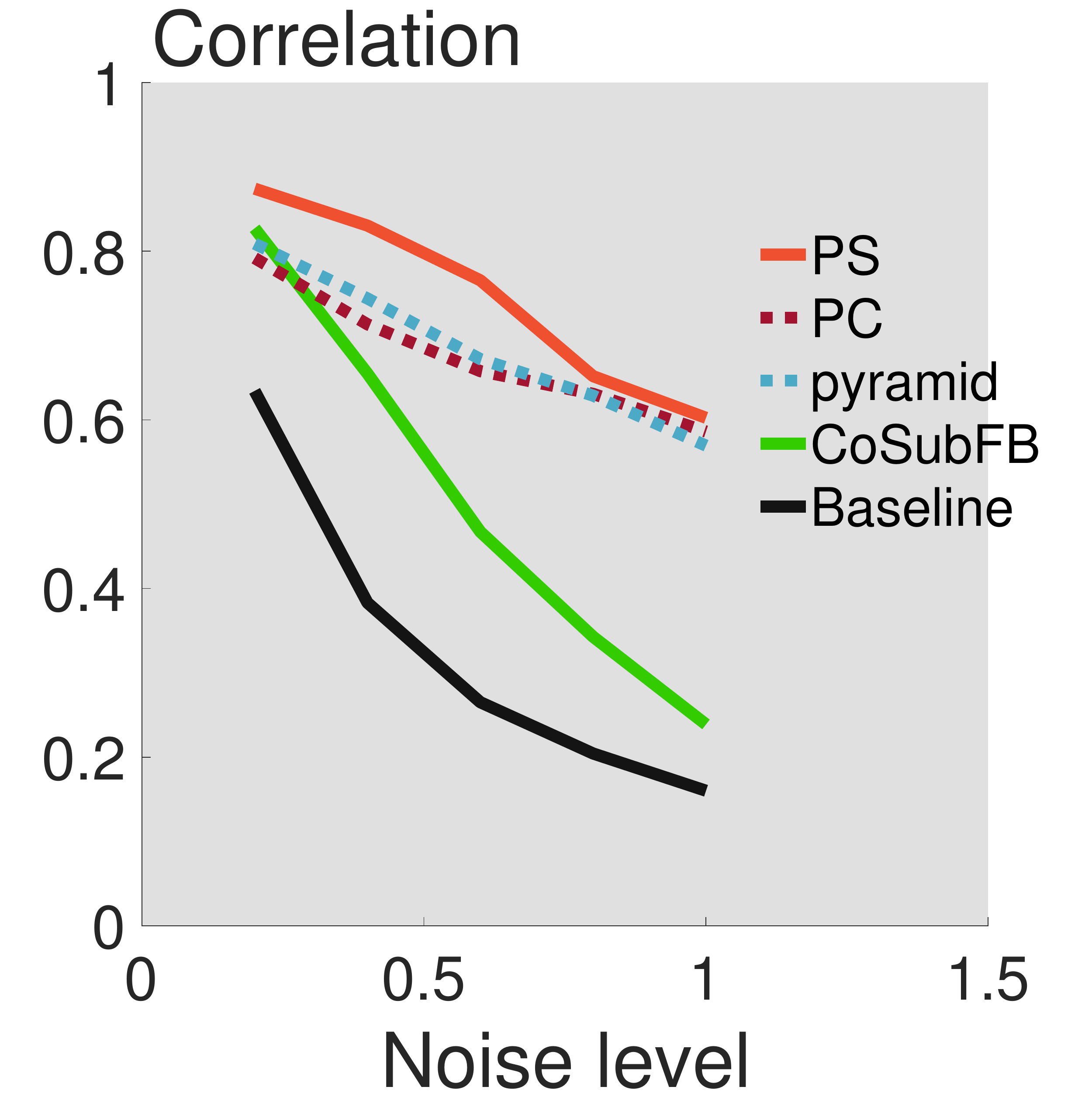}
\\
 {\small (a) Original.} & {\small (b) PS.}
\end{tabular}
  \end{center}
   \caption{\label{fig:localization} Localization performance as a function of noise level. Piecewise-smooth graph dictionary (in red) outperforms the other competitive methods.  The $x$-axis is the noise level and the $y$-axis is the signal-to-noise ratio (SNR), where higher means better.}
\end{figure}

\begin{figure*}[htb]
  \begin{center}
    \begin{tabular}{cccccc}
     \includegraphics[width=0.35\columnwidth]{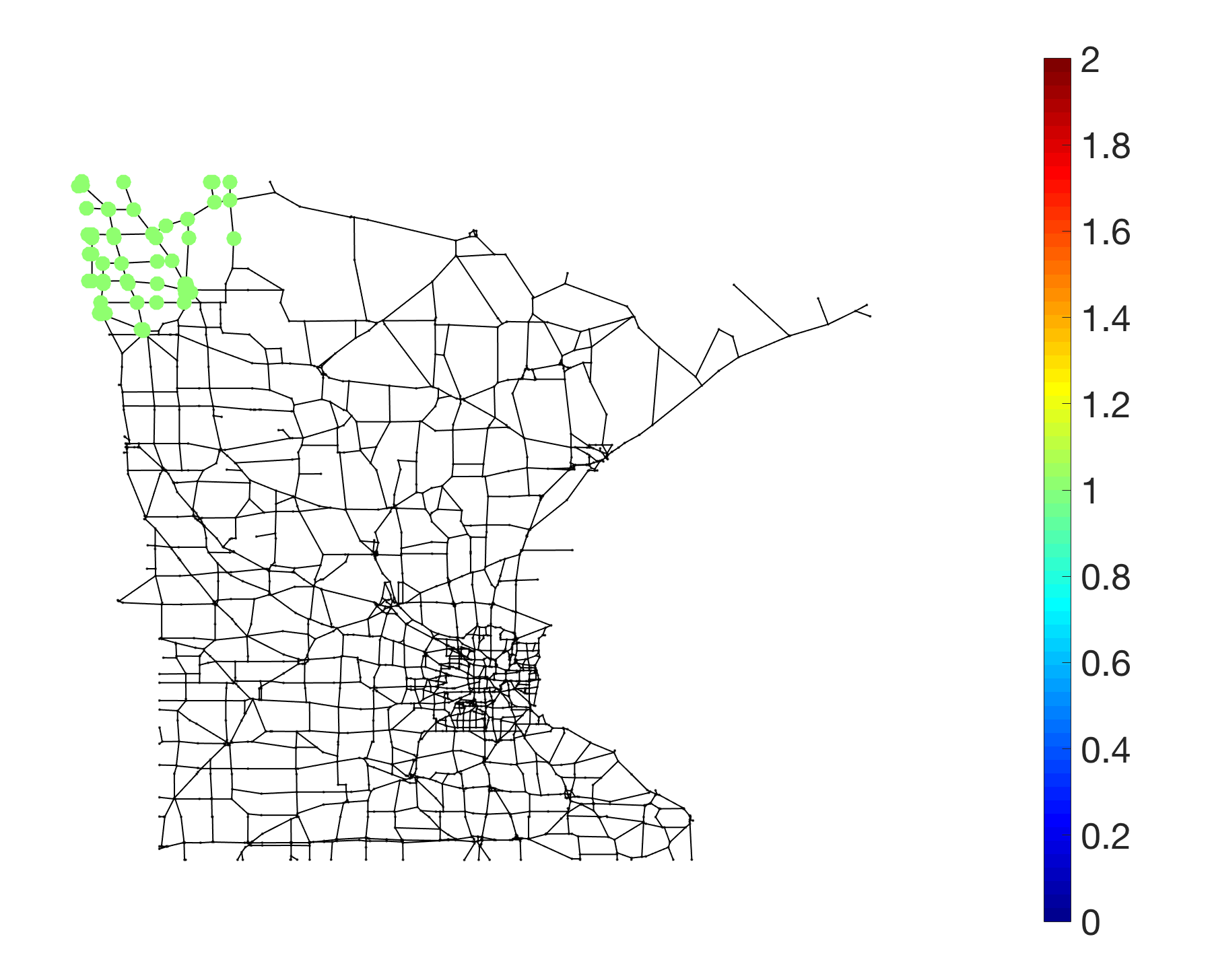}  &
         \includegraphics[width=0.35\columnwidth]{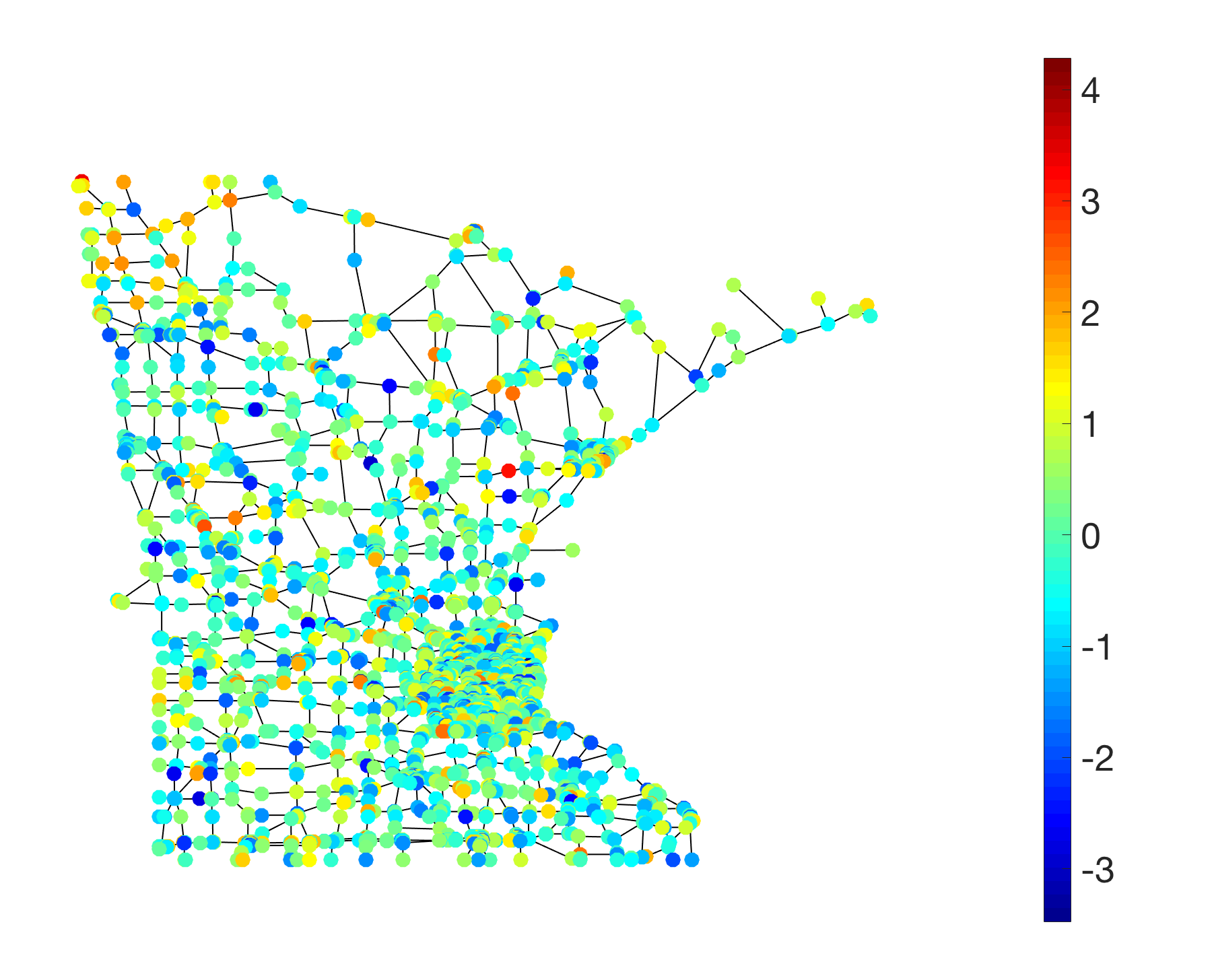} &
 \includegraphics[width=0.35\columnwidth]{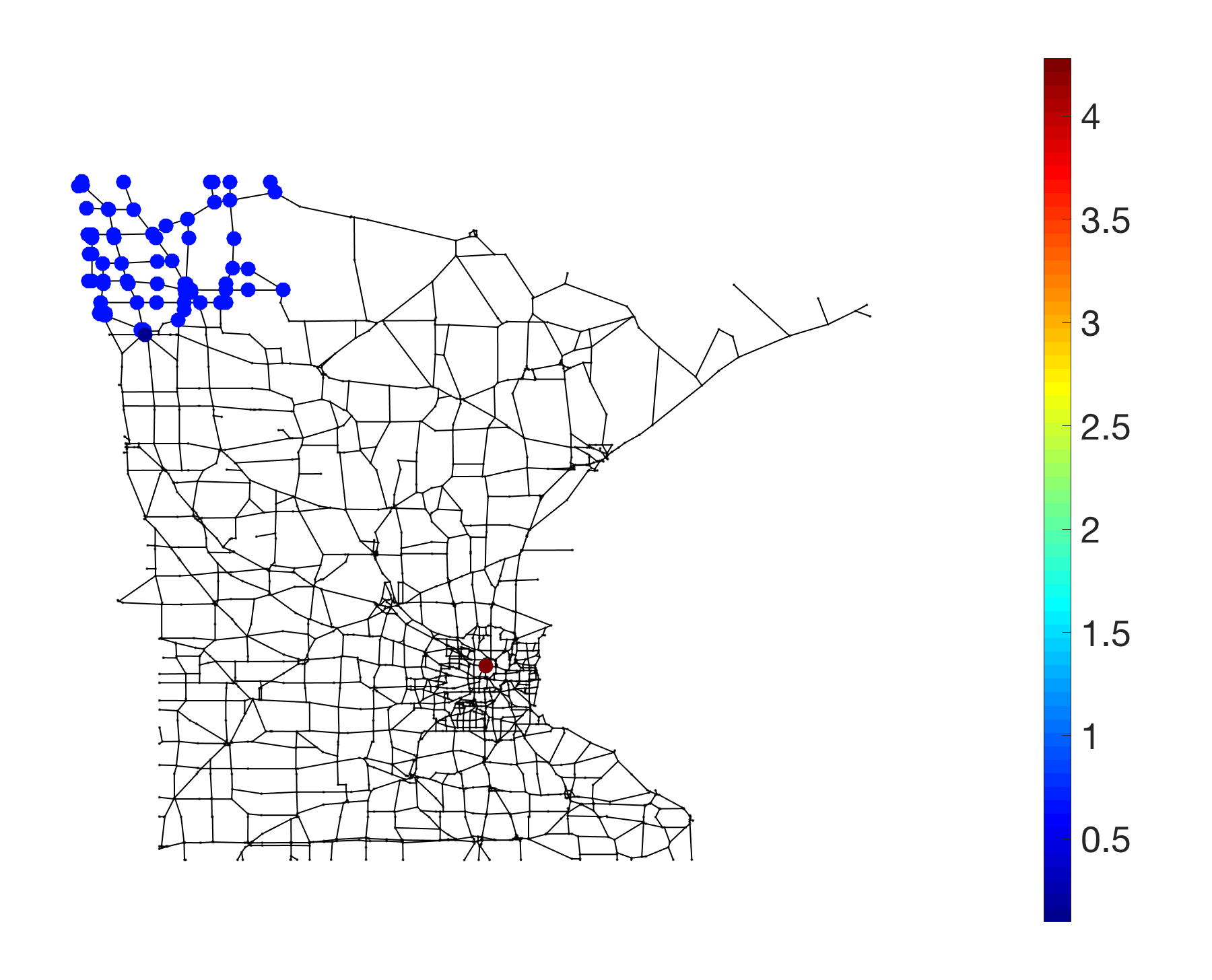} &
  \includegraphics[width=0.35\columnwidth]{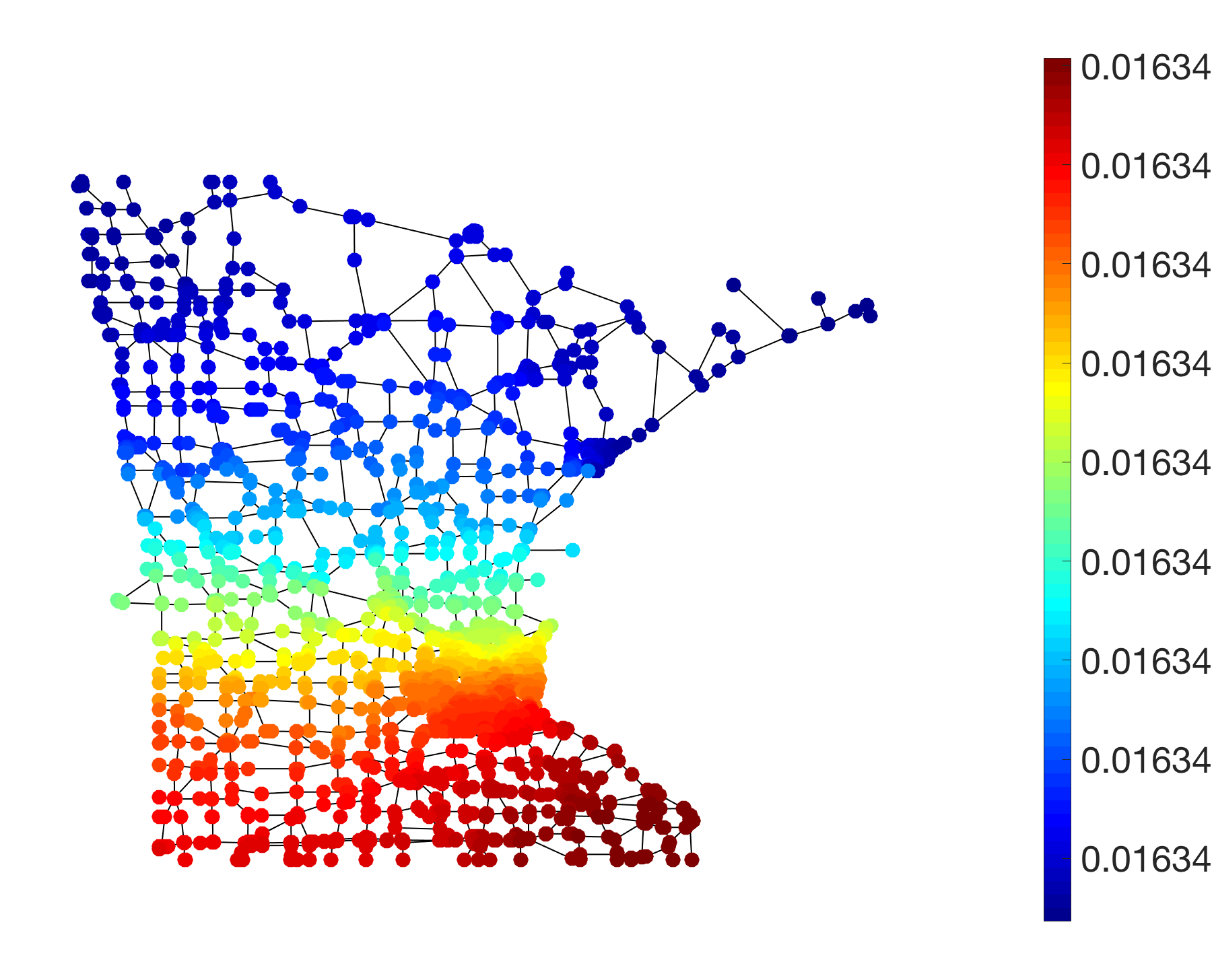} &
      \includegraphics[width=0.35\columnwidth]{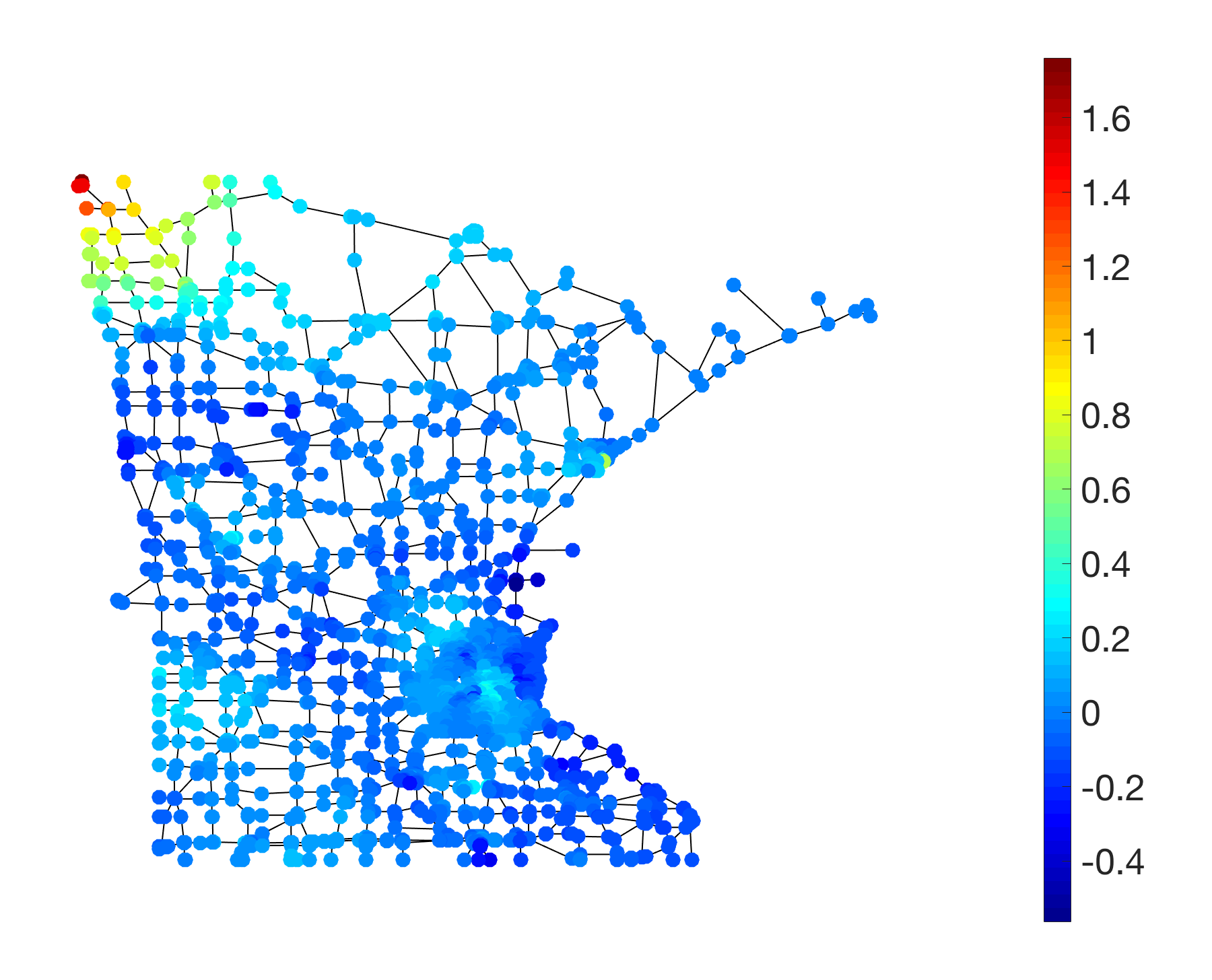}
\\
 {\small (a) Original.} & {\small (b) Noisy.}  & {\small (c) PS.} & {\small (d) CSFT.}  & {\small (e) SGWT.} 
\end{tabular}
  \end{center}
   \caption{\label{fig:localization_plot}  Localization visualization. }
\end{figure*}

\subsection{Localization}
One functionality of a graph dictionary is to detect localized graph signals~\cite{ChenYZSK:17}; applications include localizing virus attacks in cyber-physical systems, localizing stimuli in brain connectivity networks and mining traffic events in city street networks. 
We here consider  simulations on the Minnesota road networks. We generate one-piece graph signals with Gaussian noises. Given the noisy graph signals, we use graph dictionary to remove noises and reconstruct a denoised graph signal to localize the underlying activated pieces. We average over 20 random trials.

Figure~\ref{fig:localization} shows the localization performance, where the $x$-axis is the noise level and the  $y$-axis is either SNR or correlation. In both cases, higher value means better. The baseline (dark curve) means that we naively use the noisy graph signal as the reconstruction. We see that the piecewise-smooth graph dictionary outperforms the others in terms of both metrics, especially when the noise level is low; when the noise level is high, piecewise-constant graph dictionary, piecewise-smooth graph dictionary and multiscale pyramid transform perform similarly.

Figure~\ref{fig:localization_plot} compares reconstructions.   Figure~\ref{fig:localization_plot} (a) shows the original one-piece graph signal, (b) shows the noisy  graph signal,  while
(c), (d) and (e) show the denoised  graph signals by using the piecewise-smooth graph dictionary, the subgraph-based filter bank and the spectral graph wavelet transform, respectively. We see that the piecewise-smooth graph dictionary localizes the underlying piece well, the spectral graph wavelet transform does a reasonable job, but the subgraph-based filter bank provides an over-smooth reconstruction and fails.

\section{ Conclusions and Future Works}
\label{sec:conclusions}
In this paper, we model complex and irregular data, such as urban data supported on the city street networks and profile information supported on the social networks, as piecewise-smooth graph signals. We propose a well-structured and storage-friendly graph dictionary to  represent those graph signals.  To ensure a good representation, we consider the graph multiresolution analysis. To implement this, we propose the coarse-to-fine approach, which iteratively partitions a graph into two subgraphs until we reach individual nodes. This approach efficiently implements the graph multiresolution analysis and the induced graph dictionary promotes sparse representations for piecewise-smooth graph signals. Finally, we test the proposed graph dictionary on the tasks of approximation and localization. The empirical results validate that the proposed graph dictionary outperforms eight other representation methods on various datasets. Future works may include develop sampling, recovery, denoising and detection strategies based on the proposed  piecewise-smooth graph signal model.

\bibliographystyle{IEEEbib}
\bibliography{bibl_jelena}

\appendix
\section{Appendices}

\subsection{Iterated Graph Filter Bank}
\label{sec:one_layer_haar}
In this section, we generalize the classical filter banks to the graph domain and point out why the graph filter banks are hard to implement. Suppose we have an ordering of nodes $\{ v_1, v_2, \hdots, v_{N} \}$, such that two consecutive nodes $v_{2k-1}, v_{2k}$ are connected for $k = 1, 2, \hdots, K$, where $K \leq \lfloor N/2 \rfloor$. We group all pairs of $v_{2k-1}, v_{2k}$ to form a series of connected and nonoverlapping subgraphs. 
The basis vectors of the $k$th subgraph are
\begin{eqnarray}
\vv^{(1)}_k  & = &  \frac{1}{\sqrt{2}} \left( \one_{v_{2k-1}} +  \one_{v_{2k}} \right) \in \R^N,
\\
\uu^{(1)}_k  & = &  \frac{1}{\sqrt{2}} \left( \one_{v_{2k-1}} -  \one_{v_{2k}} \right)  \in \R^N,
\end{eqnarray}
where the subscript $k$ is the index of the subgraph and the superscript $1$ indicates the root layer,  the low-pass basis vector $\vv^{(1)}_k$ considers the average of two nodes within this subgraph and the high-pass basis sequence $\uu^{(1)}_k$ considers the difference between two nodes within this subgraph.  We collect all the low-pass basis vectors and high-pass basis vectors to form a low-pass subspace and a high-pass subspace, respectively,
\begin{eqnarray*}
V^{(1)}  =  {\rm span} \left(  \{ \vv^{(1)}_k \}_{k=1}^{K} \right)~{\rm and}~ U^{(1)} = {\rm span} \left( \{ \uu^{(1)}_k \}_{k=1}^{K}  \right).
\end{eqnarray*}
Different from the discrete-time scenario, $V^{(1)} \oplus U^{(1)}$ may not span the entire $\R^N$ space, as a few nodes may be isolated due to the ordering. Let the residual subspace be $R^{(1)}  =  {\rm span} \left(  \{ \one_{v_k} \}_{k=2K+1}^{N} \right)$, where each basis vector only activates an individual node. Now $V^{(1)} \oplus U^{(1)} \oplus R^{(1)} = \R^N$. For any graph signal $\x \in \R^N$, the reconstruction is
\begin{eqnarray*}
  \x =
\underbrace{ \sum_{k = 1}^{K}   
\langle \x, \vv^{(1)}_k \rangle   \vv^{(1)}_k }_{\x_{V^{(1)}}} + 
\underbrace{\sum_{k = 1}^{K}   \langle \x, \uu^{(1)}_k \rangle  \uu^{(1)}_k}_{\x_{U^{(1)}}} + 
\underbrace{ \sum_{k = 2K+1}^{N}  \langle \x, \one_{v_k} \rangle \one_{v_n}  }_{\x_{R^{(1)}}},
\end{eqnarray*}
where $\x_{V^{(1)}} \in V^{(1)}$ is the low-pass projection, $\x_{U^{(1)}} \in U^{(1)}$ is the high-pass projection and $\x_{R^{(1)}} \in R^{(1)}$ handles the residual condition.

To summarize, based on a well-designed ordering, we partition the entire graph into a series of nonoverlapping subgraphs and then design the Haar-like basis vectors on graphs. For discrete-time signals whose underlying graph is a directed line graph,  the ordering is provided by time and each subgraph contains two consecutive time stamps.  As described in Section~\ref{sec:FB},  because of the nice ordering by time, all the basis vectors can be efficiently obtained by filtering following by downsampling; however, this is not true for arbitrary graphs.

\begin{figure}[t]
  \begin{center}
     \includegraphics[width= 0.95\columnwidth]{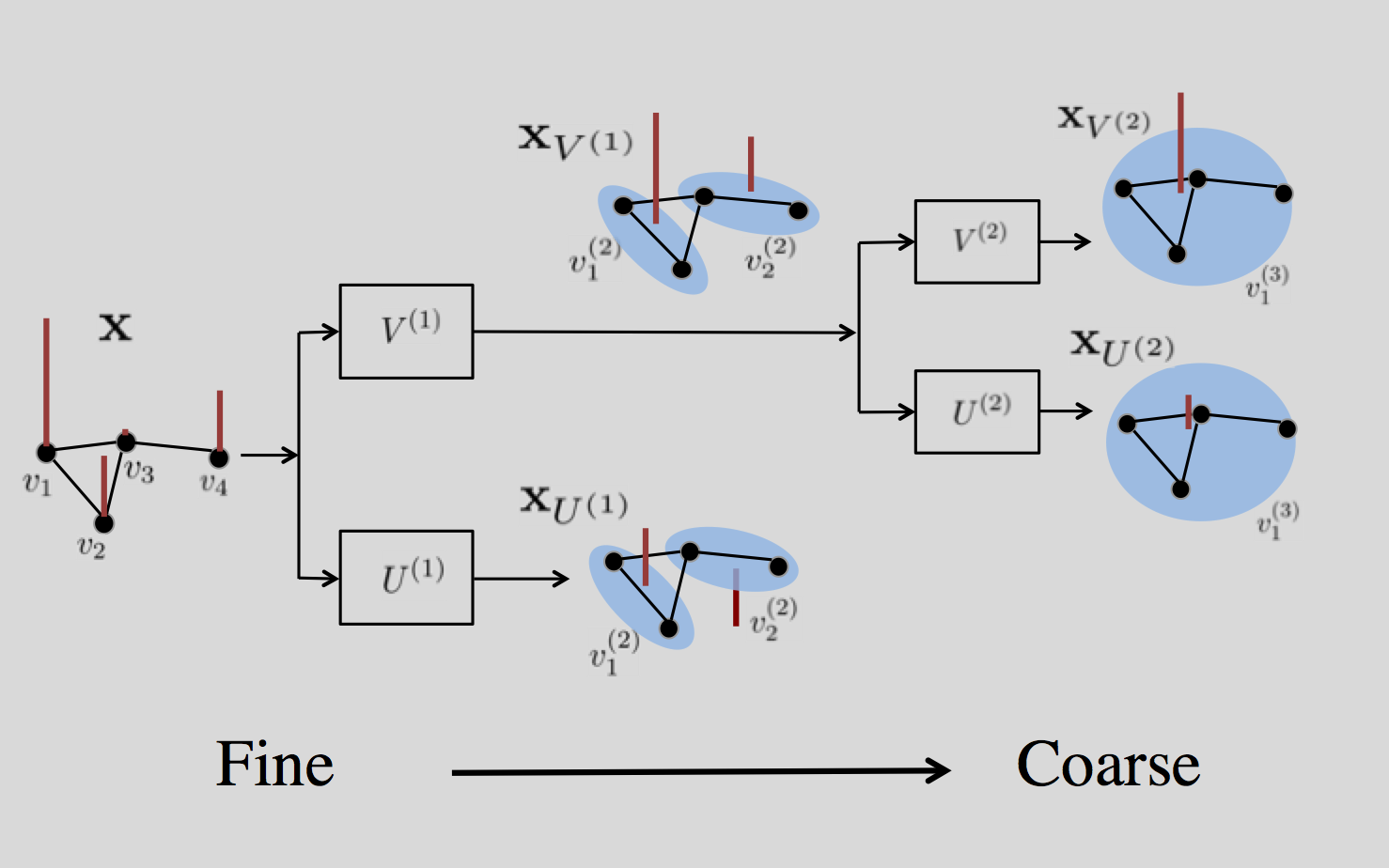}
  \end{center}
  \caption{\label{fig:local2global} As a fine-to-coarse approach, the analysis part of iterated graph filter banks implement the graph multiresolution analysis (Definition~\ref{df:MRA}). There is no residual in this case. }
\end{figure}

Following the classical discrete-time signal processing,
we can iteratively decompose the low-pass subspace and obtain smoother and smoother subspaces, which is equivalent to coarsen in the graph vertex domain. This iterated graph filter bank  divides the vertex-spectrum plane into more tiles,  approaching to the limit of uncertainty barrier. Here we show the second layer for an example. Let a supernode (connected node set) $v^{(2)}_{k} = v_{2k-1} \cup v_{2k}$ for $k = 1, 2, \hdots, K$, where the superscript of the supernode indicates the second layer. Two supernodes  $v^{(2)}_{i}, v^{(2)}_{j}$ are connected when there exists a pair of nodes $p \in v^{(2)}_{i}, q \in v^{(2)}_{j}$ satisfying that $p,q$ are connected. Similarly to the paradigm in Section~\ref{sec:one_layer_haar}, suppose we have an ordering of $K$ supernodes $\{ v^{(2)}_{1}, v^{(2)}_{2}, \hdots, v^{(2)}_{K} \}$, such that two consecutive supernodes $v^{(2)}_{2k-1}, v^{(2)}_{2k}$ are connected for $k =  1, \hdots, K^{(2)}$, where $K^{(2)} \leq \lfloor K/2 \rfloor$. We group all $v^{(2)}_{2k-1}, v^{(2)}_{2k}$ to form a series of connected, yet nonoverlapping subgraphs.  Let $S_1, S_2 \subset \V$ be two nonoverlapping supernodes. We define the low-pass and high-pass Haar template basis vector are, respectively, 
\begin{eqnarray*}
 g(S_1, S_2) & = &  \sqrt{ \frac{S_1| |S_2|}{|S_1|+|S_2|}} \left(  \frac{ {\bf 1}_{S_1} }{|S_1|}  + \frac{ {\bf 1}_{S_2 }}{|S_2|}  \right)  \in \R^N,
 \\
  h(S_1, S_2) & = &  \sqrt{ \frac{S_1| |S_2|}{|S_1|+|S_2|}} \left(  \frac{ {\bf 1}_{S_1} }{|S_1|}  - \frac{ {\bf 1}_{S_2 }}{|S_2|}  \right)  \in \R^N.
\end{eqnarray*}
Following from the template, the basis vectors of the $k$th subgraph are
\begin{eqnarray*}
\vv^{(2)}_k  & = &  g( v^{(2)}_{2k-1}, v^{(2)}_{2k}),
\\
\uu^{(2)}_k  & = & h( v^{(2)}_{2k-1}, v^{(2)}_{2k}).
\end{eqnarray*}
We collect all the low-pass basis vectors and high-pass basis vectors in the second layer to form a low-pass subspace and a high-pass subspace, respectively,
\begin{eqnarray*}
V^{(2)}  =  {\rm span} \left(  \{ \vv^{(2)}_k \}_{k=1}^{K^{(2)}} \right)~{\rm and}~ U^{(2)} = {\rm span} \left( \{ \uu^{(2)}_k \}_{k=1}^{K^{(2)}}  \right).
\end{eqnarray*}
Let the residual subspace be $R^{(2)}  =  {\rm span} \left(  \{ \one_{v^{(2)}_n} \}_{n=2K^{(2)}+1}^{K} \right)$, where each basis vector only activates an individual supernode. Now $V^{(2)} \oplus U^{(2)} \oplus R^{(2)} = V^{(1)}$. 

For any graph signal $\x \in \R^N$, the reconstruction is
\begin{eqnarray*}
  \x  & = &  \underbrace{ \sum_{k = 1}^{K^{(2)}}   
\langle \x, \vv^{(2)}_k \rangle   \vv^{(2)}_k }_{ \in V^{(2)}} + \underbrace{  \sum_{k = 1}^{K^{(2)}}   \langle \x, \uu^{(2)}_k \rangle  \uu^{(2)}_k  }_{ \in U^{(2)}} + 
\underbrace{  \sum_{k = 1}^{K}   \langle \x, \uu^{(1)}_k \rangle  \uu^{(1)}_k  }_{ \in U^{(1)}}  
\\
&&
+ \underbrace{  \sum_{k = 2K+1}^{N}  \langle \x, \one_{v_k} \rangle \one_{v_k}  }_{ \in R^{(1)}} + \underbrace{ \sum_{k = 2K^{(2)+1}}^{K}  \langle \x, \one_{v^{(2)}_k} \rangle \one_{v^{(2)}_k} }_{ \in R^{(2)}}.
\end{eqnarray*}
We can keep decomposing the low-pass subspace until there is only one constant basis vector. During the iterated decomposition, we keep coarsening in the graph vertex domain, leading to larger supernodes and more global-wise basis vectors; we thus call this a~\emph{fine-to-coarse approach}; see Figure~\ref{fig:local2global}. 

Let the decomposition depth be $L$. By induction, the general reconstruction is
\begin{eqnarray*}
  \x  & = &  \underbrace{ \sum_{k = 1}^{K^{(L)}}   
\langle \x, \vv^{(L)}_k \rangle   \vv^{(2)}_k }_{ \in V^{(L)}} 
+ \sum_{\ell=1}^{L} \underbrace{  \sum_{k = 1}^{K^{(\ell)}}   \langle \x, \uu^{(\ell)}_k \rangle  \uu^{(\ell)}_k  }_{ \in U^{(\ell)}}
\\
&&
+  \sum_{\ell=1}^{L}  \underbrace{  \sum_{k = 2K^{(\ell)}+1}^{K^{(\ell-1)}}  \langle \x, \one_{v_k^{(\ell)}} \rangle \one_{v_k^{(\ell)}}  }_{ \in R^{(\ell)}},
\end{eqnarray*}
where $v_k^{(1)} = v_k, K^{1} = K$ and $K^{0} = N$.

Note that for discrete-time signals, the ordering of time stamps is naturally provided by time, leading to straightforward downsampling and shifting, and iterated filter banks, as a fine-to-coarse approach, are efficient architectures to implement the multiresolution analysis. For graph signals, the ordering in each multiresolution level is unknown and an efficient fine-to-coarse approach to implement the graph multiresolution analysis is not straightforward any more. This is why we consider the coarse-to-fine approach in this paper; in other words,
 we convert the problem of node ordering to the problem of graph partitioning, which is more efficient and straightforward.

\subsection{Proof of Theorem~\ref{thm:basis} }
\label{app:basis}
\begin{proof}
First, we show each vector has norm one.
\begin{eqnarray*}
&& \left\| \sqrt{ \frac{|S_1| |S_2|}{|S_1|+|S_2|}} \left(  \frac{1}{ | S_1 |} {\bf 1}_{S_1} - \frac{1}{| S_2 |} {\bf 1}_{S_2}\right)  \right\|_2^2 
\\
&  \stackrel{(a)}{=} & 
 \left\| \sqrt{ \frac{|S_1| |S_2|}{|S_1|+|S_2|}} \frac{  {\bf 1}_{S_1}  }{ | S_1 |}  \right\|_2^2 + \left\| \sqrt{ \frac{|S_1| |S_2|}{|S_1|+|S_2|}} \frac{  {\bf 1}_{S_2}  }{ | S_2 |}  \right\|_2^2 
\\
& = & 1,
\end{eqnarray*}
where $(a)$ follows from that $S_1 \cap S_2 = \emptyset$.
Second, we show each vector is orthogonal to the other vectors. We have
\begin{eqnarray*}
{\bf 1}^T \w = \sqrt{ \frac{|S_1| |S_2|}{|S_1|+|S_2|}} \left( \sum_{i \in S_1} \frac{1}{ | S_1 |}  -  \sum_{i \in S_2} \frac{1}{| S_2 |}  \right) = 0.
\end{eqnarray*}
Thus, each vector is orthogonal to the first vector, ${\bf 1}_{\V}/\sqrt{|{\V}| }$. Each other individual vector is generated from two node sets. Let $S_1, S_2$ generate $\w_i$ and $S_3, S_4$ generate $\w_j$. Due to the construction, there are only two conditions, two node sets of one vector belong to one node set of the other vector, and all four node sets do not share element with each other. For the first case, without losing generality, let $\left( S_3 \cup S_4 \right) \cap S_1=  S_3 \cup S_4$, we have
\begin{eqnarray*}
 && \w_i^T \w_j 
 \\
& = &  \sqrt{ \frac{|S_1| |S_2|}{|S_1|+|S_2|}
 \frac{|S_3| |S_4|}{|S_3|+|S_4|}} \left( \sum_{i \in S_3} \frac{1}{ | S_3 |}  -  \sum_{i \in S_4} \frac{1}{| S_4 |} \right) 
\\ 
& = & 0.
\end{eqnarray*}
For the second case, the inner product between  $\w_i$ and $\w_j $ is zero because their supports do not match.
Third, we show that $\W$ spans $\R^N$. Since we recursively partition the node set until the cardinalities of all the node sets are smaller than 2, there are $N$ vectors in $\W$.
\end{proof}

\subsection{Proof of Theorem~\ref{thm:partition} }
\label{app:partition}
\begin{proof}
We first show that $\V_1, \V_2$ are connected and then bound the cardinality difference. Since the original graph is connected, $D_{v_i, v_j}$ is finite, where $v_i, v_j$ are two hubs. In Step 4, we partition the nodes according to their distances to two hubs. Every node in the node set $S_1$ is connected to $v_j$; thus, the subgraph induced by the node set $S_1$ is connected. In Step 5, we partition the boundary set $S_2$ into connected node sets, $C_1, C_2, \cdots, C_M$, and each of them connects to $S_1$; otherwise, the maximum element in the geodesic distance matrix $\D$ is infinity. We thus have $S_1 \cup C_1 \cup C_2 \cdots \cup C_{m}$ is connected for all $m=1, \cdots, M$. When we set $m = m^*$ obtained in Step 7, we have $\V_1 = S_1 \cup C_1 \cup C_2 \cdots \cup C_{m^*}$ is connected. Similarly, we can show that $\V_2$ is also connected. 

In Step 3, we set $p$ as the median value of the differences to two hubs, which sets $|S_1|$ around $|\V_0|/2$. In Step 6, we sequentially add connected components to $S_1$ and finally choose the one, whose cardinality is closest  to $|\V_0|/2$. The last 
component added to $S_1$ is $C_{m^*}$, which ensures that 
$| |\V_1| - |\V_0|/2 | \leq |C_{m^*}|$ and $| |\V_2| - |\V_0|/2 | \leq |C_{m^*}|$.
\end{proof}

\subsection{Proof of Theorem~\ref{thm:sparse} }
\label{app:sparse}

\begin{proof}
When an edge $e \in {\rm supp}(\Delta \w)$, where $\w$ is one basis vector in the graph wavelet basis $\W$, $\Delta$ is the graph incident matrix, and supp denotes the edge indices activated by the  nonzero elements of $\Delta \w$; we call that the edge $e$ is activated by the wavelet basis vector $\w$. Since in each level, the pieces are disjoint, each edge will be activated at most once in each level; in total, each edge will be activated by at most $L$ wavelet basis vectors, where $L$ the decomposition level. Let activations($e$) be the number of wavelet basis vectors in $\W$ that activates $e$. 
\begin{eqnarray*}
\left\| \W^T \x \right\|_0  & \leq & 1+\sum_{ e \in {\rm Supp}(\Delta \w) }  {\rm activations}(e) 
\\
& \leq & 1+\left\|  \Delta \x \right\|_0  L,
\end{eqnarray*}
where $1$ comes from the activation of the first column vector, which is constant. Since we promote the bisection scheme, the decomposition level $L$ is roughly $1+\log_2 N$. 
\end{proof}

\subsection{Proof of Theorem~\ref{thm:PPL} }
\label{app:PPL}

\begin{proof}
The main idea is that we approximate a bandlimited signal in the original graph by using bandlimited signals in subgraphs. Based on the eigenvectors of graph Laplacian matrix, we define the bandlimited space, where each signal can be represented as $\x = \Vm_{(K)} \a,$ where $\Vm_{(K)}$ is the submatrix of $\Vm$ containing the first $K$ columns in $\Vm$. We can show that this bandlimited space is a subspace of the small-variation space $\{\x: \x^T \LL \x \leq  \lambda_K  \x^T \x \}$. 

\begin{eqnarray*}
&& \x^T \LL \x 
\  = \  \sum_{i,j \in \E} \W_{i,j} (x_i - x_j)^2
\\
& = & \sum_{S_c} \sum_{i,j \in \E_{S_c}} \W_{i,j} (x_i - x_j)^2  + \sum_{i,j \in (\E / \cup_{c} \E_{S_c}) } \W_{i,j} (x_i - x_j)^2
\\
& = &  \sum_{S_c} \x_{S_c}^T \LL_{S_c} \x_{S_c} + \x^T \LL_{\rm cut} \x \ \leq \    \lambda_K  \x^T \x,
\end{eqnarray*}
where $\LL_{S_c}$ is the graph Laplacian matrix of the subgraph $G_{S_c}$ and  $\LL_{S_c}$ stores the residual edges, which are cut in the graph partition algorithm.

Thus, $\{\x: \x^T \LL \x \leq  \lambda_K  \x^T \x \}$ is a subset of $\bigcup_{S_c}  \{\x_{S_c}: \x_{S_c}^T \LL_{S_c} \x_{S_c} \leq  \lambda_K  \x^T \x - \x^T \LL_{\rm cut} \x  \}$; that is, any  small-variation graph signal in the whole graph can be precisely represented by small-variation graph signals in the subgraphs.

In each local set, when we use the bandlimited space 
$\{ \x: \x = {\Vm_{S_c}}_{(K)} \a \}$ to approximate the space $\{\x_{S_c}: \x_{S_c}^T \LL_{S_c} \x_{S_c} \leq   c \x_{S_c}^T \x_{S_c} \}$, the maximum error we suffer from is $c \x_{S_c}^T \x_{S_c} / \lambda^{(S_c)}_{K+1}$, which is solved by the following optimization problem,
\begin{eqnarray*}
&&	\max_{\x}  \left\|  \x - \Vm_{S_c} \Vm_{S_c}^T \x \right\|_2^2
	\\
&&	{\rm subject~to:}~   \x^T \LL_{S_c} \x \leq   c  \x^T \x.
\end{eqnarray*}
In other words, in each local set, the maximum error to represent $ \{\x_{S_c}: \x_{S_c}^T \LL_{S_c} \x_{S_c} \leq  \lambda_K  \x^T \x - \x^T \LL_{\rm cut} \x   \}$ is 
$(\lambda_K  \x^T \x - \x^T \LL_{\rm cut} \x)/\lambda^{(S_c)}_{K+1}$. Since all the local sets share the variation budget of $\lambda_K \x^T \x$ together, the maximum error we suffer from is 
\begin{displaymath}
\epsilon_{\rm par}  = 
\frac{ \x^T \left( \lambda_K \Id -  \LL_{\rm cut}  \right) \x }{ \min_{S_c} \lambda^{(S_c)}_{K+1} \left\| \x \right\|_2^2},
\end{displaymath}
which depends on the property of graph partitioning.

In Corollary~\ref{cor:sparse}, we have shown that we need at most $2 L \left\| \Delta \x_{\PC} \right\|_0$ local sets to represent the piecewise-constant template of $\x$. Since we use at most $K$ eigenvectors in each local set, we obtain the results in Theorem~\ref{thm:PPL}.
\end{proof}

\end{document}